\documentclass[]{aastex}
\usepackage[onecolumn]{emulateapj5}
\usepackage{color,psfig,natbib,ccaption}

\captiondelim{--}  %get rid of default caption deliminator {:} in ccaption
\newcommand{\uv}{\mbox{$u$-$v$}}
\newcommand{\Ho}{\mbox{$H_0$}}
\newcommand{\etal}{\mbox{et~al.}}
\newcommand{\no}{\mbox{$n_{e  0}$}}

\newcommand{\dTo}{\mbox{$\Delta T_0$}}
\newcommand{\Xo}{\mbox{$S_{x  0}$}}
\newcommand{\Xodet}{\mbox{$S_{x  0}^{det}$}}

\newcommand{\Teo}{\mbox{$T_{e  0}$}}
\newcommand{\Lamo}{\mbox{$\Lambda_{e \mbox{\tiny H} 0}$}}
\newcommand{\LameH}{\mbox{$\Lambda_{e \mbox{\tiny H}}$}}
\newcommand{\LameHo}{\Lamo}
\newcommand{\LameHodet}{\mbox{$\Lambda_{e \mbox{\tiny H} 0}^{det}$}}
\newcommand{\Lambol}{\mbox{$\Lambda_{\mbox{\tiny bol}}$}}

\newcommand{\Om}{\mbox{$\Omega_M$}}

\newcommand{\Ol}{\mbox{$\Omega_\Lambda$}}

\newcommand{\kms}{\mbox{km s$^{-1}$}}
\newcommand{\ksM}{\mbox{km s$^{-1}$ Mpc$^{-1}$}}
\newcommand{\cgsunits}{\mbox{$\frac{\mbox{erg s}^{-1} 
	\mbox{ cm}^{-2}}{\mbox{cnt s}^{-1}}$}}

\newcommand{\kB}{\mbox{$k_{\mbox{\tiny B}}$}}

\newcommand{\sigT}{\mbox{$\sigma_{\mbox{\tiny T}}$}}
\newcommand{\Tcmb}{\mbox{$T_{\mbox{\tiny CMB}}$}}
\newcommand{\muH}{\mbox{$\mu_{\mbox{\tiny H}}$}}
\newcommand{\nH}{\mbox{$n_{\mbox{\tiny H}}$}}
\newcommand{\NH}{\mbox{$N_{\mbox{\tiny H}}$}}
\newcommand{\Da}{\mbox{$D_{\!\mbox{\tiny A}}$}}

\newcommand{\uJybm}{\mbox{$\mu$Jy beam$^{-1}$}}
\newcommand{\mJybm}{\mbox{mJy beam$^{-1}$}}

\newcommand{\Lx}{\mbox{$L_{\!\mbox{\tiny X}}$}}

\newcommand{\gsim}{\gtrsim}
\newcommand{\lsim}{\lesssim}

\newcommand{\rosat}{{\it ROSAT}}
\newcommand{\einstein}{{\it Einstein}}

\shorttitle{The Cosmic Distance Scale from SZE}
\shortauthors{Reese et~al.}

\begin{document}

\title{Determining the Cosmic Distance Scale from Interferometric
Measurements of the Sunyaev-Zel'dovich Effect}

%\shorttitle{Determining the Cosmic Distance Scale from SZE}

\author{Erik~D.~Reese\altaffilmark{1,2,3},
        John~E.~Carlstrom\altaffilmark{1},
        Marshall~Joy\altaffilmark{4},
        Joseph~J.~Mohr\altaffilmark{5},
        Laura~Grego\altaffilmark{6},
        William~L.~Holzapfel\altaffilmark{2}
}

%\email{reese@cfpa.berkeley.edu}
%\email{jc@hyde.uchicago.edu}
%\email{marshall.joy@msfc.nasa.gov}
%\email{jmohr@astro.uiuc.edu}
%\email{lgrego@cfa.harvard.edu}
%\email{swlh@cfpa.berkeley.edu}

\altaffiltext{1}{Department of Astronomy \& Astrophysics, University
of Chicago, 5640 S.\ Ellis Ave., Chicago, IL 60637}

\altaffiltext{2}{Department of Physics, University of California,
Berkeley, CA 94720}

\altaffiltext{3}{Chandra Fellow}

\altaffiltext{4}{Space Science Laboratory, SD50, NASA Marshall Space
Flight Center, Huntsville, AL 35812}

\altaffiltext{5}{Department of Astronomy and Department of Physics,
University of Illinois, Urbana, IL 61801}

\altaffiltext{6}{Harvard-Smithsonian Center for Astrophysics, 60
Garden St., Cambridge, MA 02138}

\begin{abstract}
We determine the distances to 18 galaxy clusters with redshifts
ranging from $z \sim 0.14$ to $z \sim 0.78$ from a maximum likelihood
joint analysis of 30 GHz interferometric Sunyaev-Zel'dovich effect
(SZE) and X-ray observations.  We model the intracluster medium (ICM)
using a spherical isothermal $\beta$ model.  We quantify the
statistical and systematic uncertainties inherent to these direct
distance measurements, and we determine constraints on the Hubble
parameter for three different cosmologies.  These distances imply a
Hubble constant of $60 ^{+4}_{-4\phn}\, ^{+13}_{-18}$ \ksM\ for an
$\Om = 0.3$, $\Ol = 0.7$ cosmology, where the uncertainties correspond
to statistical followed by systematic at 68\% confidence.  With a
sample of 18 clusters, systematic uncertainties clearly dominate.  The
systematics are observationally approachable and will be addressed in
the coming years through the current generation of X-ray satellites
(Chandra \& XMM-Newton) and radio observatories (OVRO, BIMA, \& VLA).
Analysis of high redshift clusters detected in future SZE and X-ray
surveys will allow a determination of the geometry of the universe
from SZE determined distances.
\end{abstract}


%
% Begin Body
%

%%%%%%%%%%%%%%%%%%%%%%%%%%%%%%%%%%%%%%%%%%
%% I.   Introduction
%%%%%%%%%%%%%%%%%%%%%%%%%%%%%%%%%%%%%%%%%%

\section{Introduction}
\label{sec:intro}

Analysis of Sunyaev-Zel'dovich effect (SZE) and X-ray data provides a
method of directly determining distances to galaxy clusters at any
redshift.  Clusters of galaxies contain hot ($\kB T_e \sim 10$ keV)
gas, known as the intracluster medium (ICM), trapped in their
potential wells.  Cosmic microwave background (CMB) photons passing
through a massive cluster interact with the energetic ICM electrons
with a $\tau \approx 0.01$ probability.  This inverse Compton
scattering preferentially boosts the energy of a scattered CMB photon
causing a small ($\lsim 1$ mK) distortion in the CMB spectrum, known
as the Sunyaev-Zel'dovich effect \citep{sunyaev1970, sunyaev1972}.
The SZE appears as a decrement for frequencies $\lsim 218$ GHz and as
an increment for frequencies $\gsim 218$ GHz.  The SZE is proportional
to the pressure integrated along the line of sight $\Delta T \propto
\int n_e T_e d\ell$.  X-ray emission from the ICM has a different
dependence on the density $S_{\mbox{\tiny X}} \propto \int n_e^2
\LameH d\ell$, where \LameH\ is the X-ray cooling function.  Taking
advantage of the different density dependencies and with some
assumptions about the geometry of the cluster, the distance to the
cluster may be determined.  SZE and X-ray determined distances are
independent of the extragalactic distance ladder and provide distances
to high redshift galaxy clusters.  The promise of direct distances has
been one of the primary motivations for SZE observations.

In the last decade, SZE detections have become routine due to advances
in both instrumentation and observational strategy.  Recent high
signal-to-noise ratio detections have been made with single dish
observations at radio wavelengths \citep{birkinshaw1994, herbig1995,
myers1997, hughes1998}, millimeter wavelengths \citep{holzapfel1997b,
holzapfel1997, pointecouteau1999, pointecouteau2001} and submillimeter
wavelengths \citep{lamarre1998, komatsu1999}.  Interferometric
observations at centimeter wavelengths are now routinely producing
high quality images of the SZE \citep{jones1993, grainge1993,
carlstrom1996, carlstrom1998, saunders1999, grainge1999, reese2000,
grego2000, carlstrom2000b, grego2001, jones2001}.

In this paper, we present a maximum likelihood joint analysis of our
30 GHz interferometric SZE observations with archival {\it R\"{o}ntgen
Satellite} (\rosat) X-ray imaging observations.  Cluster X-ray
temperatures, metallicity, and H I column densities are taken from the
literature.  The intracluster medium (ICM) is modeled as a spherical
isothermal $\beta$ model.  We refine the analysis technique described
in \citet{reese2000} and apply it to a sample of 18 clusters for which
we determine distances.  These distances are then used to measure the
Hubble constant.  This is the largest homogeneously analyzed sample of
SZE clusters with distance determinations thus far.  To date, there
are about 20 published estimates of \Ho\ based on combining X-ray and
SZE data for individual clusters \citep[see][for a review and compiled
distances]{birkinshaw1999}.  Most notably, those results include one
sample consisting of 7 nearby ($z<0.1$) galaxy clusters
\citep{mason2001, mason1999, myers1997} and a sample of 5 intermediate
redshift ($0.14<z<0.3$) clusters \citep{jones2001}.

The cluster sample selection for this paper is discussed in
\S\ref{sec:sample}.  The centimeter wave SZE system and
interferometric SZE data are described in \S\ref{subsec:sze_obs}.  A
brief overview the \rosat\ X-ray cluster data is given in
\S\ref{subsec:xray_obs}.  The analysis method, including uncertainty
estimation, is outlined in \S\ref{sec:method} along with the model
fitting results.  Distances and our determination of the Hubble
parameter appear in \S\ref{sec:dist_Ho}.  Sources of possible
systematic uncertainties are discussed in \S\ref{subsec:sys_sources}.
Section~\ref{sec:disc_concl} contains a discussion of the results and
future prospects.  Throughout this work, the galaxy cluster Cl
$0016+16$ ($z=0.546$) will be used as the example cluster to
illustrate both the analysis method and general results.  All
uncertainties are 68.3\% confidence unless explicitly stated
otherwise.

%%%%%%%%%%%%%%%%%%%%%%%%%%%%%%%%%%%%%%%%%%
%% II.  Cluster Sample
%%%%%%%%%%%%%%%%%%%%%%%%%%%%%%%%%%%%%%%%%%

\section{Cluster Sample}
\label{sec:sample}

The determination the Hubble parameter requires a large sample of
galaxy clusters free of selection effects.  For example, clusters
selected by X-ray surface brightness will preferentially include
clusters elongated along the line of sight.  A spherical analysis will
underestimate the line of sight length of the cluster, causing the
derived Hubble parameter to be biased low.  In theory, selecting by
X-ray luminosity, \Lx, alleviates the selection bias problem.  In
practice, this is complicated by the fact that X-ray surveys are
surface brightness limited; clusters just at the detection limit that
are elongated along the line of sight will be detected while clusters
just at the detection limit that are instead extended in the plane of
the sky will be missed.  Staying well above the detection limit of the
survey will alleviate this potential pitfall.  Observational
considerations for cluster sample selection include the declinations
of the clusters, the size (redshift) of the cluster, possible radio
point sources in the cluster field, and SZE brightness for which we
use $\Lx$ as an indicator.  The Owens Valley Radio Observatory (OVRO)
and Berkeley-Illinois-Maryland Association (BIMA) interferometers have
been used to observe known X-ray clusters with $z\gsim 0.14$,
declination $\gsim -15^\circ$, and $L_x \gsim 5\times 10^{44}\
h_{50}^{-2}$ erg s$^{-1}$ (0.3-3.5 keV band for \einstein\ and 0.1-2.4
keV band for \rosat).  In addition, short, preliminary observations of
many clusters are also performed to investigate possible point sources
in the field.

The OVRO/BIMA SZE imaging project initially chose targets from the
limited number of known X-ray bright clusters.  With the publishing of
X-ray cluster surveys, the OVRO/BIMA SZE imaging project chose targets
from three X-ray catalogs of galaxy clusters: the {\it Einstein
Observatory} Extended Medium Sensitivity Survey, EMSS
\citep{gioia1990, stocke1991, gioia1994, maccacaro1994}; the \rosat\
X-ray Brightest Abell Clusters, XBACs \citep{ebeling1996,
ebeling1996b}; and the \rosat\ Brightest Cluster Sample, BCS
\citep{ebeling2000a, crawford1999, ebeling1998, ebeling1997}.  We have
also recently included two more recent \rosat\ samples of distant
massive clusters to our cluster selection database: the Wide Angle
\rosat\ Pointed Survey, WARPS \citep{fairley2000, ebeling2000b,
jones1998, scharf1997}; and the MAssive Cluster Survey, MACS
\citep{ebeling2001}.  So far, we have high S/N detections in 21
clusters with redshifts $z > 0.45$.

The distance calculation requires three data sets: SZE, X-ray imaging,
and X-ray spectroscopic data.  We have obtained high signal-to-noise
detections of the SZE in 45 galaxy clusters.  The subsample of these
clusters that also have high signal-to-noise X-ray imaging data and
published electron temperatures contains 18 galaxy clusters.
Table~\ref{tab:sample} summarizes the redshifts and X-ray luminosities
for each galaxy cluster in our sample.

%%%%%
%  Table: Cluster Sample
%%%%%

%%Cluster Sample  Info
%%
%% z, L_X Te
%%
\begin{deluxetable}{llccl}%{llccll}
%\singlespace
%\footnotesize
%\rotate
\tablewidth{0pt}
%\tablenum{}
\tablecolumns{6}
%\tableheadfrac{}
\tablecaption{Cluster Sample \label{tab:sample}}
\tablehead{
\colhead{} &
\colhead{} &
\colhead{\Lx\tablenotemark{a}} &
\colhead{band} &
\colhead{}
\\
\colhead{cluster} &
\colhead{redshift} &
\colhead{($10^{44} h_{50}^{-2}$ erg s$^{-1}$)} &
\colhead{(keV)} &
\colhead{references- $z$; \Lx}%; $T_e$}
}
\startdata
MS $1137.5+6625$
	& 0.784
	& \phn 5.4
	& $0.3-3.5$
	& D99;GL94\\
MS $0451.6-0305$
	& 0.550
	& 20.0
	& $0.3-3.5$
	& GL94;GL94\\
Cl $0016+16$
	& 0.546
	& 14.6
	& $0.3-3.5$
	& DG92;GL94\\
RX J$1347.5-1145$
	& 0.451
	& 73.0
	& $0.1-2.4$
	& S95;S97\\
Abell \phn370
	& 0.374
	& 11.7\tablenotemark{b}
	& $0.1-2.4$
	& M88;AE99\\
MS $1358.4+6245$
	& 0.327
	& 10.6
	& $0.3-3.5$
	& GL94;GL94\\
Abell 1995 
	& 0.322
	& 13.4
	& $0.1-2.4$
	& P00;B00\\
Abell \phn611
	& 0.288
	& \phn 8.6
	& $0.1-2.4$
	& C95;B00\\
Abell \phn697
	& 0.282
	& 19.2
	& $0.1-2.4$	
	& C95;B00,E98\\
Abell 1835
	& 0.252
	& 32.6
	& $0.1-2.4$	
	& SR99;B00,E98\\
Abell 2261	
	& 0.224
	& 20.6
	& $0.1-2.4$	
	& C95;B00\\
Abell \phn773	
	& 0.216
	& 12.1
	& $0.1-2.4$	
	& SR99;B00,E98\\
Abell 2163 
	& 0.202	
	& 37.5
	& $0.1-2.4$	
	& SR99;E96\\
Abell \phn520 
	& 0.202
	& 14.5
	& $0.1-2.4$	
	& GL94;B00,E98\\
Abell 1689	
	& 0.183
	& 20.7
	& $0.1-2.4$	
	& SR91;E96\\
Abell \phn665	
	& 0.182
	& 15.7
	& $0.1-2.4$	
	& SR91;B00\\
Abell 2218	
	& 0.171	
	& \phn 8.2
	& $0.1-2.4$	
	& L92;B00\\
Abell 1413
	& 0.142
	& 10.9
	& $0.1-2.4$	
	& SR99;B00,E98\\
\enddata
\tablenotetext{a}{Computed for a flat $\Om = 1$ universe.}
\tablenotetext{b}{Converted the 2--10 keV flux in AE99 to the 0.1--2.4
keV band (approximate factor of 0.9 determined from cooling function
calculation).} 
\tablenotetext{\phn}{\\
	REF:
	AE99-\citealt{arnaud1999};
	B00-\citealt{bohringer2000};
	C95-\citealt{crawford1995};
	D99-\citealt{donahue1999};
	DG92-\citealt{dressler1992};
	E96-\citealt{ebeling1996};
	E98-\citealt{ebeling1998};
	GL94-\citealt{gioia1994};
	L92-\citealt{leborgne1992};
	M88-\citealt{mellier1988};
	P00-\citealt{patel2000};
	S95-\citealt{schindler1995};
	S97-\citealt{schindler1997};
	SR91-\citealt{struble1991};
	SR99-\citealt{struble1999};
}
%\tablecomment{}
\end{deluxetable}

%%%%%%%%%%%%%%%%%%%%%%%%%%%%%%%%%%%%%%%%%%
%% III. Data
%%%%%%%%%%%%%%%%%%%%%%%%%%%%%%%%%%%%%%%%%%

\section{Data}
\label{sec:obs}\label{sec:data}

Here we briefly describe the SZE and X-ray observations and data
reduction.  Table~\ref{tab:data} summarizes the observation times for
both the SZE and \rosat\ observations of each cluster in our sample.
The SZE observation times are the total on-source integration times
for the interferometric SZE data used in this analysis.  The \rosat\
observation times are the total livetimes of the pointings used in
this analysis.

%%%%%
% Table: Cluster Data
%%%%%
%%
%%Data Info
%%
%%
%%
%%

%% Thesis table  : below is the more detailed SZE summary version

\begin{deluxetable}{lrrrrrrrrrrrr}

%\singlespace
%\footnotesize
%\rotate
\tablewidth{0pt}
%\tablenum{}
\tablecolumns{12}
%\tableheadfrac{}
\tablecaption{Cluster Data\label{tab:data}}
\tablehead{
\colhead{} &
\multicolumn{9}{c}{Interferometric SZE Data} &
\colhead{} &
\multicolumn{2}{c}{\rosat\ Data}
\\
\cline{2-9}\cline{12-13}
\colhead{} &
\multicolumn{4}{c}{OVRO (hr)} &
\colhead{} &
\multicolumn{4}{c}{BIMA (hr)} &
\colhead{} &
\colhead{PSPC} &
\colhead{HRI}
\\
\cline{2-5}\cline{7-10}
\colhead{cluster} &
\colhead{1994} &
\colhead{1995} &
\colhead{1996} &
\colhead{1998} &
\colhead{} &
\colhead{1996} &
\colhead{1997} &
\colhead{1998} &
\colhead{2000} &
\colhead{} &
\colhead{(ks)} &
\colhead{(ks)}
}
\startdata
MS1137 %MS $1137.5+6625$
	& \nodata & \nodata & \nodata & \nodata &
	& \nodata & 40 & 48 & \nodata &
	& \nodata & 99.1
\\
MS0451 %MS $0451.6-0305$
	& \nodata & \nodata &  30  & \nodata &
	& \nodata & \nodata & \nodata & \nodata &
	& 15.4	& 45.9
\\
Cl0016 %Cl $0016+1609$
	& 87	& 13	& \nodata & \nodata &
	& 29\tablenotemark{a} 	& 8	& \nodata & \nodata &
	& 41.6	& 70.2
\\
R1347 %RX J$1347.5-1145$
	& \nodata & \nodata & \nodata & \nodata &
	& \nodata & \nodata & \nodata & 20.0 &
	& \nodata & 36.1
\\
A370 %Abell \phn370
	& \nodata & \nodata & 33 & \nodata &
	& \nodata & 26 & \nodata & \nodata &
	& \nodata & 31.9
\\
MS1358 %MS $1358.4+6245$
	& \nodata & \nodata & 9 & 7 &
	& \nodata & \nodata & 70 & \nodata &
	& 22.1 & 29.2
\\
A1995 %Abell 1995 	
	& \nodata & \nodata & 58 & \nodata &
	& \nodata & \nodata & 50 & \nodata &
	& \nodata & 37.6
\\
A611 %Abell \phn611
	& \nodata & 45 & 12 & \nodata &
	& \nodata & \nodata & \nodata & \nodata &
	& \nodata & 17.2
\\
A697 %Abell \phn697
	& \nodata & 47 & \nodata & \nodata &
	& \nodata & \nodata & \nodata & \nodata &
	& \nodata & 27.8
\\
A1835 %Abell 1835	
	& \nodata & \nodata & \nodata & \nodata &
	& \nodata & \nodata & 27 & \nodata &
	& 8.5 & 2.8
\\
A2261 %Abell 2261	
	& \nodata & 40 & \nodata & \nodata &
	& \nodata & \nodata & 3 & \nodata &
	& \nodata & 16.1
\\
A773 %Abell \phn 773	
	& 57 & 9 & \nodata & \nodata &
	& 5\tablenotemark{a} & \nodata & 18 & \nodata &
	& \nodata & 16.5
\\
A2163 %Abell 2163 	
	& \nodata & \nodata & 25 & 12 &
	& 2 & 10 & 11 & \nodata &
	& 11.7 & 35.8
\\
A520 %Abell \phn520
	& \nodata & \nodata & \nodata & 7 &
	& 13\tablenotemark{a} & 23 & 20 & \nodata &
	& 4.7 & 12.6
\\
A1689 %Abell 1689	
	& \nodata & 26 & \nodata & \nodata &
	& \nodata & 16 & \nodata & \nodata &
	& 13.5 & 22.5
\\
A665 %Abell \phn665
	& \nodata & \nodata & \nodata & \nodata &
	& 38\tablenotemark{a} & 24 & \nodata & \nodata &
	& 37.0 & 98.3
\\
A2218 %Abell 2218	
	& \nodata & 64 & 6 & \nodata &
	& 20\tablenotemark{a} & 12 & \nodata & \nodata &
	& 42.5 & 35.5
\\
A1413 %Abell 1413	
	& \nodata & \nodata & \nodata & \nodata &
	& 11 & 17 & \nodata & \nodata &
	& 7.5 & 18.6
\enddata
\tablenotetext{a}{Contains 1996 BIMA data with delay loss problem;
data only used to make images and not in the analysis.}

\end{deluxetable}

%%%%%%%%%%
%% III.1 Interferometric SZE Data
%%%%%%%%%%
\subsection{Interferometric SZE Data}
\label{subsec:sze_obs}

The extremely low systematics of interferometers make them well suited
to study the weak SZE signal.  A unique feature of interferometers is
their ability to separate the diffuse, negative SZE emission from
small scale, positive point source emission through the spatial
filtering of the interferometer.  Interferometers also provide a well
defined angular and spectral filter, which is important in the
analysis of the SZE data discussed in \S\ref{sec:method}.

\subsubsection{Centimeter-Wave System and Observing Strategy}
\label{subsubsec:cmsystem}

Over the past several summers, we outfitted the
Berkeley-Illinois-Maryland Association (BIMA) millimeter array in Hat
Creek, California, and the Owens Valley Radio Observatory (OVRO)
millimeter array in Big Pine, California, with centimeter wavelength
receivers.  Our receivers use cooled ($\sim 10$ K) High Electron
Mobility Transistor (HEMT) amplifiers \citep{pospieszalski1995}
operating over 26-36 GHz with characteristic receiver temperatures of
$T_{rx}\sim $11-20 K over the 28-30 GHz band used for the observations
presented here.  When combined with the BIMA or OVRO systems, these
receivers obtain typical system temperatures scaled to above the
atmosphere of $T_{sys}\sim 35$-$45$ K.  Most telescopes are placed in
a compact configuration to maximize sensitivity on angular scales
subtended by distant clusters ($\sim$ 1\arcmin), but telescopes are
always placed at longer baselines for simultaneous detection of point
sources.  Every half hour we observe a bright quasar, commonly called
a phase calibrator, for about two minutes to monitor the system phase
and gain.  The total integration time for each cluster field is given
in Table~\ref{tab:data} for both OVRO and BIMA.

An interferometer samples the Fourier transform of the sky brightness
distribution multiplied by the primary beam rather than the direct
image of the sky.  The SZE data files include the positions in the
Fourier domain, which depend on the arrangement of the telescopes in
the array and the declination of the source, the real and imaginary
Fourier components, and a measure of the noise in the real and
imaginary components.  The Fourier conjugate variables to right
ascension and declination are commonly called $u$ and $v$,
respectively, and the Fourier domain is commonly referred to as the
\uv\ plane.  The real and imaginary Fourier component pairs as a
function of $u$ and $v$ are called visibilities.

The finite size of each telescope dish results in an almost Gaussian
response pattern, known as the primary beam.  The product of the
primary beam and the sky brightness distribution is equivalent to a
convolution in the Fourier domain.  The primary beams are measured
using holography data for both OVRO and BIMA.  The main lobe of the
primary beams are well fit by a Gaussian with a full width at half
maximum (FWHM) of $4\arcmin\! .2$ for OVRO and $6\arcmin\! .6$ for
BIMA at 28.5 GHz.  However, we use the measured primary beam profiles
for our analysis.

The primary beam sets the field of view.  The effective resolution,
called the synthesized beam, depends on the sampling of the \uv\ plane
and is therefore a function of the configuration of the telescopes and
the declination of the source.  The cluster SZE signal is largest on
the shortest baselines (largest angular scales).  The shortest
possible baseline is set by the diameter of the telescopes, $D$.  Thus
the system is not sensitive to angular scales larger than about
$\lambda/2 D$, which is $\sim 2\arcmin\! .8$ for BIMA observations and
$\sim 1\arcmin\! .7$ for OVRO observations.  The compact configuration
used for our observations yields significant SZE signal at these
angular scales, but filters out signal on larger angular scales.
Because of the spatial filtering by the interferometer, it is
necessary to fit models directly to the data in the \uv\ plane, rather
than to the deconvolved image.

Interferometers simultaneously measure both the cluster signal and the
point sources in the field.  The SZE signal is primarily present in
the short baseline data while the response of an interferometer to a
point source is independent of the baseline.  Therefore, observations
with a range of baselines allow us to separate the extended cluster
emission from point source emission.  We show an example of this after
first presenting details of deconvolved 30 GHz images \citep[see ][
for additional examples]{jones1993, carlstrom2000b}.

\subsubsection{Data Reduction}
\label{subsubsec:szedata_reduce}

The data are reduced using the MIRIAD \citep{sault1995} software
package at BIMA and using MMA \citep{scoville1993} at OVRO.  In both
cases, data are excised when one telescope is shadowed by another,
when cluster data are not straddled by two phase calibrators, when
there are anomalous changes in instrumental response between
calibrator observations, or when there is spurious correlation.  For
absolute flux calibration, we use observations of Mars and adopt the
brightness temperature from the Rudy (1987)\nocite{rudy1987} Mars
model.  For observations not containing Mars, calibrators in those
fields are bootstrapped back to the nearest Mars calibration (see
\citealp{grego2001} for more details).  The observations of the phase
calibrators over each summer give us a summer-long calibration of the
gains of the BIMA and OVRO interferometers.  They both show very
little gain variation, changing by less than 1\% over a many-hour
track, and the average gains remain stable from day to day.  In fact,
the gains are stable at the $\sim 1$\% level over a period of months.

\subsubsection{Data Visualization: 30 GHz Images}
\label{subsubsec:sze_visualize}

Here we present deconvolved images of our 30 GHz interferometric
observations.  However, we stress that these images are made to
demonstrate the data quality.  The model fitting is performed in the
Fourier plane, where the noise characteristics of the data and the
spatial filtering of the interferometer are well understood.  The SZE
and X-ray image overlays of Figure~\ref{fig:image} show that the
region of the cluster sampled by the interferometric SZE observations
and the X-ray observations is similar for the clusters in our sample.
In addition, the interferometer measures a range of angular scales,
which is not apparent from the images in Figure~\ref{fig:image}.
Images showing examples of our SZE data at varying resolutions appear
in \citet{carlstrom1996} for Cl0016 and \citet{carlstrom2000b} for
R1347.

Point sources are identified from SZE images created with DIFMAP
\citep{pearson1994} using only the long baseline data ($\gsim 2000\
\lambda$) and natural weighting ($\sigma^{-2}$ weight).  Approximate
positions and fluxes for each point source are obtained from this
image and used as inputs for the model fitting discussed in
\S\ref{subsec:model_fit}.  The data are separated by observatory,
frequency, and year to allow for temporal and spectral variability of
the point source flux.  The positions and fluxes of the detected point
sources from the model fitting are summarized in
Table~\ref{tab:pt_sources}.  Also listed are the corresponding 1.4 GHz
fluxes for these sources from the NRAO VLA Sky Survey (NVSS)
\citep{condon1998} and the 5 GHz and 15 GHz fluxes for sources in the
three cluster fields surveyed by \citet{moffet1989}.  The uncertainty
in the positions of the point sources is $\sim \pm 3\arcsec$ at 68.3\%
confidence based on model fits of point sources described in
\S\ref{subsec:model_fit}.  Figure~\ref{fig:nvss_pt} shows the 30 GHz
high resolution ($\geq 2000\ \lambda$) maps (color scale) with NVSS 1.4
GHz contours.  The color scale wedge above each image shows the range
in the map in units of mJy beam$^{-1}$.  Contours are multiples of
twice the NVSS rms (rms $\sim 0.45$ \mJybm).  The FWHM of the 30 GHz
synthesized beam is shown in the lower left hand corner of each panel
and the 45\arcsec\ FWHM beam of the NVSS survey is shown in the lower
right hand corner of each panel.  Table~\ref{tab:imstat} summarizes
the sensitivity and the FWHM of the synthesized beam of the high
resolution maps used to find point sources in each field and shown in
Figure~\ref{fig:nvss_pt}.

%%%%%
% Table: Radio Point Sources
%%%%%

%%Radio Point Sources
%%
%%Field RA DEC F_30 F_28.5 F_15 F_5 F_1.4
%%
\begin{deluxetable}{lccccccc}

\tablewidth{0pt}
%\tablenum{}
\tablecaption{Radio Point Sources \label{tab:pt_sources}}
\tablehead{
  \colhead{} &
  \colhead{R.A.\tablenotemark{a} (J2000)} &
  \colhead{Dec.\tablenotemark{a} (J2000)} &
  \colhead{$F_{30.0}$} & 
  \colhead{$F_{28.5}$} &
  \colhead{$F_{15}$\tablenotemark{b}} & 
  \colhead{$F_{5}$\tablenotemark{b}} & 
  \colhead{$F_{1.4}$\tablenotemark{c}}\\
  \colhead{Field} &
  \colhead{(h\ \  m\ \  s)} & 
  \colhead{(d\ \  \arcmin\ \ \arcsec )} & 
  \colhead{(mJy)} &
  \colhead{(mJy)} & 
  \colhead{(mJy)} & 
  \colhead{(mJy)} & 
  \colhead{(mJy)}
}
\tablecolumns{8}
\startdata
MS1137  & \nodata & \nodata &
	\nodata & \nodata 
	& \nodata & \nodata & \nodata\\	
MS0451  & $04$ $54$ $22.1$ & $-03$ $01$ $25$ &
	$\phn 1.41^{+0.26}_{-0.26}$  & $\phn 1.86 ^{+0.26}_{-0.26}$
	& \nodata & \nodata & $\phn 14.9^{+0.7}_{-0.7}$ \\
Cl0016  & $00$ $18$ $31.1$ & $+16$ $20$ $45$ &
	\nodata & $\phn 9.11 ^{+1.97}_{-1.97}$
	& $25.0^{+1.5}_{-1.5}$ & $84.5^{+1.1}_{-1.1}$ & $269.3^{+8.1}_{-8.1}$\\
R1347   & $13$ $47$ $30.7$ & $-11$ $45$ $09$ &
	\nodata & $10.81 ^{+0.19}_{-0.19}$
	& \nodata & \nodata & $\phn 47.6 ^{+1.9}_{-1.9}$\\
A370	& $02$ $39$ $55.5$ & $-01$ $34$ $06$ &
	$\phn 0.84^{+0.09}_{-0.09}$ & $\phn 0.77^{+0.07}_{-0.07}$ 
	& \nodata & \nodata & $\phn 11.7^{+1.1}_{-1.1}$\\
MS1358  & $13$ $59$ $50.6$ & $+62$ $31$ $05$ &
	\nodata & $\phn 1.61^{+0.17}_{-0.17}$
	& \nodata & \nodata & \nodata\\	
A1995   & $14$ $53$ $00.5$ & $+58$ $03$ $19$ &
	$\phn 0.58^{+0.05}_{-0.05}$ & $\phn 0.58^{+0.04}_{-0.04}$ 
	& \nodata & \nodata & $\phn\phn 8.9 ^{+0.9}_{-0.9}$\\
A611	& \nodata & \nodata &
	\nodata & \nodata 
	& \nodata & \nodata & \nodata\\	
A697	& \nodata & \nodata &
	\nodata & \nodata 
	& \nodata & \nodata & \nodata\\	
A1835   & $14$ $01$ $02.0$ & $+02$ $52$ $42$ &
	\nodata & $\phn 2.76 ^{+0.14}_{-0.14}$
	& \nodata & \nodata & $\phn 41.4 ^{+1.9}_{-1.9}$\\	
	& $14$ $01$ $00.5$ & $+02$ $51$ $53$ &
	\nodata & $\phn 1.16 ^{+0.15}_{-0.15}$
	& \nodata & \nodata & \nodata\\	
A2261   & $17$ $22$ $17.1$ & $+32$ $09$ $14$ &
	$10.10 ^{+0.24}_{-0.24}$ & $10.80 ^{+0.24}_{-0.24}$ 
	& \nodata & \nodata & $\phn 24.3 ^{+1.6}_{-1.6}$\\	
A773	& \nodata & \nodata &
	\nodata & \nodata 
	& \nodata & \nodata & \nodata\\	
A2163   & $16$ $15$ $43.3$ & $-06$ $08$ $40$ &
	$\phn 1.44 ^{+0.12}_{-0.12}$ & $\phn 1.44 ^{+0.08}_{-0.08}$
	& \nodata & \nodata & \nodata\\	
A520	& $04$ $54$ $01.1$ & $+02$ $57$ $47$ &
	\nodata & $\phn 7.97 ^{+0.23}_{-0.23}$  
	& \nodata & \nodata & $\phn\phn 6.7 ^{+0.5}_{-0.5}$\\	
	& $04$ $54$ $17.0$ & $+02$ $55$ $32$ &
	\nodata & $\phn 1.01 ^{+0.10}_{-0.10}$ 
	& \nodata & \nodata & $\phn 15.3 ^{+1.1}_{-1.1}$\\	
	& $04$ $54$ $20.3$ & $+02$ $54$ $56$ &
	\nodata & $\phn 1.03 ^{+0.12}_{-0.12}$
	& \nodata & \nodata & $\phn 27.8 ^{+1.6}_{-1.6}$\\		
A1689   & $13$ $11$ $31.6$ & $-01$ $19$ $33$ &
	$\phn 1.33 ^{+0.10}_{-0.10}$  & $\phn 1.51 ^{+0.09}_{-0.09}$
	& \nodata & \nodata & $\phn 61.0 ^{+2.5}_{-2.5}$\\	
	& $13$ $11$ $30.1$ & $-01$ $20$ $37$ &
	$\phn 0.45 ^{+0.09}_{-0.09}$ & $\phn 0.42 ^{+0.09}_{-0.09}$  
	& \nodata & \nodata & $\phn 10.9 ^{+0.6}_{-0.6}$\\	
A665	& $08$ $31$ $30.9$ & $+65$ $52$ $35$ &
	\nodata & $\phn 4.83 ^{+0.28}_{-0.28}$  
	& $12.7^{+0.3}_{-0.3}$ & $25.7^{+1.6}_{-1.6}$
	& $\phn 31.1^{+1.3}_{-1.3}$\\	
A2218   & $16$ $35$ $22.1$ & $+66$ $13$ $23$ &
	$\phn 4.29 ^{+0.21}_{-0.21}$  & $\phn 4.43 ^{+0.20}_{-0.20}$
	& $\phn 5.0^{+0.6}_{-0.6}$ & $\phn 2.8^{+0.2}_{-0.2}$& \nodata\\
	& $16$ $35$ $47.7$ & $+66$ $14$ $46$ &
	$\phn 1.36 ^{+0.10}_{-0.10}$  & $\phn 1.59 ^{+0.11}_{-0.11}$  
	& $\phn 2.4^{+0.6}_{-0.6}$& $\phn 3.7^{+0.3}_{-0.3}$ & $\phn 18.0 ^{+1.8}_{-1.8}$\\	
	& $16$ $36$ $16.0$ & $+66$ $14$ $23$ &
	$\phn 2.41 ^{+0.29}_{-0.29}$  & $\phn 3.13 ^{+0.30}_{-0.30}$ 
	& $\phn 2.0^{+0.5}_{-0.5}$&$\phn 4.2^{+0.1}_{-0.1}$& $\phn 13.3 ^{+0.6}_{-0.6}$\\
A1413   & $11$ $55$ $08.7$ & $+23$ $26$ $17$ &
	\nodata & $\phn 2.01 ^{+0.23}_{-0.23}$
	& \nodata & \nodata & $\phn 28.8 ^{+1.3}_{-1.3}$\\	
\enddata
\tablenotetext{a}{Positions from SZE observations.}
\tablenotetext{b}{From Moffet \& Birkinshaw (1989)\nocite{moffet1989}.}
\tablenotetext{c}{From Condon \etal\ (1998)\nocite{condon1998}.}
%\tablecomment{}

\end{deluxetable}

%%%%%
% Figure 1: 
%%%%%
%%%%%
% NVSS overlay on high resolution 30 GHz images
%%%%%

%%
% Figure 1
%%

\begin{figure*}[!ptbh]
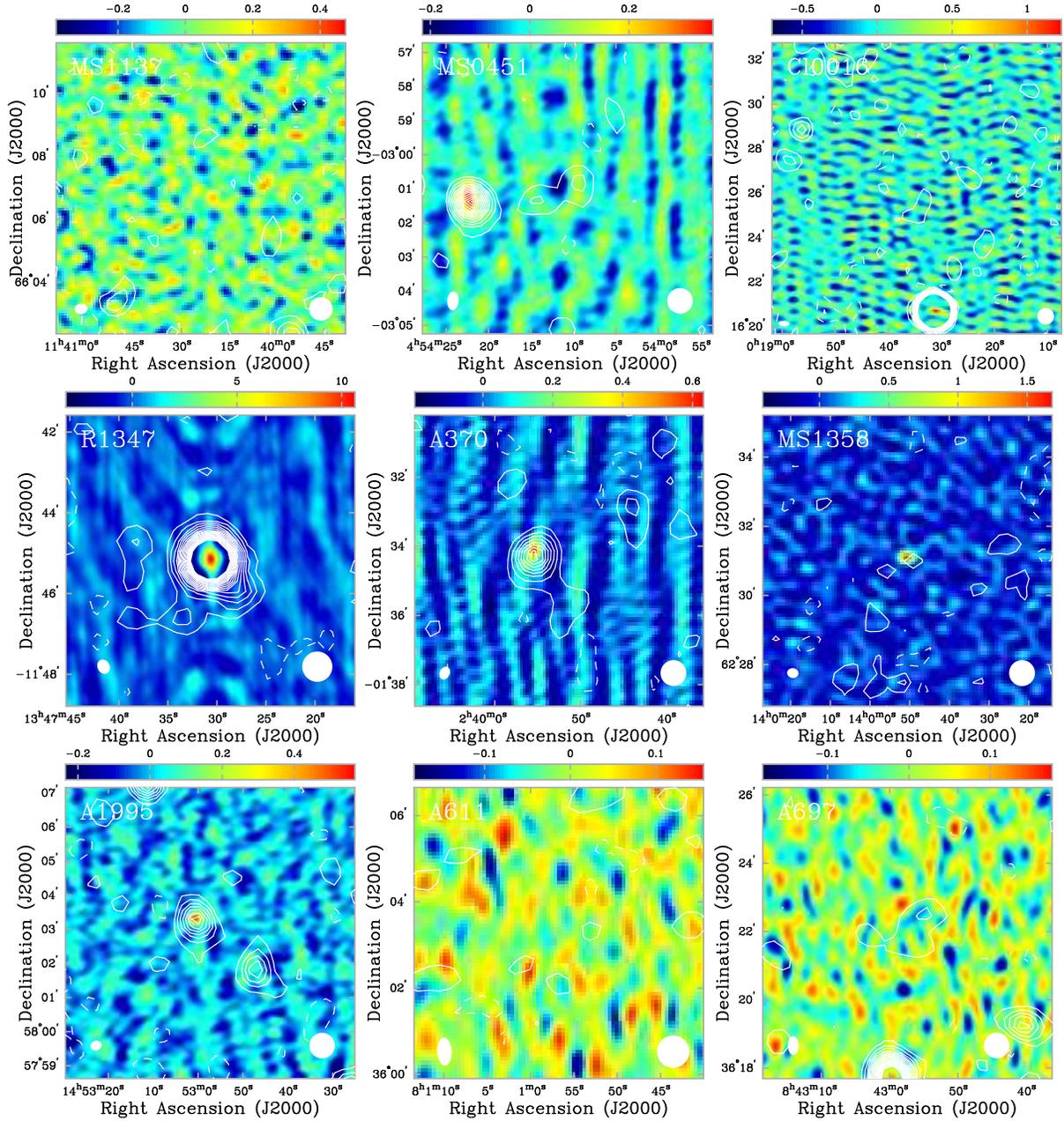

%\epsfxsize = 7.5 in
%%  \epsfxsize = 7.5 in
%%  \epsfysize = 6.0 in
%\centerline{\epsfbox{f?.eps}} 
\centerline{
  \psfig{figure=f1a_color.ps,height=2.2in}
  \psfig{figure=f1b_color.ps,height=2.2in}
  \psfig{figure=f1c_color.ps,height=2.2in}
}
\centerline{
  \psfig{figure=f1d_color.ps,height=2.2in}
  \psfig{figure=f1e_color.ps,height=2.2in}
  \psfig{figure=f1f_color.ps,height=2.2in}
}
\centerline{
  \psfig{figure=f1g_color.ps,height=2.2in}
  \psfig{figure=f1h_color.ps,height=2.2in}
  \psfig{figure=f1i_color.ps,height=2.2in}
}
\caption[High resolution 30 GHz and 1.4 GHz overlays]{High resolution
($\geq 2000\ \lambda$) 30 GHz (color scale) with NVSS 1.4 GHz contours.
The color scale wedge above each image shows the range in the 30 GHz map in
units of mJy beam$^{-1}$.  Contours are multiples of twice the NVSS
rms of $\sim 0.45$ mJy beam$^{-1}$.  The FWHM of the 30 GHz
synthesized beam is shown in the lower left hand corner of each panel
and the 45\arcsec\ FWHM beam of the NVSS survey is shown in the lower
right hand corner of each panel.  The 30 GHz image statistics are
summarized in Table~\ref{tab:imstat}.
% Bogus line spaces so get rid of 1 line of text below this figure
% (not sure why !p is not working...)
\\
\phn 
\\
}
\label{fig:nvss_pt}
\end{figure*}

\begin{figure*}[!pthb]
\centerline{
  \psfig{figure=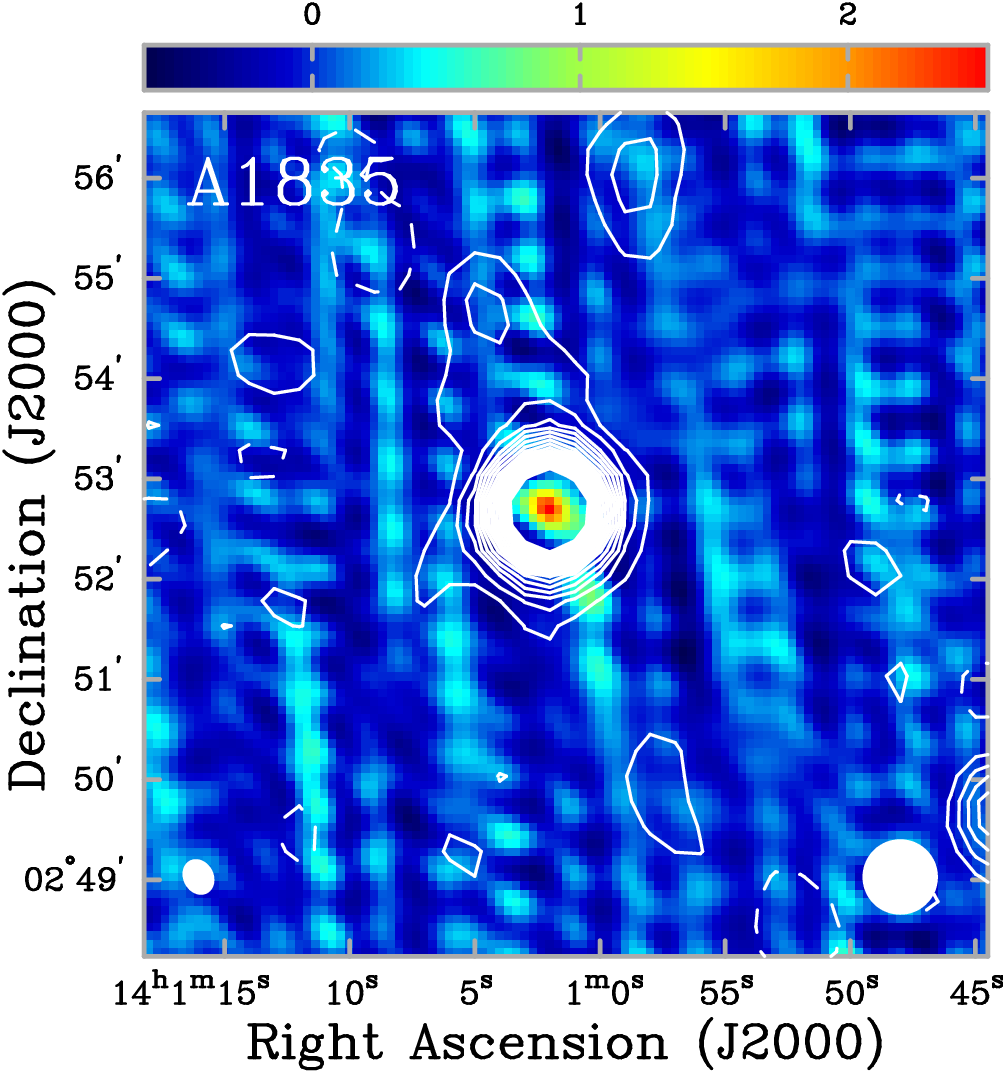,height=2.2in}
  \psfig{figure=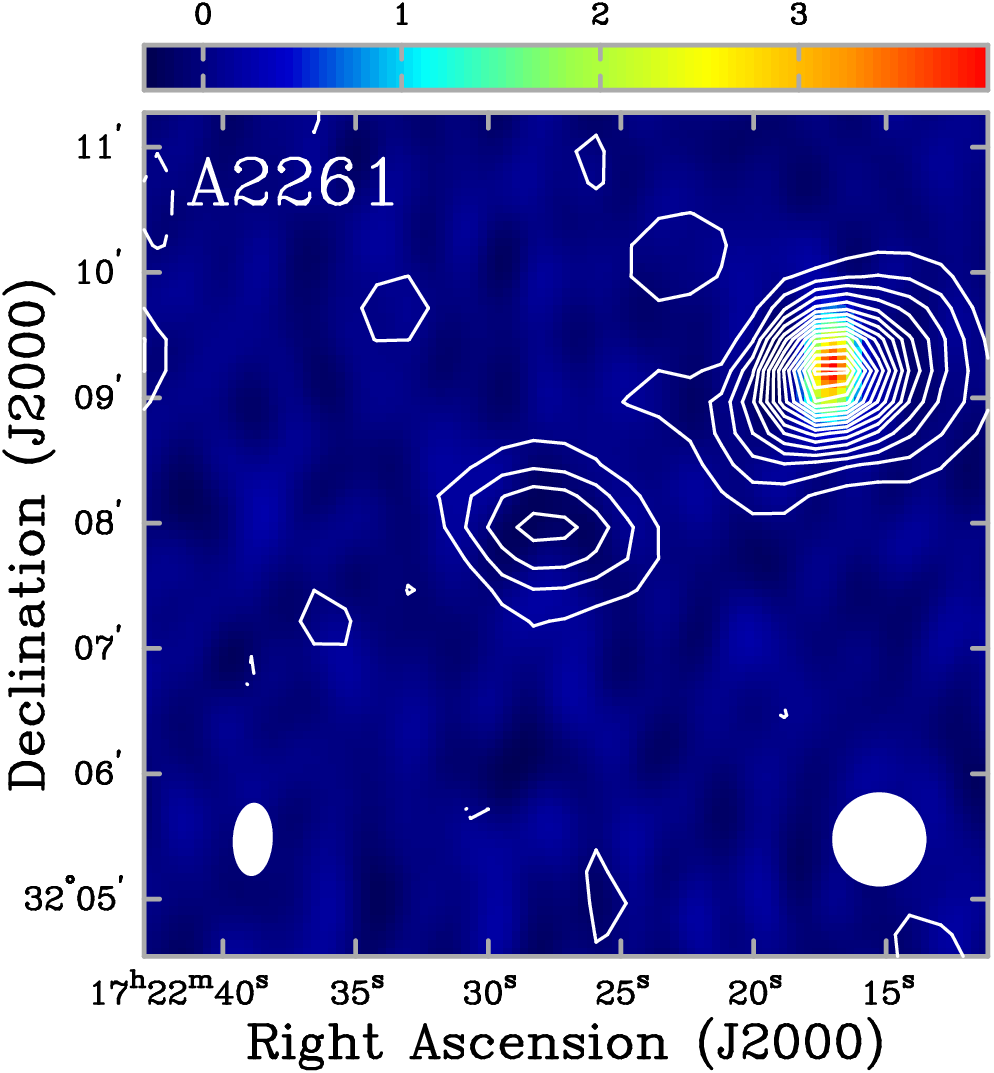,height=2.2in}
  \psfig{figure=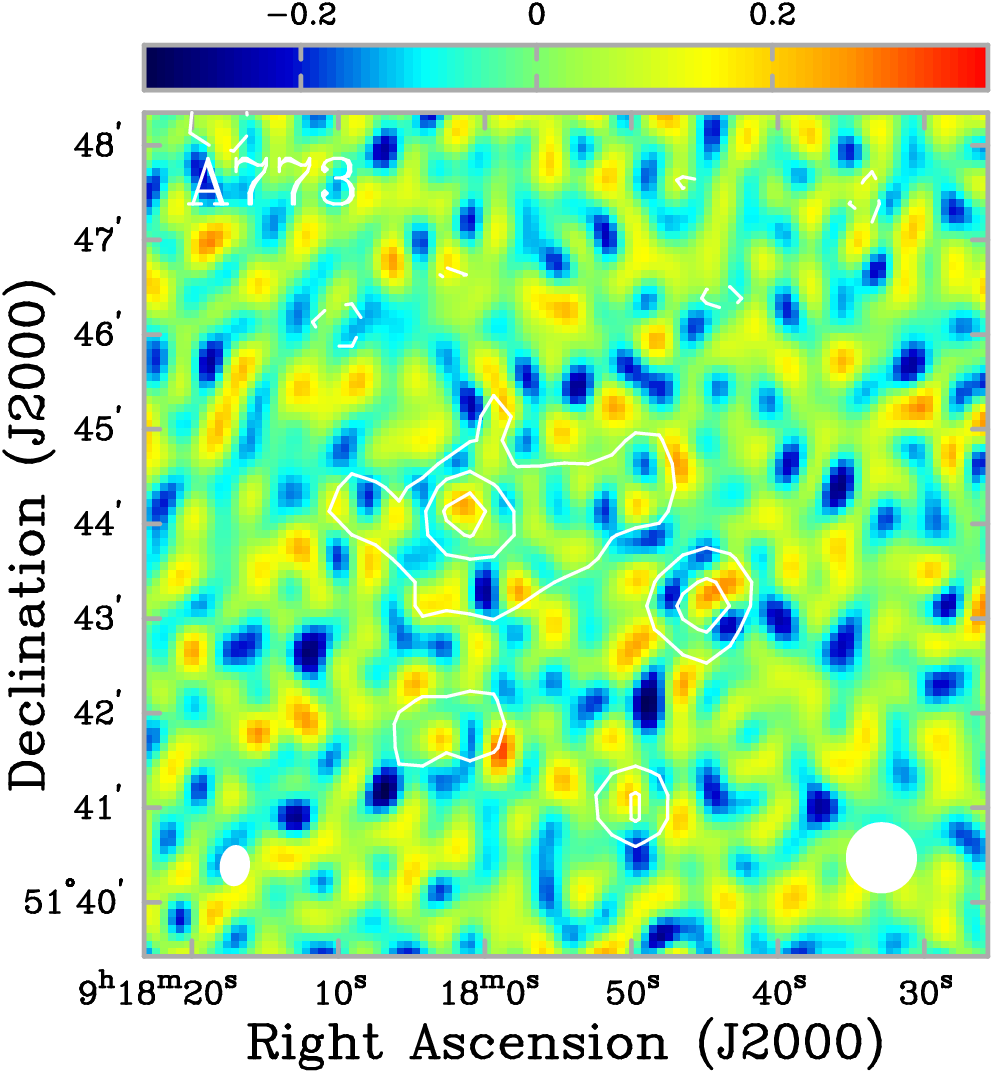,height=2.2in}
}
\centerline{
  \psfig{figure=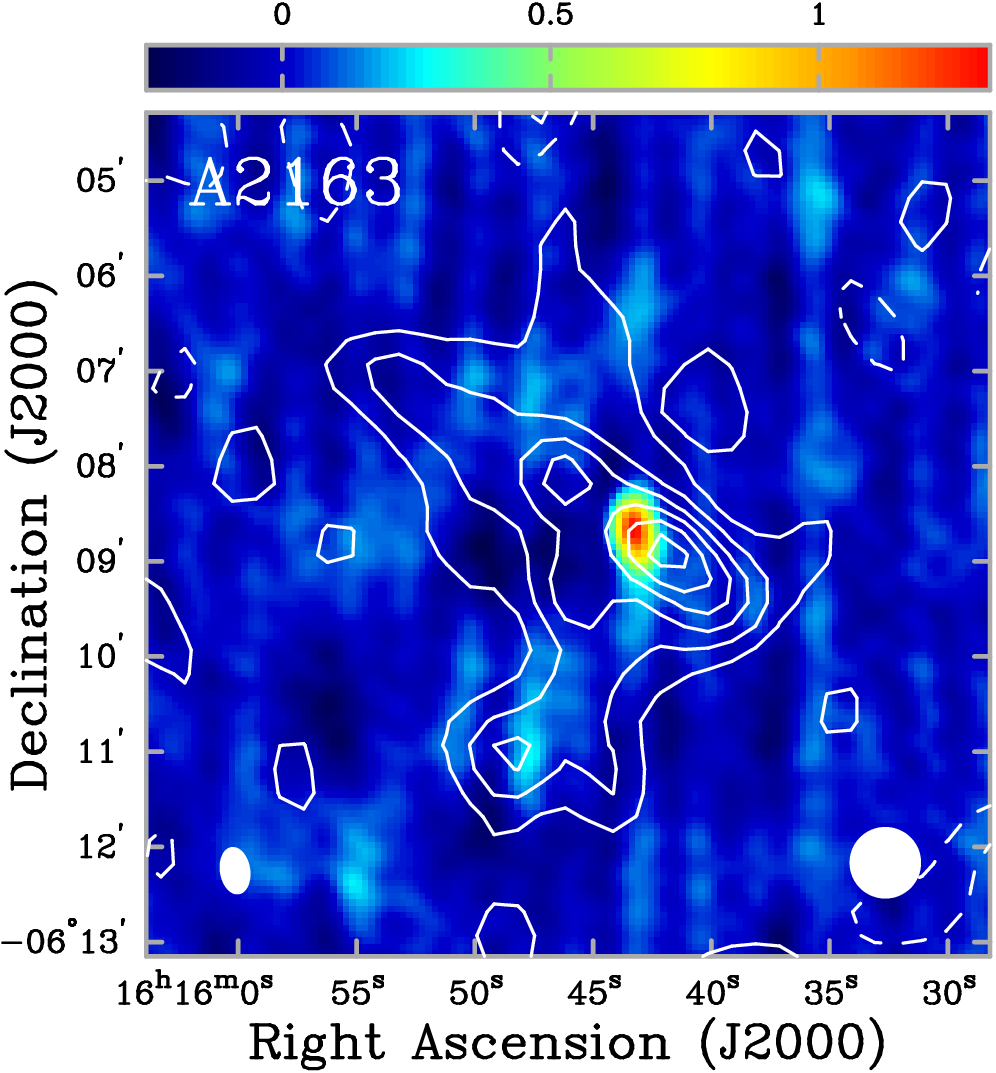,height=2.2in}
  \psfig{figure=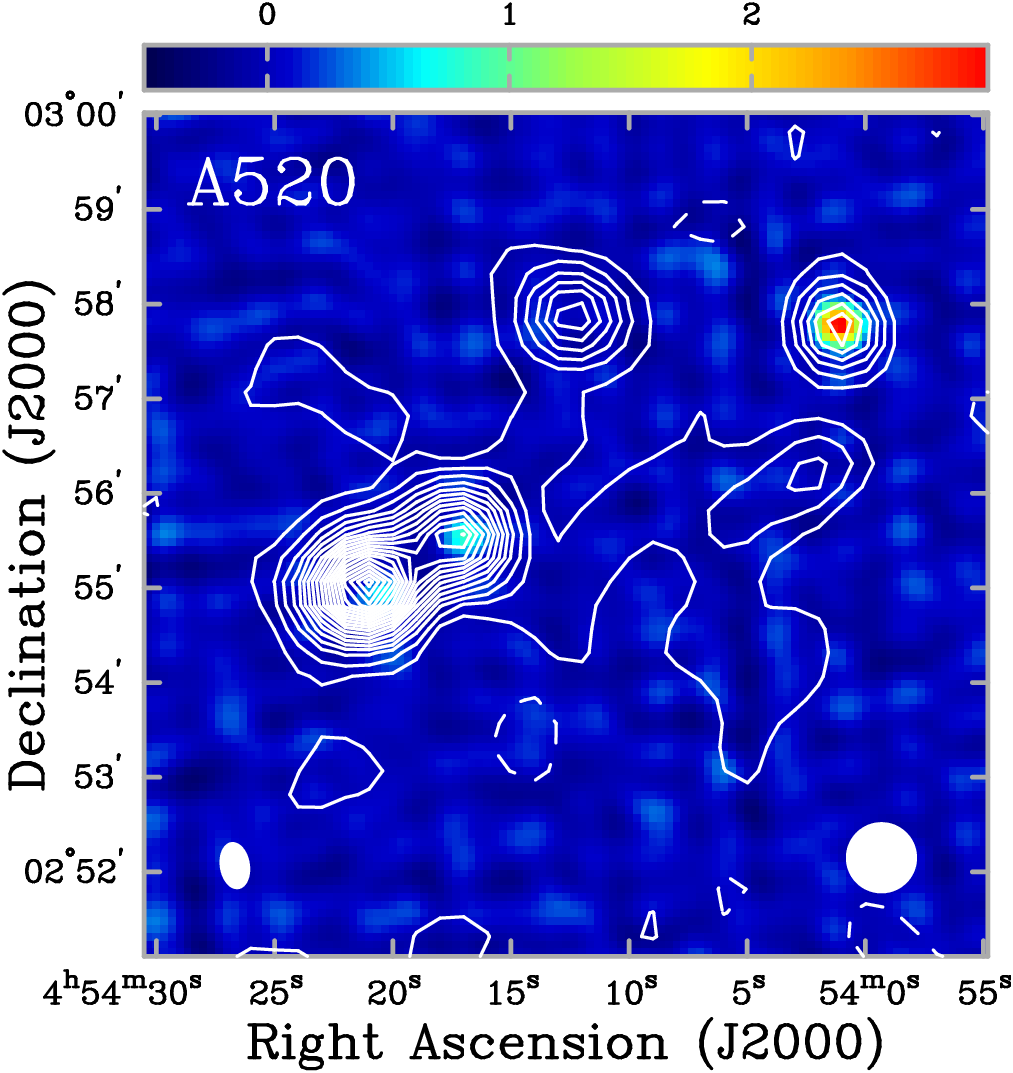,height=2.2in}
  \psfig{figure=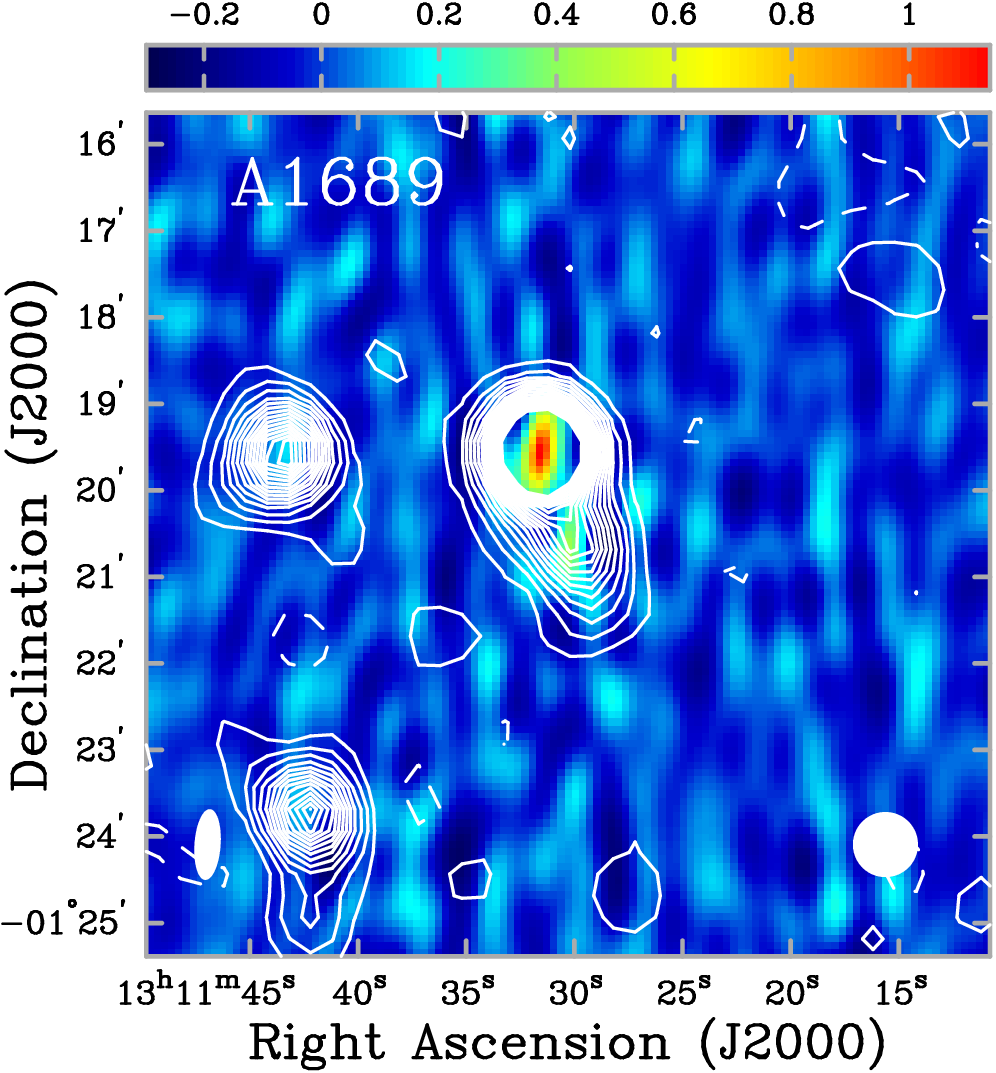,height=2.2in}
}
\centerline{
  \psfig{figure=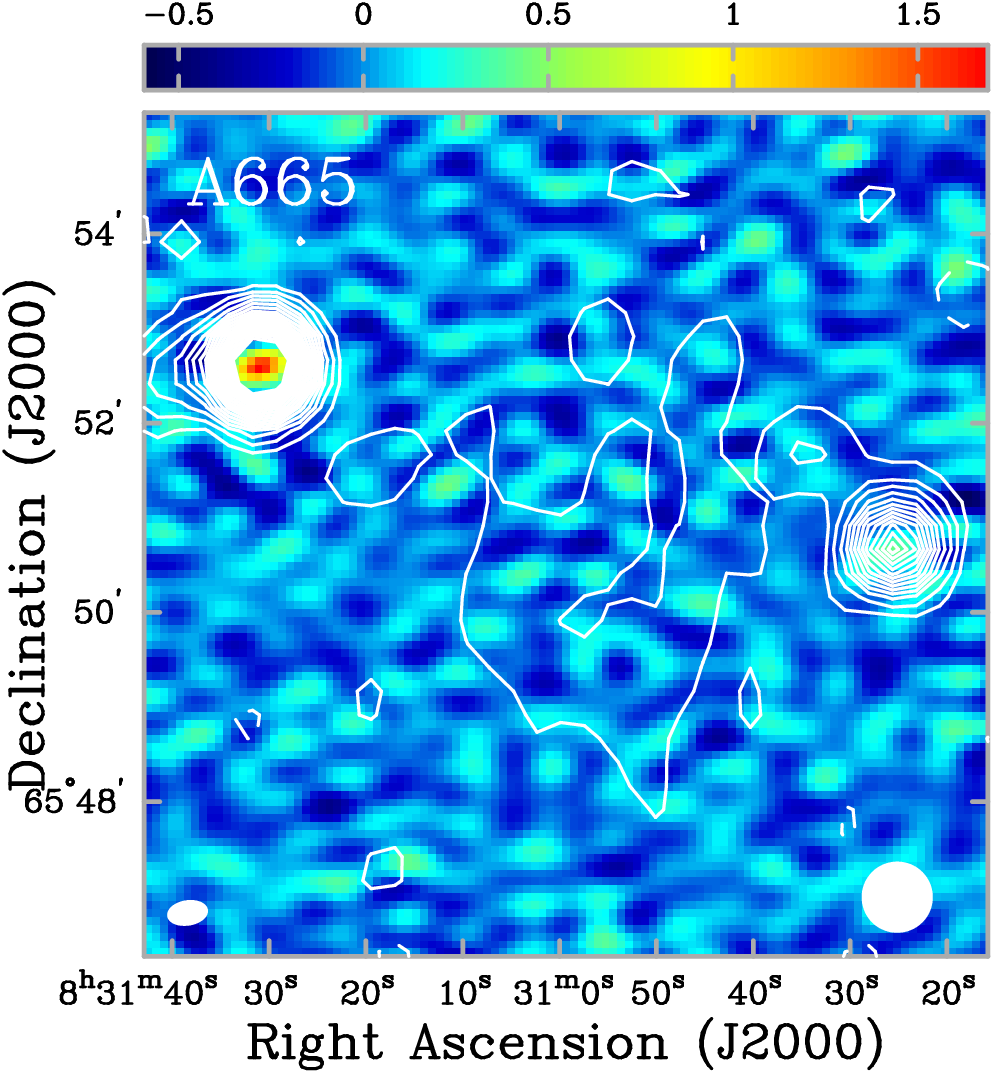,height=2.2in}
  \psfig{figure=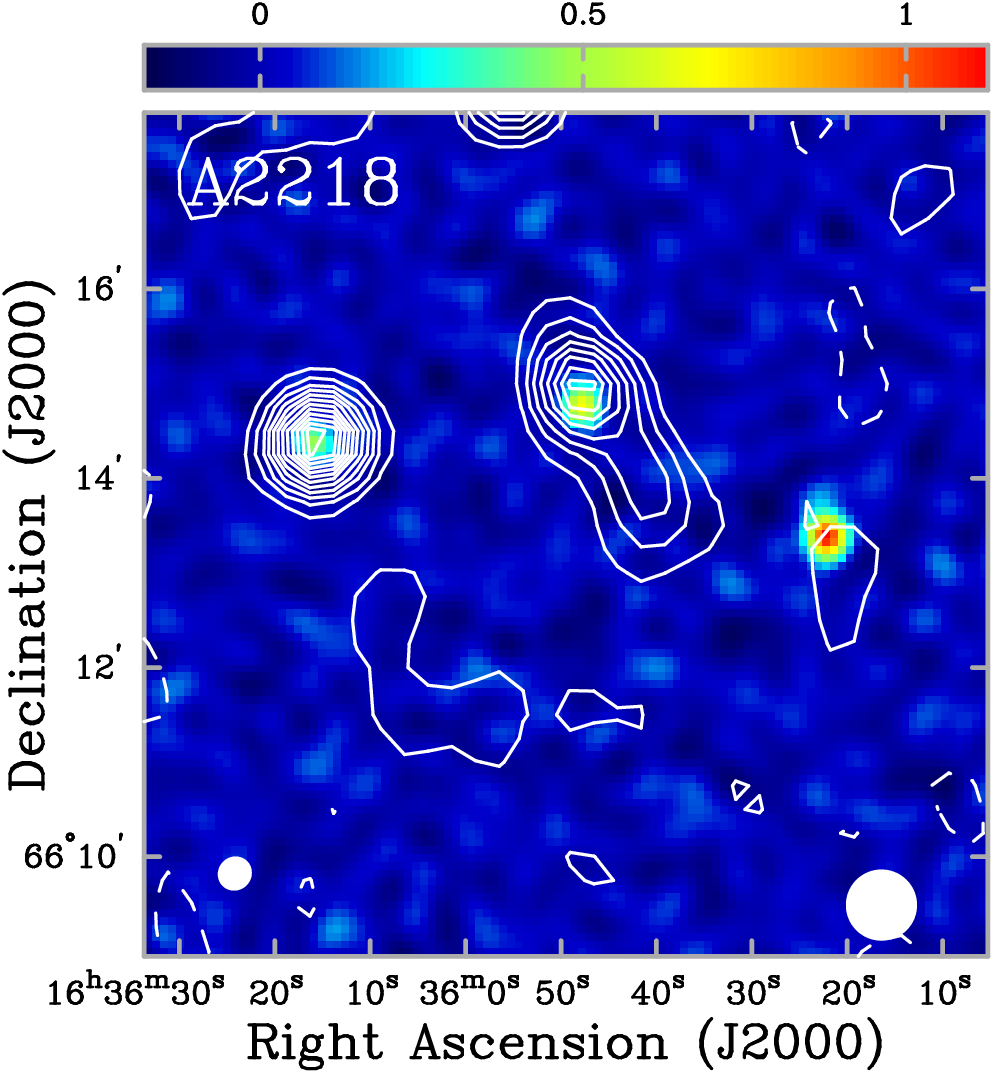,height=2.2in}
  \psfig{figure=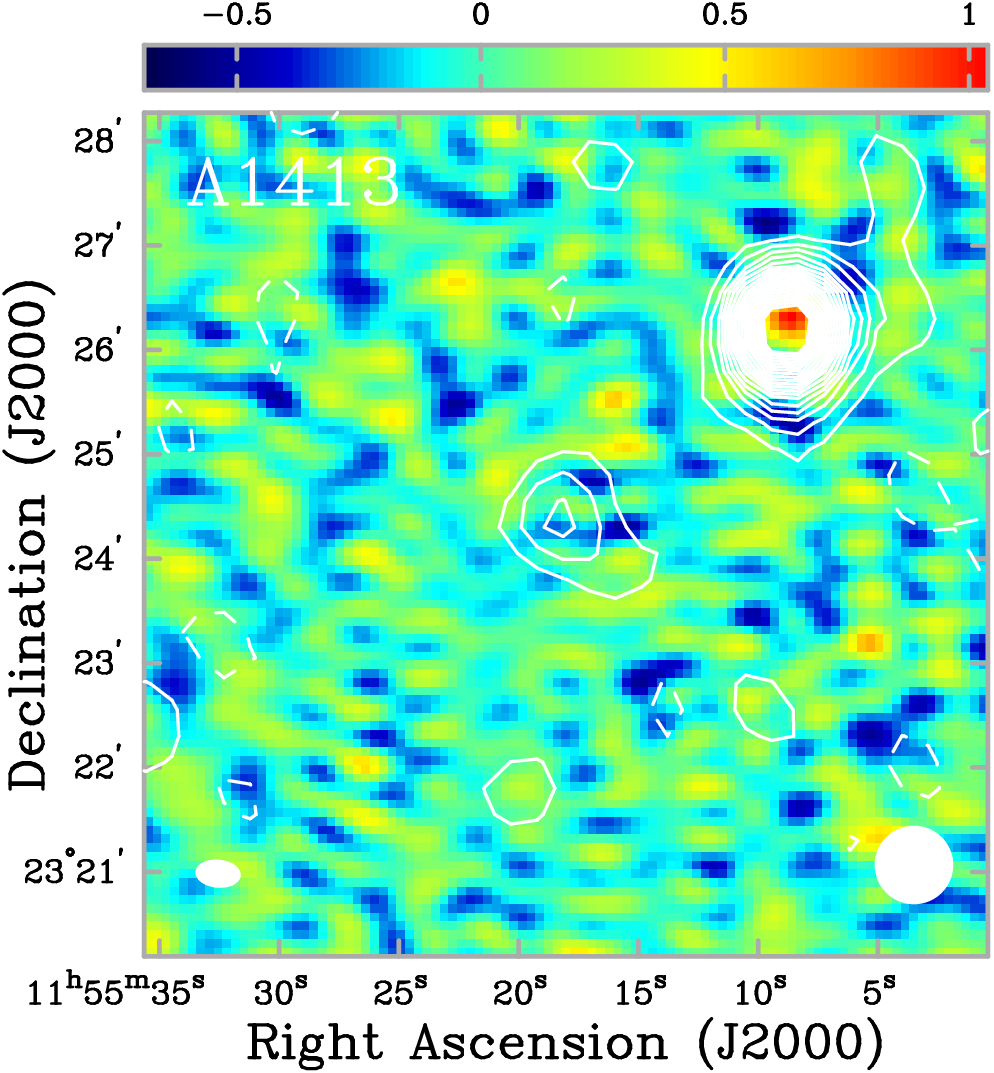,height=2.2in}
}
\contcaption{Cont.}
\end{figure*}

%%%%%
% Table: SZE Image Statistics
%%%%%
 
%% rms info for OVRO/BIMA data sets
%%
%%	       tapered	       high res
%%          ----------------  ---------
%% cluster, rms, beam, rms_K, rms, beam
%%
%%

\begin{deluxetable}{llcccclcc}

%\singlespace
%\footnotesize
%\rotate
\tablewidth{0pt}
%\tablenum{}
\tablecolumns{9}
%\tableheadfrac{}
\tablecaption{SZE Image Statistics\label{tab:imstat}}
\tablehead{
\colhead{} &
\multicolumn{4}{c}{Tapered\tablenotemark{a}} &
\colhead{} &
\multicolumn{3}{c}{High Resolution\tablenotemark{b}}
\\
\cline{2-5}\cline{7-9}
\colhead{} &
\colhead{} &
\colhead{$\sigma$} &
\colhead{Beam} &
\colhead{$\sigma_{\mbox{\tiny RJ}}$} &
\colhead{} &
\colhead{} &
\colhead{$\sigma$} &
\colhead{Beam}
\\
\colhead{Cluster} &
\colhead{Observatory} &
\colhead{(\uJybm)} &
\colhead{(arcsec)} &
\colhead{($\mu$K)} &
\colhead{} &
\colhead{Observatory} &
\colhead{(\uJybm)} &
\colhead{(arcsec)}
}
\startdata
MS1137
	& \ \ \ BIMA
	& 120 
	& $90 \times \phn 94$
	& 21
	&& \ \ \ BIMA
	& 105
	& $20 \times 24$
\\
MS0451
	& \ \ \ OVRO
	& \phn 90
	& $44 \times \phn 69$
	& 45
	&& \ \ \ OVRO
	& \phn 60
	& $19 \times 25$
\\
Cl0016
	& \ \ \ BIMA
	& 250
	& $81 \times 101$
	& 46
	&& \ \ \ BIMA
	& 220
	& $14 \times 30$
\\
R1347
	& \ \ \ BIMA
	& 307\tablenotemark{c}
	& $93 \times \phn 94$\tablenotemark{c}
	& 53\tablenotemark{c}
	&& \ \ \ BIMA
	& 245
	& $17 \times 27$
\\
A370
	&  \ \ \ OVRO
	& \phn 60
	& $56 \times \phn 86$
	& 19
	&& \ \ \ OVRO
	& \phn 70
	& $17 \times 23$
\\
MS1358
	& \ \ \ BIMA
	& 140
	& $96 \times \phn 98$
	& 22
	&&\ \ \ BIMA
	& 120
	& $17 \times 20$
\\
A1995 
	&  \ \ \ BIMA
	& 134
	& $70 \times \phn 77$
	& 37
	&& \ \ \ OVRO
	& \phn 65
	& $17 \times 20$
\\
A611
	&  \ \ \ OVRO
	& \phn 60
	& $48 \times \phn 58$
	& 32
	&& \ \ \ OVRO
	& \phn 45
	& $20 \times 39$
\\
A697
	&  \ \ \ OVRO
	& \phn 65
	& $50 \times \phn 53$
	& 37
	&& \ \ \ OVRO
	& \phn 50
	& $19 \times 33$
\\
A1835
	&  \ \ \ BIMA
	& 213
	& $87 \times 121$
	& 30
	&& \ \ \ BIMA
	& 190
	& $18 \times 22$
\\
A2261	
	&  \ \ \ OVRO
	& \phn 85
	& $49 \times \phn 53$
	& 49
	&& \ \ \ OVRO
	& \phn 75
	& $19 \times 35$
\\
A773	
	&  \ \ \ BIMA
	& 260
	& $91 \times \phn 99$
	& 43
	&& \ \ \ OVRO
	& \phn 90
	& $19 \times 26$
\\
A2163 
	&  \ \ \ BIMA
	& 300
	& $90 \times 104$
	& 48
	&& \ \ \ OVRO
	& \phn 85
	& $19 \times 30$
\\
A520 
	&  \ \ \ BIMA
	& 180
	& $90 \times 101$
	& 30
	&& \ \ \ OVRO
	& \phn 80
	& $12 \times 16$
\\
A1689	
	&  \ \ \ BIMA
	& 320
	& $93 \times \phn 94$
	& 55
	&& \ \ \ OVRO
	& \phn 72 
	& $18 \times 49$
\\
A665	
	&  \ \ \ BIMA
	& 160
	& $93 \times \phn 99$
	& 26
	&& \ \ \ BIMA
	& 150
	& $16 \times 26$
\\
A2218	
	&  \ \ \ BIMA
	& 200
	& $93 \times \phn 99$
	& 31
	&& \ \ \ OVRO
	& \phn 50
	& $21 \times 22$
\\
A1413
	&  \ \ \ BIMA
	& 250
	& $93 \times \phn 99$
	& 44
	&& \ \ \ BIMA
	& 210
	& $16 \times 26$
\enddata
\tablenotetext{a}{Gaussian taper with FWHM of 2000 $\lambda$ for OVRO
data and 1000 $\lambda$ for BIMA data.}
\tablenotetext{b}{Using only data with $\sqrt{(u^2 + v^2)} > 2000\
\lambda$.} 
\tablenotetext{c}{Used Gaussian taper with FWHM of 1500 $\lambda$.}
%\tablecomment{}

\end{deluxetable}

Figure~\ref{fig:image} shows the deconvolved SZE image contours
overlaid on the X-ray images for each cluster in our sample.  Negative
contours are shown as solid lines and the contours are multiples of
twice the rms of each image.  The images for MS0451 and Cl0016 have
been published previously \citep{reese2000}, but we include them here
so that the entire sample appears together.  We use DIFMAP
\citep{pearson1994} to produce the naturally weighted SZE images.  If
any point sources are detected in the cluster field, they are
subtracted from the data and a Gaussian taper applied to the
visibilities to emphasize brightness variations on cluster scales
before the image is deconvolved (CLEANed).  The half-power radius of
the Gaussian taper applied varies between 1000 $\lambda$ and 2000
$\lambda$, depending of the observatory and telescope configurations
used during the observations.  Typically a 1000 $\lambda$ half-power
radius taper is applied to BIMA data and a 2000 $\lambda$ half-power
radius taper is applied to OVRO data.  The full-width at half maximum
(FWHM) of the synthesized (restoring) beam is shown in the lower left
hand corner of each image.  Table~\ref{tab:imstat} summarizes the rms
sensitivities and the FWHM's of the synthesized beams of the tapered
maps shown in Figure~\ref{fig:image}, as well as the corresponding
statistics for the high resolution ($\geq 2000\ \lambda$) images.  In
addition, Table~\ref{tab:imstat} lists the Rayleigh-Jeans (RJ)
brightness sensitivities for each tapered, deconvolved image.

\subsubsection{Point Source Identification Using Spatial Filtering}
\label{subsubsec:pt_remove}

%%%%%
% SZE and X-ray image overlays
%%%%%

%%
% Figure 2
%%
\begin{figure*}[!ptbh]
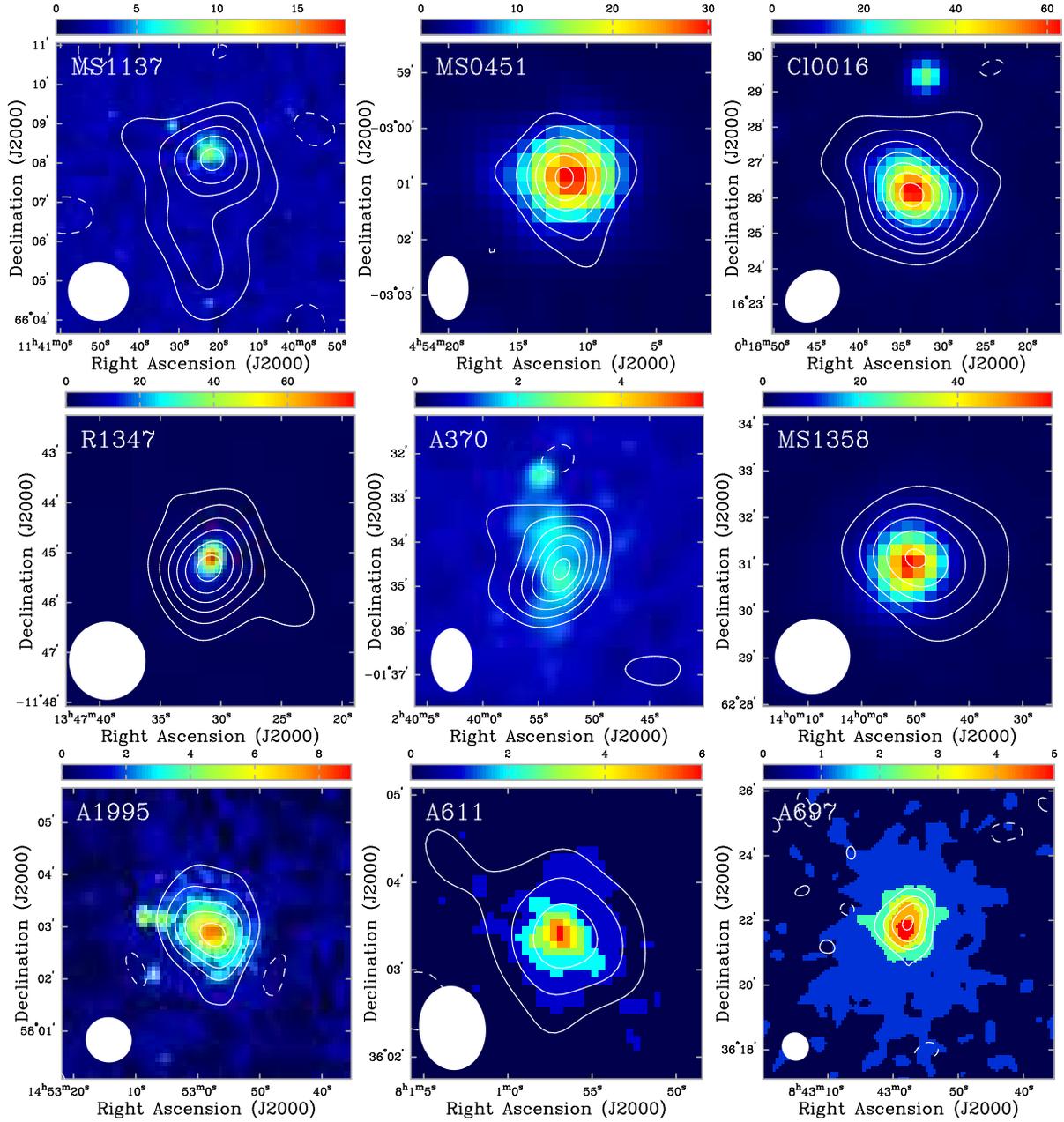

%\epsfxsize = 7.5 in
%%  \epsfxsize = 7.5 in
%%  \epsfysize = 6.0 in
%\centerline{\epsfbox{f?.eps}} 
\centerline{
  \psfig{figure=f2a_color.ps,height=2.2in}
  \psfig{figure=f2b_color.ps,height=2.2in}
  \psfig{figure=f2c_color.ps,height=2.2in}
}
\centerline{
  \psfig{figure=f2d_color.ps,height=2.2in}
  \psfig{figure=f2e_color.ps,height=2.2in}
  \psfig{figure=f2f_color.ps,height=2.2in}
}
\centerline{
  \psfig{figure=f2g_color.ps,height=2.2in}
  \psfig{figure=f2h_color.ps,height=2.2in}
  \psfig{figure=f2i_color.ps,height=2.2in}
}
\caption[SZE (contours) and X-ray (color scale) images]{SZE (contours) and
X-ray (color scale) images of each cluster in our sample.  Negative contours
are shown as solid lines.  The contours are multiples of 2 $\sigma$
and the FWHM of the synthesized beams are shown in the bottom left
corner.  The X-ray color scale images are raw counts images smoothed with
Gaussians with $\sigma = 15$\arcsec\ for PSPC data and $\sigma =
5$\arcsec\ for HRI data.  There is a color scale mapping for the counts
above each image.  The 30 GHz image statistics are summarized in
Table~\ref{tab:imstat}.}
\label{fig:image}
\end{figure*}

\begin{figure*}[!ptbh]
\centerline{
  \psfig{figure=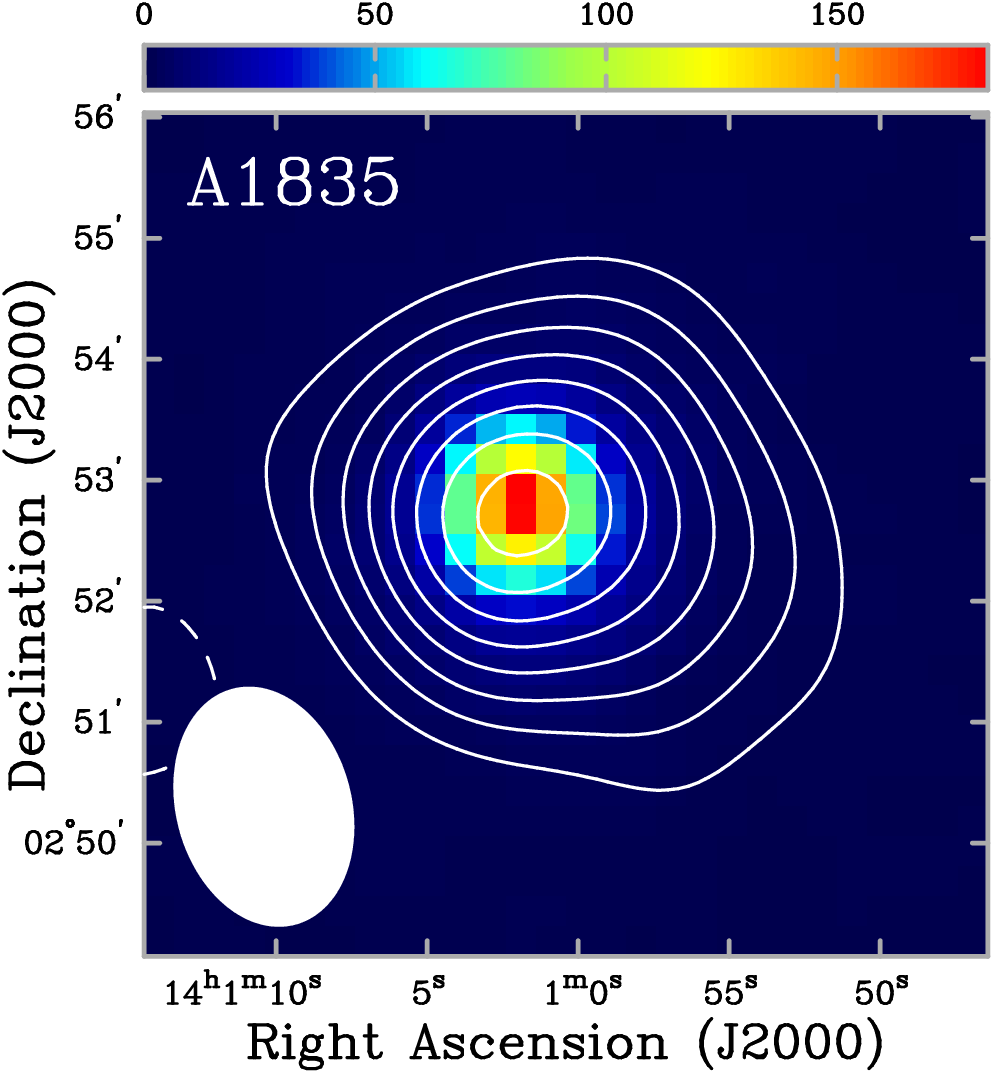,height=2.2in}
  \psfig{figure=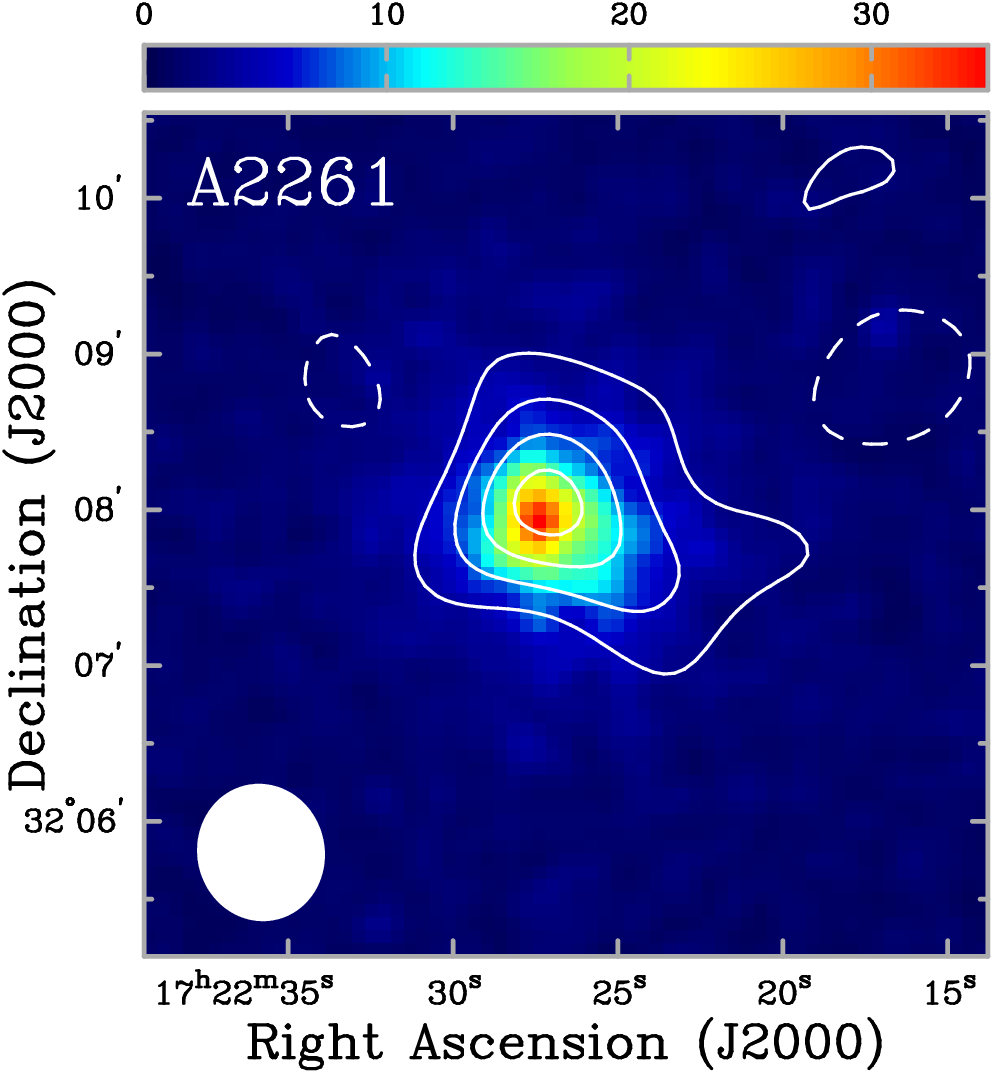,height=2.2in}
  \psfig{figure=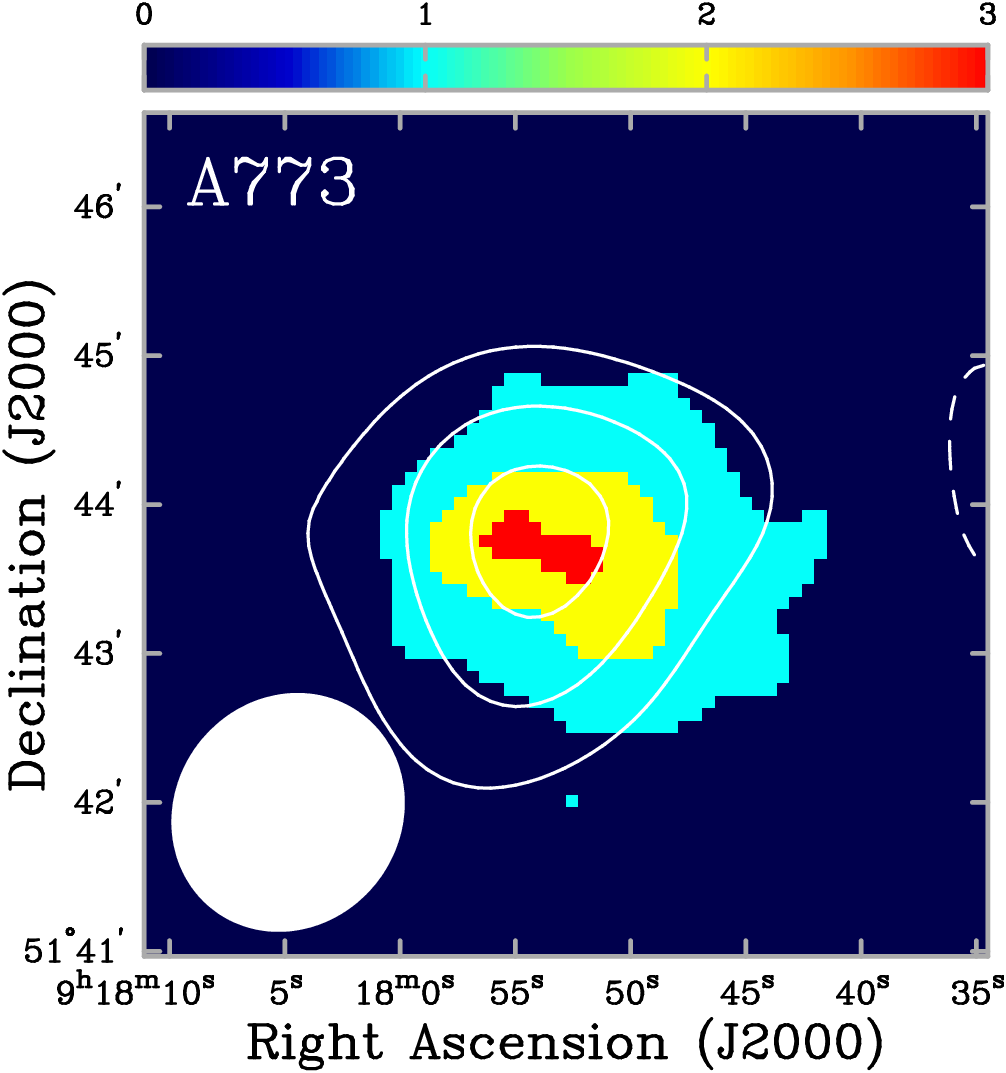,height=2.2in}
}
\centerline{
  \psfig{figure=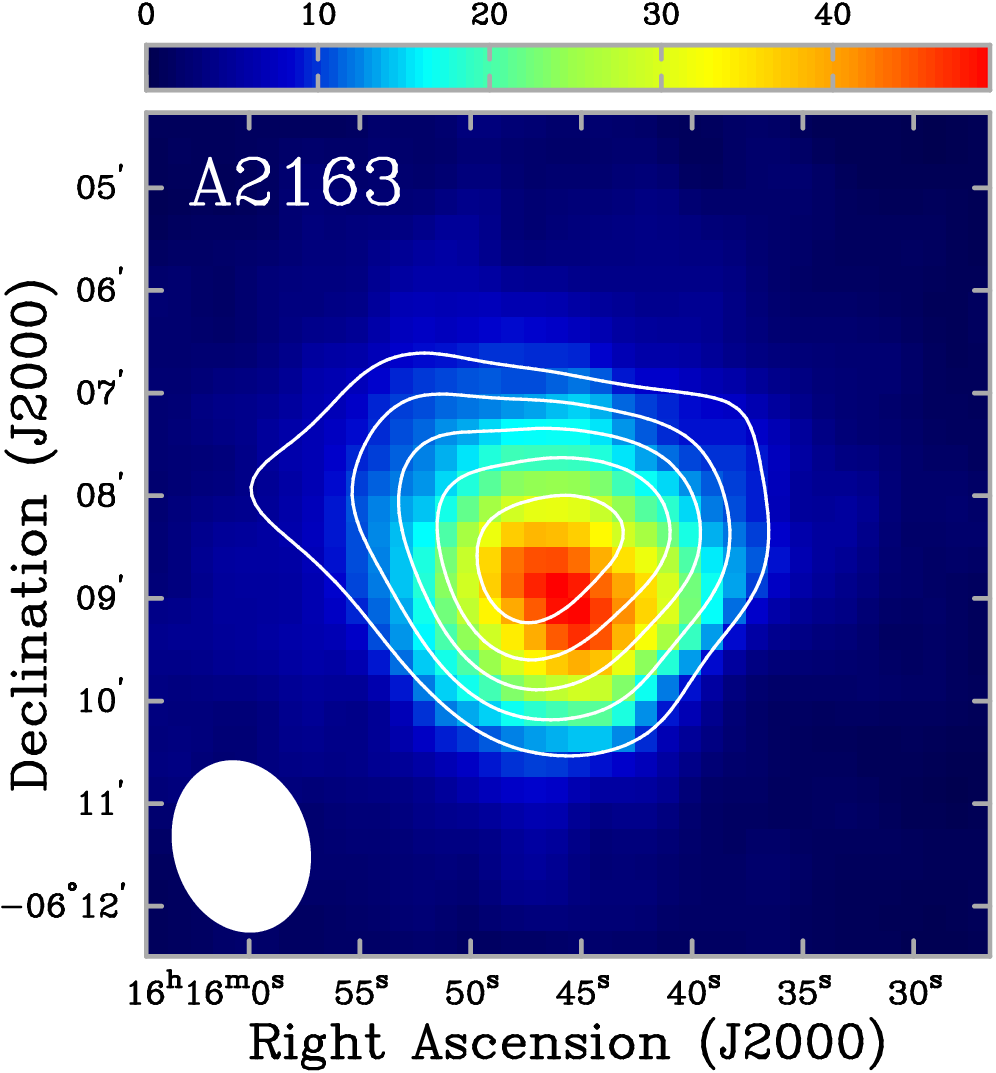,height=2.2in}
  \psfig{figure=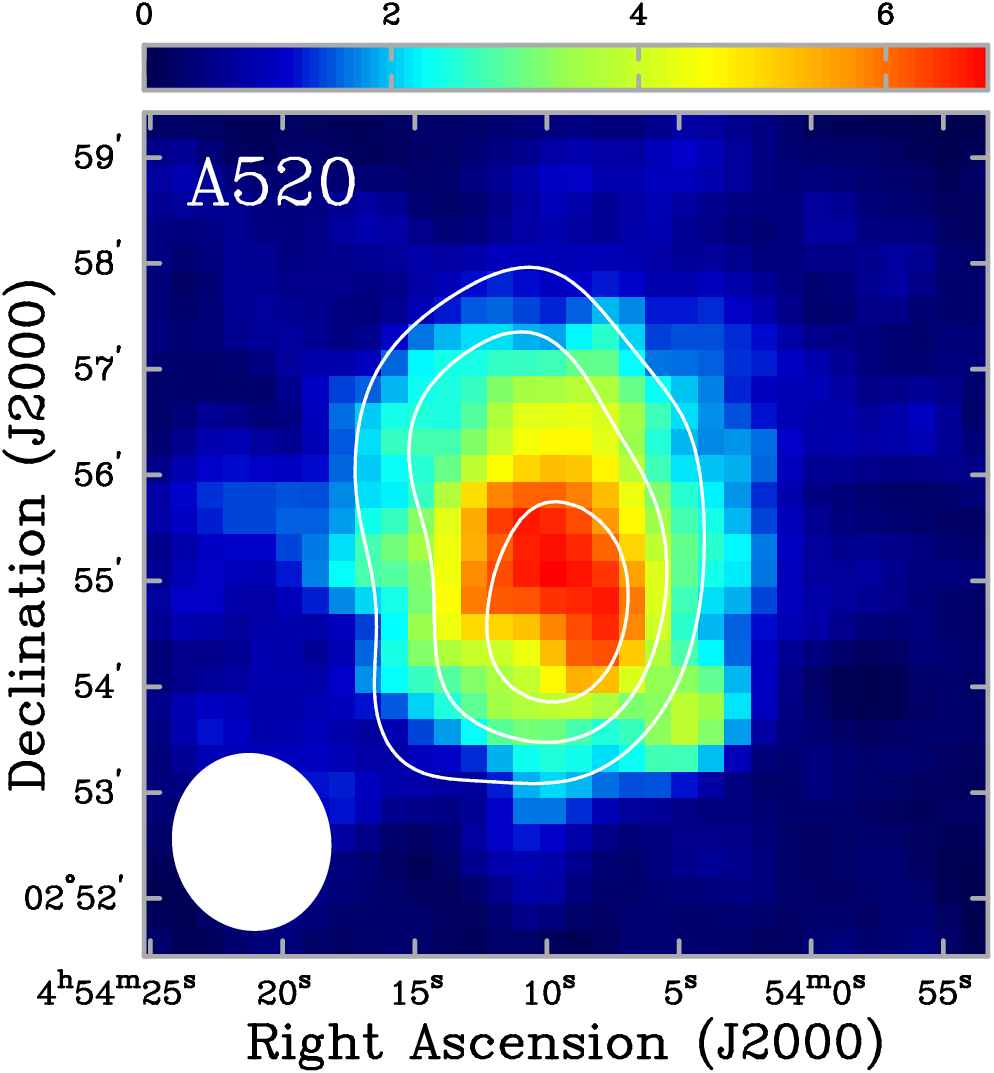,height=2.2in}
  \psfig{figure=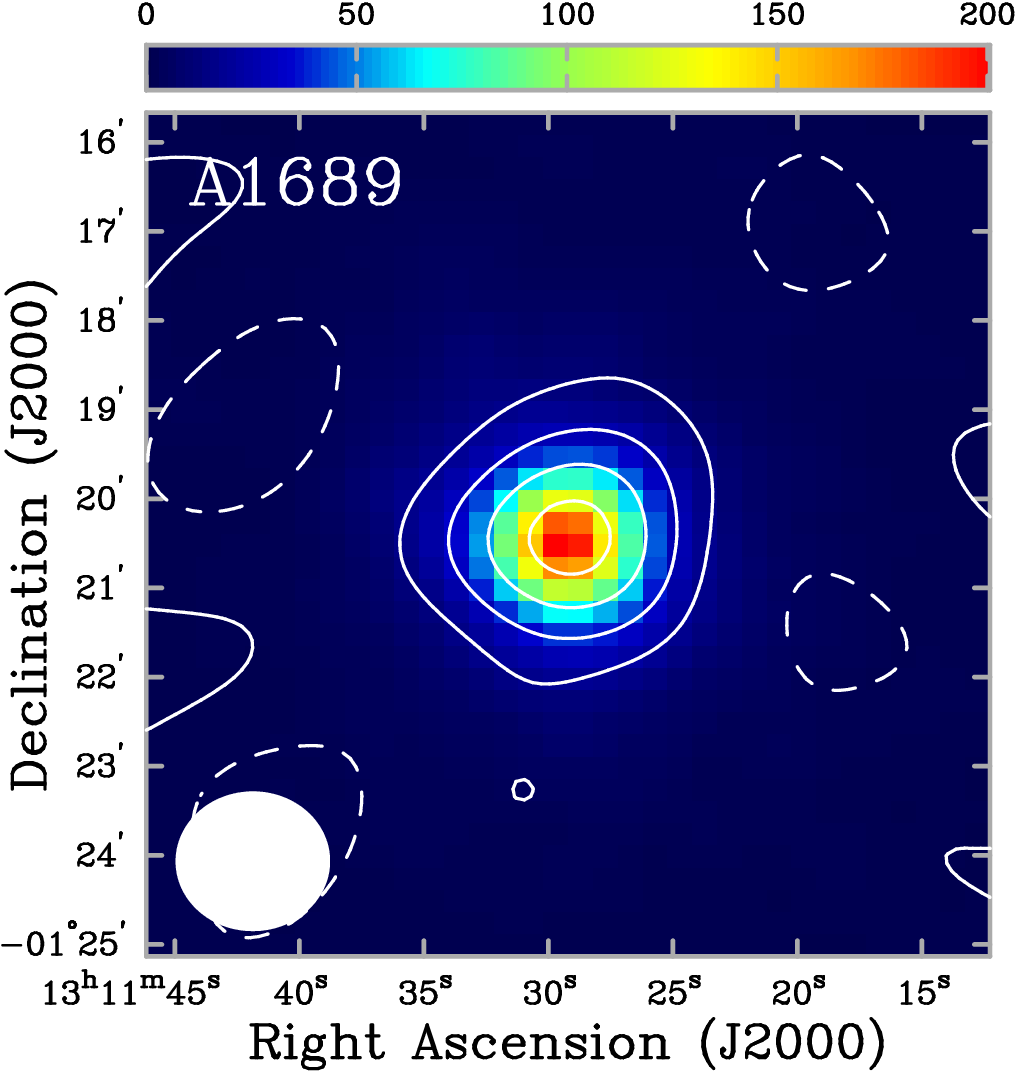,height=2.2in}
}
\centerline{
  \psfig{figure=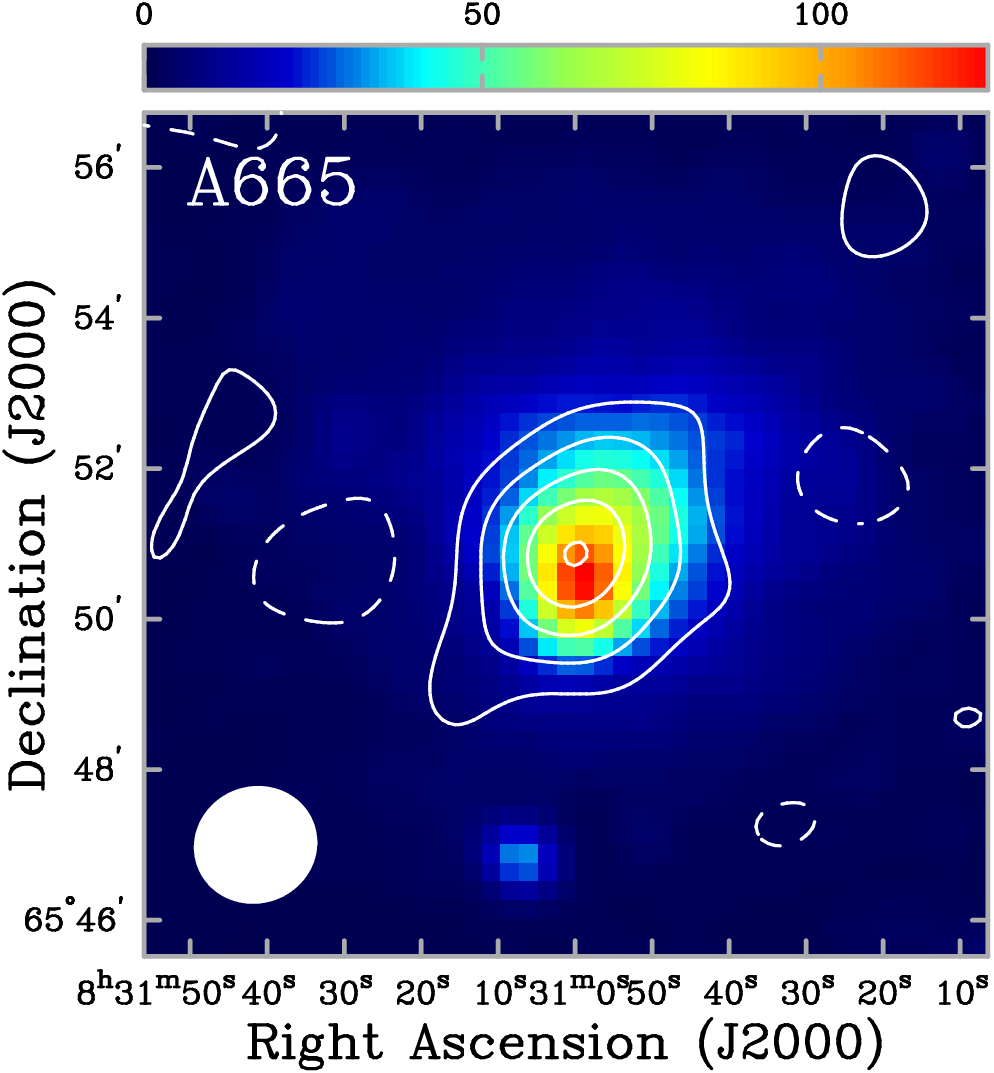,height=2.2in}
  \psfig{figure=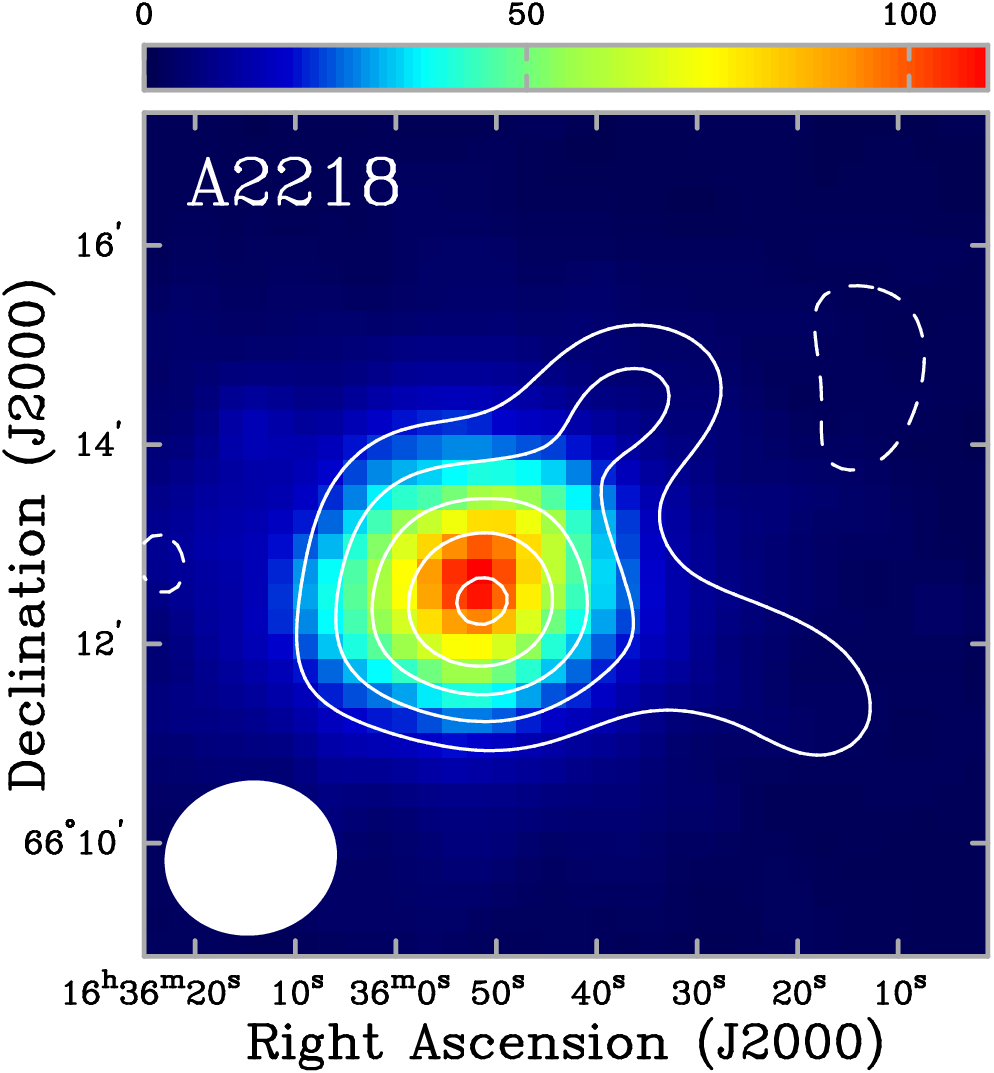,height=2.2in}
  \psfig{figure=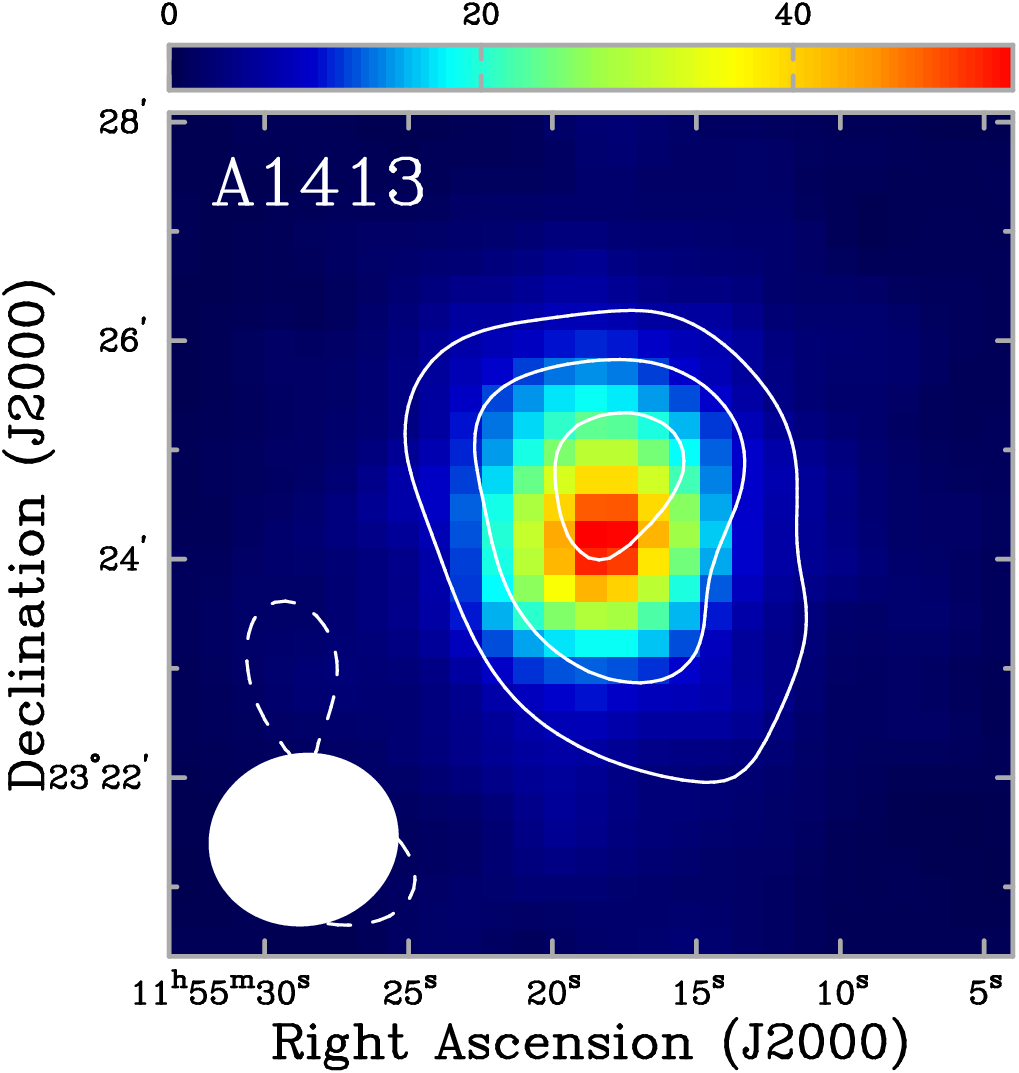,height=2.2in}
}
\contcaption{Cont.}
\end{figure*}

\begin{figure*}[ptbh]
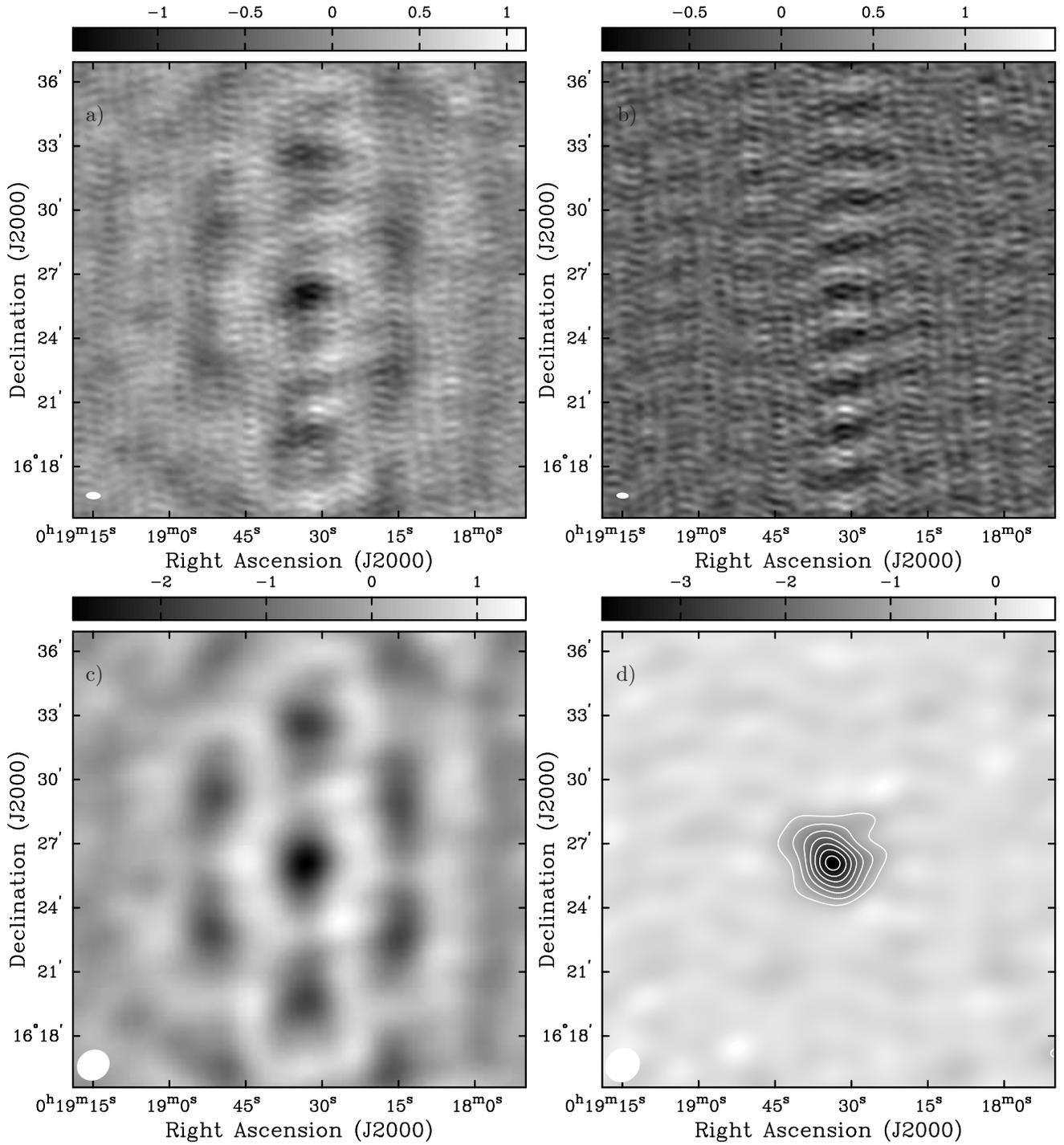

\centerline{
  \psfig{figure=f3a.ps,height=3.7in}  % was 3.7
  \psfig{figure=f3b.ps,height=3.7in}
}
\centerline{
  \psfig{figure=f3c.ps,height=3.7in}
  \psfig{figure=f3d.ps,height=3.7in}
}
\caption[Spatial filtering of the interferometer: Cl0016]{Panels
illustrating how the spatial filtering of the interferometer is used
to disentangle the point source emission from the SZE emission.  Each
panel is roughly $20\arcmin$ on a side with the FWHM of the
synthesized beam shown in the lower left hand corner.  The mapping of
the color scale is shown above each panel in units of mJy beam$^{-1}$.
Panels: $a$) ``natural''; $b$) high resolution ($\geq 2000\ \lambda$);
$c$) tapered map after point source removal; and $d$) deconvolved SZE
image (contours are multiples of 2 $\sigma$).  See text for details.

%
%%%Add panel labels
\vspace{-7.65in}
\hspace{0.7in} a) \hspace{3.25in} b)

\vspace{3.5in}
\hspace{0.7in} c) \hspace{3.25in} d)

% Now move back to where we were before the negative vspace; otherwise
% we will have some text super-imposed on our figures!
\vspace{4.25in}
}
\label{fig:cl0016_filter}
\end{figure*}

The identification and removal of point sources by taking advantage of
the spatial filtering of the interferometer is illustrated in the
panels of Figure~\ref{fig:cl0016_filter} for the BIMA SZE Cl0016 data.
During the maximum likelihood joint analysis (see
\S\ref{subsec:model_fit}), radio point sources identified from this
procedure are modeled and fit for directly in the \uv\ plane.  Each
panel covers the same angular region, roughly $20\arcmin$ on a side,
and each panel shows the FWHM of the synthesized beam in the lower
left hand corner.  Above each image is the color scale mapping showing
the flux density of the map in units of mJy beam$^{-1}$.  Panel $a$
shows the ``natural'' image, which includes all of the data.  There is
smooth, extended, negative emission in the center of the map; this is
the SZE decrement of Cl0016.  There is also a bright spot roughly
2$\arcmin$ south of the cluster that may be a point source.  The large
scale symmetric pattern is the synthesized beam of the low resolution
data (compare to panel $c$); even when all baselines are considered,
the SZE signal dominates.  Figure~\ref{fig:cl0016_filter}$b$ shows the
high resolution map using data with projected baselines $\geq 2000\
\lambda$ only.  The point source shows up easily now with the
characteristic shape of the synthesized beam for these data.  We
remove the point source by CLEANing.  A Gaussian \uv\ taper
(half-power radius of 1000 $\lambda$) is then applied to the full data
set to emphasize the short baselines, corresponding to the angular
scales typical of galaxy clusters, and shown in
Figure~\ref{fig:cl0016_filter}$c$.  The cluster is apparent as is the
symmetric pattern of the synthesized beam.  Deconvolving (CLEANing)
the tapered image results in panel $d$, which appeared in
Figure~\ref{fig:image} overlaid on X-ray data.  The contours are
multiples of twice the rms of the map.

	%%%%%
% Cl0016 spatial filtering exposition through a series of 4 figures
%%%%%

% Natural image
% High Resolution Image
% Image w/taper and point source removed
% Deconvolved w/taper
%%
% Figure 3
%%

%%%%%%%%%%
%% III.2  ROSAT X-ray Data
%%%%%%%%%%
\subsection{\rosat\ X-ray Data}
\label{subsec:xray_obs}

We use archival {\it R\"{o}ntgen Satellite} (\rosat) data from both
the Position Sensitive Proportional Counter (PSPC) and High-Resolution
Imager (HRI) instruments.  The live times of the observations we use
are listed in Table~\ref{tab:data} for both PSPC and HRI observations.

\subsubsection{Data Reduction}
\label{subsubsec:xdata_reduce}

We use the Snowden Extended Source Analysis Software (ESAS)
\citep{snowden1994, snowden1998} to reduce the data.  We use the ESAS
software to generate a raw counts image, a noncosmic background image,
and an exposure map for the HRI (0.1-2.4 keV) data and for each of the
Snowden bands R4-R7 (PI channels $52-201$; approximately $0.5-2.0$
keV) for the PSPC data, using a master veto rate (a measure of the
cosmic-ray and $\gamma$-ray backgrounds) of 200 counts s$^{-1}$ for
the PSPC data.  We examine the light curve data of both instruments
looking for time intervals with anomalously high count rates
(short-term enhancements) and for periods of high scattered solar
X-ray contamination.  Contaminated and anomalously high count rate
data are excised.  The Snowden software produces $512 \times 512$
pixel images with $14.^{\!\!\! \prime\prime} 947$ pixels for the PSPC
and $5.^{\!\!\! \prime\prime} 0$ pixels for the HRI.  For the PSPC,
final images for all of the R4-R7 bands together are generated by
adding the raw counts images and the background images.  Each Snowden
band has a slightly different effective exposure map and there is an
energy dependence in the point spread function (PSF).  Thus, we
generate a single exposure image and a single PSF image by combining
cluster photon-weighted averages of the four exposure images and the
four PROS
\citep{worrall1992, conroy1993} generated on-axis PSF images.  The
cluster photon-weighting is determined using the background subtracted
detected photons within a circular region centered on the cluster.
The region selected to construct the weights is the largest circular
region encompassing the cluster which contains no bright point
sources, typically a 15 pixel radius.

\subsubsection{X-ray Images and Data Properties}
\label{subsubsec:ximages}

We show smoothed X-ray raw counts images in Figure~\ref{fig:image}
(color) with SZE image contours overlaid.  PSPC images are shown when
available and HRI images otherwise.  Table~\ref{tab:data} summarizes
the on-source integration time of the \rosat\ observations of the
clusters in our sample for both the PSPC and HRI.  These images
roughly contain a few thousand cluster counts.  PSPC images are
smoothed with Gaussians with $\sigma = 15$\arcsec\ and HRI images with
$\sigma = 5$\arcsec.  The color scale wedge above each figure shows the
mapping between color and detector counts.

\subsubsection{X-ray Spectral Data}
\label{subsubsec:xray_spec}

We used published temperatures, metallicities, and H I column
densities from observations with the Advanced Satellite for Cosmology
and Astrophysics (ASCA).  Temperatures and metallicities for most of
the clusters in our sample appear in \citet{allen1998},
\citet{allen1998b}, and \citet{allen2000}.  When there is a detailed
account of the analysis for a particular cluster we use those results
instead.  When fitted metallicities are unavailable, we adopt a 0.2
solar metallicity with a 100\% uncertainty.  We use fitted H I column
densities when available, otherwise those from 21 cm surveys of our
galaxy \citep{dickey1990} are adopted.  We assign a conservative 50\%
uncertainty to the column densities adopted from 21 cm surveys of our
galaxy.  Table~\ref{tab:param_xspec} summarizes our adopted electron
temperatures, metallicities, and column densities with references for
the sources of this information.

%%%%%
% Table: X-ray Spectral Information
%%%%%

%%X-ray Spectral Info
%%
%% Te, metalicity, N_H
%%

\begin{deluxetable}{lcccl}
%\singlespace
%\footnotesize
%\rotate
\tablewidth{0pt}
%\tablenum{}
\tablecolumns{5}
%\tableheadfrac{}
\tablecaption{X-ray Spectral Information\label{tab:param_xspec}}
%%\tablecaption{Cluster Parameters: X-ray Spectra\label{tab:param_xspec}}
\tablehead{
\colhead{} &
\colhead{$kT_e$} &
\colhead{} &
\colhead{\NH} &
\colhead{}
\\
\colhead{cluster} &
\colhead{(keV)} &
\colhead{[Fe$/$H]} &
\colhead{($\times 10^{20}$ cm$^{-2}$)} &
\colhead{ref--$T_e$; [Fe$/$H]; \NH}
}
\startdata
MS1137
	& $\phn\phn 5.7^{+1.3\phn}_{-0.7\phn}$
	& $0.43^{+0.26}_{-0.24}$
	& $1.00$\tablenotemark{a}  %%/1.21\tablenotemark{b}
	& D99;D99;D99\\
MS0451  
	& $\phn10.4^{+1.0\phn}_{-0.8\phn}$		
	& $0.15^{+0.07}_{-0.07}$ 
	& $\phn 3.00^{+0.40}_{-0.30}$
	& D96;D96;D96\\
Cl0016  
	& $\phn 7.55^{+0.72}_{-0.58}$	
	& $0.07^{+0.11}_{-0.07}$ 
	& $\phn 5.59^{+0.41}_{-0.36}$
	& HB98;HB98;HB98\\
R1347
	& $\phn\phn 9.3^{+0.7\phn}_{-0.6\phn}$	
	& $0.33^{+0.06}_{-0.06}$ 
	& $10.00^{+4.00}_{-4.00}$
	& S97;S97;S97\\
A370
	& $\phn\phn 6.6^{+0.7\phn}_{-0.5\phn}$	
	& $0.3 ^{+0.1}_{-0.1}$
	& 3.1\tablenotemark{b}
	& O98;O98;G\\
MS1358
	& $\phn 7.48 ^{+0.50}_{-0.42}$
	& $0.32 ^{+0.09} _{-0.09}$
	& 1.93\tablenotemark{b}
	& AF98;AF98b;G\\
A1995 
	& $\phn 8.59 ^{+0.86}_{-0.67}$
	& $0.14 ^{+0.07}_{-0.07}$
	& $\phn 5.0 ^{+1.6}_{-1.6}$
	& P00;P00;P00\\
A611
	& $\phn\phn 6.6 ^{+0.6\phn}_{-0.6\phn}$
	& 0.20\tablenotemark{c}
	& 4.99\tablenotemark{b}
	& HPC;--;G\\
A697
	& $\phn\phn 9.8 ^{+0.7\phn}_{-0.7\phn}$
	& 0.20\tablenotemark{c}
	& 3.41\tablenotemark{b}
	& HPC;--;G\\
A1835
	& $\phn 8.21 ^{+0.19}_{-0.17}$
	& $0.35 ^{+0.04}_{-0.03}$	
	& 2.32\tablenotemark{b}
	& AF98;AF98b;G\\
A2261	
	& $\phn 8.82^{+0.37}_{-0.32}$
	& $0.32^{+0.06}_{-0.05}$
	& 3.28\tablenotemark{b}
	& AF98;AF98b;G\\
A773	
	& $\phn 9.29^{+0.41}_{-0.36}$
	& $0.21^{+0.05}_{-0.05}$
	& 1.44\tablenotemark{b}
	& AF98;AF98b;G\\
A2163 
	& $\phn 12.2 ^{+1.1\phn}_{-0.7\phn}$
	& $0.40^{+0.09}_{-0.08}$
	& $16.50^{+0.90}_{-1.14}$
	& M96;EAB95;EAB95\\
A520 
	& $\phn 8.33 ^{+0.46}_{-0.40}$
	& $0.14^{+0.06}_{-0.06}$
	& 7.80\tablenotemark{b}
	& AF98;AF98b;G\\
A1689	
	& $\phn 9.66 ^{+0.22}_{-0.20}$
	& $0.29 ^{+0.03}_{-0.03}$
	& 1.82\tablenotemark{b}
	& AF98;AF98b;G\\
A665	
	& $\phn 9.03^{+0.35}_{-0.31}$
	& $0.22^{+0.04}_{-0.05}$
	& 4.24\tablenotemark{b}
	& AF98,AF98b;G\\
A2218	
	& $\phn 7.05^{+0.22}_{-0.21}$
	& $0.18^{+0.04}_{-0.04}$
	& 3.24\tablenotemark{b}
	& AF98;AF98b;G\\
A1413
	& $\phn 7.54 ^{+0.17} _{-0.16}$
	& $0.28 ^{+0.03}_{-0.03}$
	& 2.19\tablenotemark{b}
	& AF98;AF98b;G\\
\enddata
\tablenotetext{a}{Adopted value (see ref D99) and adopted 50\% uncertainty.}
\tablenotetext{b}{Galactic value \citep{dickey1990} with adopted 50\%
uncertainty.} 
%%\tablenotetext{b}{W3nH weighted average value.}
\tablenotetext{c}{Adopted value with assumed 100\% uncertainty.}
\tablenotetext{\phn}{NOTE:  Uncertainties are 68\%
	confidence.\\REF:
	AF98-\citealt{allen1998};
	AF98b-\citealt{allen1998b};
	D96-\citealt{donahue1996};
	D99-\citealt{donahue1999};
	EAB95-\citealt{elbaz1995};
	G-\citealt{dickey1990};
	HPC-Hughes, private communication;
	HB98-\citealt{hughes1998};
	M96-\citealt{markevitch1996};
	O98-\citealt{ota1998};
	P00-\citealt{patel2000};
	S97-\citealt{schindler1997}
}
%\tablecomment{}
\end{deluxetable}

Temperatures for many of our clusters also appear in
\citet{mushotzky1997}.  Multiple temperature determinations agree
within the 1 $\sigma$ intervals for most of the clusters in our
sample.  The measurements overlap within $2\ \sigma$ in the worst
cases, i.e., for the clusters, MS1358, A1995, A2163, A1689, and A1413.
The temperatures, metallicity, column densities, and redshift of the
cluster are used to determine the X-ray cooling functions and the
conversion factor between detector counts and cgs units, $\Sigma$ (see
\S\ref{subsubsec:emiss} below for details).  The cooling functions and
conversion factors are summarized in Table~\ref{tab:emiss}.

%%%%%
% Table: X-ray Cooling Functions
%%%%%
%%
%%X-ray Emissivities
%%
%%
%%

%%
%  This new table has the values after the correction in the
%  Raymond-Smith calculation (IB!=JB):  old values appear in a similar
%  table following this one, now labeled ``old''
%%
\begin{deluxetable}{lcccccccc}
%%%\begin{deluxetable}{lccccccccl}
%\footnotesize
%\rotate
\tablewidth{0pt}
%\tablenum{}
\tablecolumns{9}
%\tableheadfrac{}
\tablecaption{X-ray Cooling Functions
	\label{tab:emiss}}
\tablehead{
\colhead{} &
\colhead{} &
\multicolumn{2}{c}{PSPC} &
\colhead{} &
\multicolumn{2}{c}{HRI} &
\colhead{} 
\\
\cline{3-4} \cline{6-7}
\colhead{cluster} &
\colhead{$\LameHo$\tablenotemark{a}} &
\colhead{$\LameHodet$\tablenotemark{b}} &
\colhead{$\Sigma$\tablenotemark{c}} &
\colhead{} &
\colhead{$\LameHodet$\tablenotemark{b}} &
\colhead{$\Sigma$\tablenotemark{c}} &
\colhead{$\frac{\mbox{$n_e$}}{\nH}=
	  \frac{\mbox{$\mu_{\mbox{\tiny H}}$}}{\mbox{$\mu_e$}}$} & 
\colhead{$\Lambol$\tablenotemark{d}} 
}
\startdata
MS1137  & 7.751 & \nodata & \nodata &
	& 1.765 & 2.461
	& 1.202 & 2.146
\\
MS0451  & 6.948 & 3.263 & 1.373 &
	& 1.470 & 3.050
	& 1.198 & 2.702 
\\
Cl0016  & 6.922 & 3.003 & 1.489 &
	& 1.289 & 3.471
	& 1.197 & 2.260 
\\
R1347   & 6.922 & \nodata & \nodata &
	& 1.167 & 4.089
	& 1.201 & 2.643
\\
A370	& 6.790 & \nodata & \nodata &
	& 1.627 & 3.037
	& 1.200 & 2.223 
\\
MS1358	& 6.717 & 3.785 & 1.336 &
	& 1.793 & 2.821 
	& 1.200 & 2.371 
\\
A1995	& 6.434 & \nodata & \nodata &
	& 1.434 & 3.395 
	& 1.198 & 2.448
\\
A611	& 6.511 & \nodata & \nodata &
	& 1.489 & 3.395
	& 1.199 & 2.175
\\
A697	& 6.334 & \nodata & \nodata &
	& 1.570 & 3.148
	& 1.199 & 2.647
\\
A1835	& 6.462 & 3.839 & 1.344 &
	& 1.775 & 2.909
	& 1.201 & 2.496
\\
A2261   & 6.359 & \nodata & \nodata &
	& 1.649 & 3.150
	& 1.200 & 2.570
\\
A773	& 6.171 & \nodata & \nodata &
	& 1.870 & 2.715
	& 1.199 & 2.581
\\
A2163   & 6.135 & 2.555 & 1.998 &
	& 1.021 & 5.000
	& 1.202 & 3.064
\\
A520	& 6.119 & 3.209 & 1.586 &
	& 1.337 & 3.807
	& 1.198 & 2.411 
\\
A1689   & 6.158 & 3.899 & 1.335 &
	& 1.835 & 2.836
	& 1.200 & 2.673 
\\
A665	& 6.102 & 3.616 & 1.428 &
	& 1.570 & 3.289
	& 1.199 & 2.550 
\\
A2218	& 6.112 & 3.789 & 1.378 &
	& 1.691 & 3.086
	& 1.198 & 2.237 
\\
A1413   & 6.133 & 4.000 & 1.343 &
	& 1.856 & 2.894
	& 1.200 & 2.361
\enddata
\tablenotetext{a}{Units are $\times 10^{-24}$ erg s$^{-1}$ cm$^3$.
The emissivity in the cluster frame integrated over the ROSAT band
(0.5-2.0 keV) redshifted to the cluster frame.}
\tablenotetext{b}{Units are $\times 10^{-13}$ cnts s$^{-1}$
cm$^5$. The emissivity in the detector frame accounting for the
response of the instrument.}
\tablenotetext{c}{Units are $\times 10^{-11}$ \cgsunits. The
conversion of detector units to cgs units including the $(1+z)$ factor
between energy and counts.}
\tablenotetext{d}{Units are $\times 10^{-23}$ erg s$^{-1}$ cm$^3$.
The bolometric emissivity.}
%\tablecomment{}
\end{deluxetable}

\subsubsection{X-ray Cooling Function}
\label{subsubsec:emiss}

The X-ray cooling function enters the distance calculation linearly
and indirectly as the conversion between detector counts and cgs
units (see \S\ref{subsec:Da_calc}).  We use a Raymond-Smith (1977)
\nocite{raymond1977} spectrum to describe the hot ICM, which includes
contributions from electron-ion thermal bremsstrahlung, line emission,
recombination, and two photon processes.  We replace the
non-relativistic bremsstrahlung calculation in the Raymond-Smith model
with the relativistic calculation of \citet{gould1980}.  A discussion
of this calculation appears in \cite{reese2000}.  

The cooling function results for the \rosat\ data used in our analysis
are summarized in Table~\ref{tab:emiss}, where $\LameHo$ is the
cooling function in cgs units, $\LameHodet$ is the cooling function in
detector units, $\Sigma$ is the conversion between counts and cgs
units, and $\Lambol$ is the bolometric cooling function.  The cooling
functions with relativistic corrections are typically 1.05 times the
Raymond-Smith ``uncorrected'' value for the clusters in our sample.

%%%%%%%%%%%%%%%%%%%%%%%%%%%%%%%%%%%%%%%%%%
%% IV.  Method
%%%%%%%%%%%%%%%%%%%%%%%%%%%%%%%%%%%%%%%%%%

\vspace{1in}
\section{Method}
\label{sec:method}
%%%%%%%%%%
%% IV.1 Angular Diameter Distance Calculation
%%%%%%%%%%
\subsection{Angular Diameter Distance Calculation}
\label{subsec:Da_calc}

The calculation begins by constructing a model for the cluster gas
distribution.  We use a spherical isothermal $\beta$ model to describe
the ICM.  With this model, the cluster's extent along the line of
sight is the same as that in the plane of the sky.  This is clearly
invalid in the presence of cluster asphericities.  Thus cluster
geometry introduces an important uncertainty in SZE and X-ray derived
distances.  In general, clusters are dynamically young, are
aspherical, and rarely exhibit projected gas distributions which are
circular on the sky \citep{mohr1995}.  We currently cannot disentangle
the complicated cluster structure and projection effects, but
numerical simulations provide a good base for understanding these
difficulties.  The effects of asphericity contribute significantly to
the distance uncertainty for each cluster, but are not believed to
result in any significant bias in the Hubble parameter derived from a
large sample of clusters \citep{sulkanen1999}.

The spherical isothermal $\beta$ model has the form
\citep{cavaliere1976, cavaliere1978}
\begin{equation}
n_e({\mathbf{r}}) = \no \left ( 1 + \frac{r^2}{r_c^2} \right )^{-3\beta/2},
\label{eq:iso_beta}
\end{equation}
where $n_e$ is the electron number density, $r$ is the radius from the
center of the cluster, $r_c$ is the core radius of the ICM, and
$\beta$ is the power law index.  With this model, the SZE signal is
\begin{equation}
\Delta T = f_{(x, T_e)} \Tcmb \Da \! \int\!\! d\zeta \, \sigT n_e \frac{\kB
	T_e}{m_e c^2} = \dTo \left ( 1 +
	\frac{\theta^2}{\theta_c^2} \right )^{(1-3\beta)/2},
	\label{eq:szsignal}
\end{equation}
where $\Delta T$ is the thermodynamic SZE temperature
decrement/increment, $f_{(x, T_e)}$ is the frequency dependence of the
SZE with $x = h\nu/k\Tcmb$, $\Tcmb$ (=2.728 K; \citealp{fixsen1996})
is the temperature of the CMB radiation, \kB\ is the Boltzmann
constant, $\sigT$ is the Thompson cross section, $m_e$ is the mass of
the electron, $c$ is the speed of light, \dTo\ is the central
thermodynamic SZE temperature decrement/increment, $\theta$ is the
angular radius in the plane of the sky and $\theta_c$ the
corresponding angular core radius, and the integration is along the
line of sight $\ell=\Da\zeta$.  The frequency dependence of the
thermal SZE is
\begin{equation}
f_{(x, T_e)}= \left ( x \frac{e^x + 1}{e^x - 1} - 4  \right ) \left ( 1 +
\delta_{\mbox{\tiny SZE}}(x,T_e) \right ) 
\label{eq:fx_thermal}
\end{equation}
where $\delta_{\mbox{\tiny SZE}}(x,T_e)$ is the relativistic
correction to the frequency dependence.  In the non-relativistic and
Rayleigh-Jeans (RJ) limits, $f_{(x, T_e)} \rightarrow -2$.  We apply
the relativistic corrections $\delta_{\mbox{\tiny SZE}}(x,T_e)$ to
fifth order in $kT_e/m_e c^2$ \citep{itoh1998}, which agrees with
other works \citep{rephaeli1995, rephaeli1997, stebbins1997,
challinor1998, sazonov1998, sazonov1998b, molnar1999, dolgov2001} for
clusters with $\kB T_e \leq 15$ keV, satisfied by all the clusters in
our sample.  This correction decreases the magnitude of $f_{(x, T_e)}$
by $\lsim 5$\% (typically 3\%) for the clusters considered here.

The X-ray surface brightness is 
\begin{equation}
S_x = \frac{1}{4\pi (1+z)^4} \Da \! \int \!\! d\zeta \, n_e \nH \LameH
	\:\;\! = \Xo \left ( 1 + \frac{\theta^2}{\theta_c^2} \right
	)^{(1-6\beta)/2},  \label{eq:xsignal}
\end{equation}
where $S_x$ is the X-ray surface brightness in cgs units (erg s$^{-1}$
cm$^{-2}$ arcmin$^{-2}$), $z$ is the redshift of the cluster, \nH\ is
the hydrogen number density of the ICM, $\LameH = \LameH (T_e,
\mbox{abundance})$ is the X-ray cooling function of the ICM in the
cluster rest frame in cgs units (erg cm$^3$ s$^{-1}$) integrated over
the redshifted \rosat\ band, and \Xo\ is the X-ray surface brightness
in cgs units at the center of the cluster.  Since the X-ray
observations are in instrument counts, we also need the conversion
factor between detector counts and cgs units, $\Sigma$ ($\Xo = S_{x
0}^{\mbox{\tiny det}} \Sigma$), discussed in detail in
\citet{reese2000} along with a description of the calculation of
\LameH, which includes relativistic corrections \citep{gould1980} to
the \citet{raymond1977} spectrum.  The normalizations, \dTo\ and \Xo,
used in the fit include all of the physical parameters and geometric
terms that come from the integration of the $\beta$ model along the
line of sight.

One can solve for the angular diameter distance by eliminating \no\
(noting that $\nH = n_e \mu_e / \muH$ where $n_j \equiv \rho/\mu_j
m_p$ for species $j$) yielding
\begin{equation}
\Da = \frac{(\dTo)^2}{\Xo} \left( \frac{m_e c^2}{\kB \Teo} \right)^2
	\frac{\Lamo \mu_e / \muH}{4\pi^{3/2} \, f_{(x, T_e)}^2 \,
	T_{\mbox{\tiny CMB}}^2 \, \sigma_{\mbox{\tiny T}}^2 \,
	(1+z)^4} \frac{1}{\theta_c} 
	\left [
	\frac{\Gamma(\frac{3}{2}\beta)}{\Gamma(\frac{3}{2}\beta-\frac{1}{2})}
	\right ]^2 \frac{\Gamma(3\beta-\frac{1}{2})}{\Gamma(3\beta)}
	\label{eq:Da}
\end{equation}
where $\Gamma(x)$ is the Gamma function.  Similarly, one can eliminate
$\Da$ instead and solve for the central density \no.

More generally, the angular diameter distance is
\begin{equation}
\Da = \frac{(\dTo)^2}{\Xo} \left ( \frac{m_e c^2}{\kB \Teo} \right
	)^2 \frac{\Lamo \mu_e / \muH}{4\pi \, f_{(x, T_e)}^2 \,
	T_{\mbox{\tiny CMB}}^2 \, \sigma_{\mbox{\tiny T}}^2 \,
	(1+z)^4} \frac{1}{\theta_c} 
	\frac{\int \left (
	\frac{\mbox{$n_e$}}{\no} \right )^2 \frac{\LameH}{\LameHo} d\eta
	\left. \right |_{R=0}} {\left [ \int \frac{\mbox{$n_e$}}{\no}
	\frac{\mbox{$T_e$}}{\Teo} d\eta \left. \right |_{R=0} \right ]^2}, 
   \label{eq:Da_general}
\end{equation}
where $\theta_c$ is the characteristic angular scale of the galaxy
cluster whose exact meaning depends on the ICM model (the core radius
for the isothermal $\beta$ model) and $\eta \equiv \zeta / \theta_c
\equiv \ell/r_c$ is the line of sight length in units of the
characteristic radius, $r_c = \theta_c \Da$.  For simplicity in
notation, we have assumed that the density and temperature models are
normalized at the central value (denoted with 0) though any location
for the normalization is allowed.  The above integrals are along the
central line of sight, denoted as zero projected radius $R=0$.  The
$\Gamma$ functions and the factor of $\pi^{1/2}$ in equation~(\ref{eq:Da})
come from the integration of the $\beta$ model along the central line
of sight for both the SZE and X-ray models.

%%%%%%%%%%
%% IV.2 Joint SZE & X-ray Model Fitting
%%%%%%%%%%
\subsection{Joint SZE and X-ray Model Fitting}
\label{subsec:model_fit}

The SZE and X-ray emission both depend on the properties of the ICM,
so a joint fit to the interferometric SZE data and the PSPC and HRI
X-ray data provides the best constraints on those properties.  Each
data set is assigned a collection of parameterized models.  Typically,
SZE data sets are assigned a $\beta$ model and point sources and X-ray
images are assigned a $\beta$ model and a cosmic X-ray background
model.  This set of models is combined for each data set to create a
composite model which is then compared to the data.

Our analysis procedure is described in detail in \citet{reese2000}.
The philosophy behind the analysis is to keep the data in a reduced
but ``raw'' state and run the model through the observing strategy to
compare directly with the data.  In particular, the interferometric
SZE observations provide constraints in the Fourier (\uv) plane, so we
perform our model fitting in the \uv\ plane, where the noise
properties of the data and the spatial filtering of the interferometer
are well defined.  The SZE model is generated in the image plane,
multiplied by the primary beam, and fast Fourier transformed to
produce model visibilities.  We then interpolate the model
visibilities to the $u$ and $v$ of each data visibility and compute
the Gaussian likelihood.  For X-ray data, the model is convolved with
the appropriate point spread function and the Poisson likelihood is
computed pixel by pixel, ignoring the masked point source regions.

Each data set is independent, and likelihoods from each data set can
simply be multiplied together to construct the joint likelihood.
Likelihood ratio tests can then be performed to get confidence regions
or compare two models.  Rather than working directly with likelihoods,
$\mathcal{L}$, we work with $S \equiv -2\ln(\mathcal{L})$.  We then
construct a $\Delta \chi^2$-like statistic from the log likelihoods,
$\Delta S \equiv S_n - S_{min}$ where $S_{min}$ is the minimum of the
$S$ function and $S_n$ is the $S$ statistic where $n$ parameters
differ from the parameters at $S_{min}$.  The statistic $\Delta S$ is
sometimes referred to as the Cash (1979)\nocite{cash1979} statistic
and tends to a $\chi^2$ distribution with $n$ degrees of freedom for
large $n$ \citep[][for example]{kendall1979}.  This $\Delta S$
statistic is equivalent to the likelihood ratio test and is used to
generate confidence regions and confidence intervals.  For one
interesting parameter, the 68.3\% ($\sim 1\sigma$) confidence level
corresponds to $\Delta S = 1.0$.

\subsubsection{Model Fitting Uncertainty Estimation}
\label{subsubsec:uncertain}

Uncertainties in the angular diameter distance from the fit parameters
are calculated by varying the interesting parameters to explore the
$\Delta S$ likelihood space.  The most important parameters in this
calculation are \dTo, \Xo, $\beta$, and $\theta_c$.  Radio point
sources and the cosmic X-ray background affect \dTo\ and \Xo,
respectively.  As a compromise between precision and computation time,
we systematically vary \dTo, \Xo, $\beta$, and $\theta_c$ allowing the
X-ray backgrounds for the PSPC and HRI to float independently while
fixing the positions of the cluster (both SZE and X-ray), and the
positions and flux densities of any radio point sources in the SZE
cluster fields.  We describe our estimation of the effects of point
sources below.

From this four dimensional $\Delta S$ hyper-surface, we construct
confidence intervals for each parameter individually as well as
confidence intervals for \Da\ due to \Xo, \dTo, $\beta$, and
$\theta_c$ jointly.  To compute the 68.3\% confidence region we find
the minimum and maximum values of the parameter within a $\Delta S$ of
1.0.  We emphasize that these uncertainties are meaningful only within
the context of the spherical isothermal $\beta$ model.

\vspace{0.1in}
\centerline{\it Measured Radio Point Sources}
\vspace{0.1in}

Two methods of estimating the effect of the measured radio point
sources in the cluster field are examined, one which is reasonably
quick and one that is more rigorous.  For the quick estimate, we first
determine the 1 $\sigma$ confidence limits on the flux density of each
point source by varying the point source flux density while keeping
the ICM parameters fixed at their best fit values.  These are the
uncertainties listed in Table~\ref{tab:pt_sources}, after correcting
for the primary beam attenuation appropriate for each point source's
distance from the pointing center.  We then determine the change in
the central decrement over the 68.3\% confidence region for the point
source flux densities by fixing the point source flux density at the
$\pm 1\ \sigma$ values and varying \dTo\ while fixing the ICM shape
parameters at their best fit values.  This is done for each point
source in the field and all combinations of the $\pm 1\ \sigma$ flux
densities for fields with multiple point sources.  We adopt the
maximum percentage change in \dTo\ as our uncertainty from radio point
sources on the central decrement.  The above procedure will be
referred to as the quick estimate of the effects of measured radio
point sources.  We tested this estimate against marginalizing over the
point source flux density by varying $\theta_c$, $\beta$, \Xo, \dTo,
and point source flux for each point source, while fixing the X-ray
background (simply saves computation time by isolating the point
source flux issue).  From the marginalized likelihood function we
find the best fit and 68.3\% uncertainty on \dTo\ and \Da.  The
uncertainty from measured radio point sources, $\sigma_{pt}$, is
computed assuming the uncertainties add in quadrature from
\begin{equation}
\sigma_{pt}^2 = \sigma_{mar}^2 - \sigma_{fix}^2,
\label{eq:sigpt}
\end{equation}
where $\sigma_{mar}$ and $\sigma_{fix}$ are the uncertainties from the
marginalized grids and the initial, point source fixed grids,
respectively.  Marginalizing over point source flux density was
performed on two clusters with one point source each, A2261 and
MS1358, and one cluster with two point sources, A1835.  The quick
estimation of the effects of point sources agrees to within 2\% on
\Da\ (1\% on \dTo) with the marginalized likelihood analysis, just
slightly over estimating the uncertainty due to point sources.  The
marginalization procedure is computationally intensive.  Therefore, to
save computation time, we use the quick procedure to estimate the
effects of detected point sources on the central decrement.

As an additional test, we explore the maximum likelihood parameter
space (varying $\theta_c$, $\beta$, \Xo, and \dTo) with the point
source fluxes fixed at the $\pm 1\ \sigma$ values for our three test
case clusters: A2261, MS1358, and A1835.  The effects of point sources
on the central decrement from this study agree within a few percent
with both the quick and marginalized procedures.  This is what was
originally done for MS0451 and Cl0016 \citep{reese2000}, which is now
shown to give essentially the same result as marginalizing over the
point source flux.  For all clusters, we use the updated, quick
estimates of the effects of measured point sources.

%%%%%%%%%%
%% IV.3 Model Fitting Results
%%%%%%%%%%
\subsection{Model Fitting Results}
\label{subsec:fit_results}

We apply the analysis procedure described above to all 18 of our galaxy
clusters.  The results from our maximum likelihood joint fit to the
SZE and X-ray data are summarized in Table~\ref{tab:fitparam}, which
shows the best-fit ICM shape parameters and the uncertainties on each
parameter from the model fit.

\begin{figure*}[ptbh]
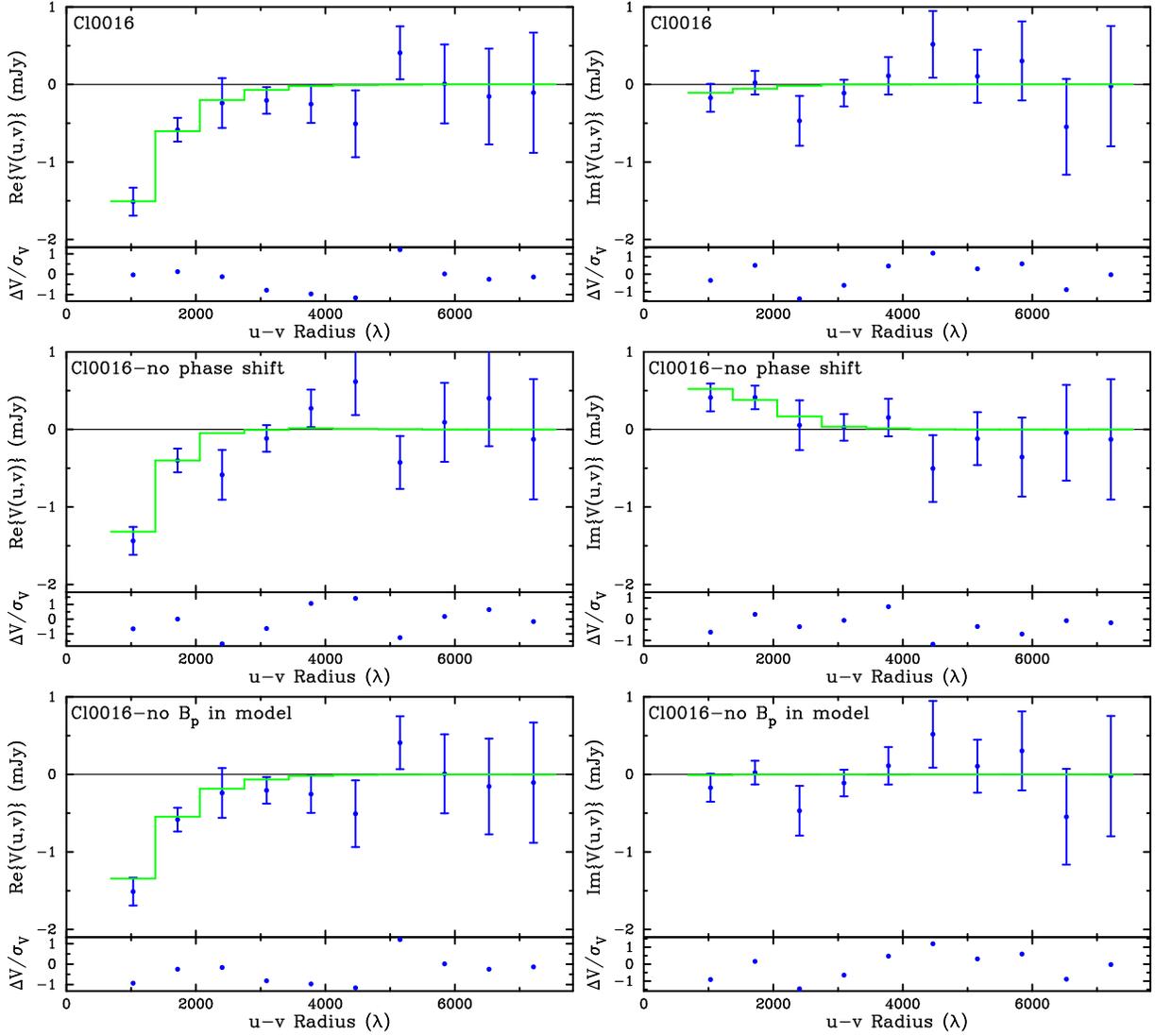

%\epsfxsize = 7.5 in
%%  \epsfxsize = 7.5 in
%%  \epsfysize = 6.0 in
%\centerline{\epsfbox{f?.eps}} 
\centerline{
  \psfig{figure=f4a_color.ps,height=1.9in}
  \psfig{figure=f4b_color.ps,height=1.9in}
}
\centerline{
  \psfig{figure=f4c_color.ps,height=1.9in}
  \psfig{figure=f4d_color.ps,height=1.9in}
}
\centerline{
  \psfig{figure=f4e_color.ps,height=1.9in}
  \psfig{figure=f4f_color.ps,height=1.9in}
}
\caption[SZE \uv\ radial profiles for Cl0016 illustrating the effects
of shifting the phase center and the primary beam] {The real and
imaginary components of the complex visibilities plotted as a function
of radius in the \uv\ plane for Cl0016.  The points with error bars
are the data and the best fit model from the joint SZE and X-ray
analysis is the solid line, averaged the same way as the data.
Residuals are shown in units of the standard deviation.  Shown are the
phase shifted and primary beam corrected (upper panel), not phase
shifted (middle panel), and not corrected for primary beam (lower
panel) versions.  Not shifting the phase center to the center of the
cluster (middle panel) shows an imaginary component from this offset.
Not applying the primary beam attenuation to the model after shifting
the phase to the center of the cluster (lower panel) shows the
expected zero imaginary component; a real and symmetric image should
have a real only Fourier transform.  The asymmetry induced by the
primary beam correction for the off-center cluster introduces a small
imaginary component (see upper panel).}
\label{fig:uvr_pbshift}
\end{figure*}

So far, we have only shown the SZE data in the form of images though
the data are recorded as visibilities.  Figure~\ref{fig:uvr_pbshift}
shows the SZE \uv\ radial profiles for Cl0016 with a series of 3
panels illustrating the features of such profiles.  These profiles are
azimuthal averages in the Fourier plane plotted as a function of the
radius in the \uv\ plane, $\sqrt{u^2 + v^2}$.  The data are the points
with error bars and the best fit $\beta$ model from the joint SZE and
X-ray analysis is the solid line averaged the same way as the data.
The point sources are subtracted directly from the visibilities before
constructing the \uv\ radial profiles.  All of these panels are shown
on the same scale for easy comparison.  Also plotted are the residuals
in units of the standard deviation, $\Delta V / \sigma$=(data $-$
model)/$\sigma$.  For a circular cluster at the phase center
(coincident with the pointing center), one expects a monotonic real
component and a zero imaginary component.  Clusters are rarely exactly
centered at the phase/pointing center of our observations.  Therefore,
we shift the phase center to the cluster center before constructing
the \uv\ radial profiles.  The phase shifted radial profiles are shown
in the upper panel of Figure~\ref{fig:uvr_pbshift} for both the real
(left) and imaginary (right) components of the complex visibilities.
The model provides a good fit to the data for a wide range of spatial
frequencies.  The middle panel shows the \uv\ radial profile when the
phase center is not shifted to the center of the cluster.  The
off-center cluster introduces corrugation in the Fourier plane
modifying the expected real component and introducing a non-zero
imaginary component.  In addition, asymmetry in the cluster will
manifest itself as a non-zero imaginary component.  Our model is
symmetric and its imaginary component should be identically zero.  The
attenuation from the primary beam introduces asymmetry, producing a
non-zero imaginary component.  This is illustrated in the lower panel
of Figure~\ref{fig:uvr_pbshift} showing the \uv\ radial profile
including the phase center shift but not including the primary beam
correction when computing the model.  Notice the model is identically
zero unlike the upper panel, where the asymmetry produced by the
primary beam on the off-center cluster shows a small imaginary
component.

%%%%%
% Table: ICM Parameters
%%%%%

%%ICM Parameters
%%
%%Cluster beta theta_c S_det S_cgs dTo Da
%%

\begin{deluxetable}{lcccccc}

\tablewidth{0pt}
%\tablenum{}
\tablecaption{ICM Parameters \label{tab:fitparam}}
\tablehead{
\colhead{} & 
\colhead{} & 
\colhead{$\theta_c$} & 
\colhead{\Xodet} & 
\colhead{\Xo} & 
\colhead{\dTo}&
\colhead{\Da}
 \\
\colhead{Cluster} & 
\colhead{$\beta$} & 
\colhead{(arcsec)} & 
\colhead{(detector)\tablenotemark{a}} & 
\colhead{(cgs)\tablenotemark{b}} & 
%\colhead{(cnt s$^{-1}$ arcmin$^{-2}$)} & 
%\colhead{(erg s$^{-1}$ cm$^{-2}$ arcmin$^{-2}$)} & 
\colhead{($\mu$K)} &
\colhead{(Mpc)}
}
\tablecolumns{7}
\startdata
MS1137	& $0.786^{+0.220}_{-0.120}$ & $\phn 19.4^{+6.4}_{-4.0}$ &
	$1.80^{+0.30}_{-0.24}$ $\times 10^{-2}$ &
	$4.43^{+0.74}_{-0.59}$ $\times 10^{-13}$ &
	$-\phn 818^{+98\phn}_{-113}$ 
	& $3179 ^{+1103}_{-1640}$  \\
MS0451	& $0.806^{+0.052}_{-0.043}$ & $\phn 34.7^{+3.9}_{-3.5}$ &
	$6.96^{+0.63}_{-0.61}$ $\times 10^{-2}$ & 
	$9.56^{+0.86}_{-0.84}$ $\times 10^{-13}$ &
	$-1431^{+98\phn}_{-93\phn}$ 
	& $1278 ^{+265\phn}_{-299\phn}$  \\
Cl0016	& $0.749^{+0.024}_{-0.018}$ & $\phn 42.3^{+2.4}_{-2.0}$ &
	$4.14^{+0.15}_{-0.19}$ $\times 10^{-2}$ & 
	$6.17^{+0.22}_{-0.28}$ $\times 10^{-13}$ &
	$-1242^{+105}_{-105}$ 
	& $2041 ^{+484\phn}_{-514\phn}$  \\
R1347	& $0.604^{+0.011}_{-0.012}$ & $\phn\phn 9.0^{+0.5}_{-0.5}$ &
	$6.70^{+0.39}_{-0.34}$ $\times 10^{-1}$ & 
	$2.74^{+0.16}_{-0.14}$ $\times 10^{-11}$ &
	$-3950^{+350}_{-350}$ 
	& $1221 ^{+368\phn}_{-343\phn}$  \\
A370	& $0.518^{+0.090}_{-0.080}$ & $\phn 39.5^{+10.5}_{-10.5}$ &
	$8.88^{+1.41}_{-0.99}$ $\times 10^{-3}$ &
	$2.70^{+0.43}_{-0.30}$ $\times 10^{-13}$ &
	$-1253^{+218}_{-533}$ 
	& $4352 ^{+1388}_{-1245}$ \\
MS1358	& $0.622^{+0.015}_{-0.015}$ & $\phn 18.2^{+1.4}_{-1.5}$ &
	$1.27^{+0.11}_{-0.08}$ $\times 10^{-1}$ &
	$1.70^{+0.15}_{-0.11}$ $\times 10^{-12}$ &
	$-\phn 784^{+90\phn}_{-90\phn}$ 
	& $\phn 866 ^{+248\phn}_{-310\phn}$ \\
A1995	& $0.770^{+0.117}_{-0.063}$ & $\phn 38.9^{+6.9}_{-4.3}$ &
	$3.18^{+0.24}_{-0.21}$ $\times 10^{-2}$ &
	$1.08^{+0.08}_{-0.07}$ $\times 10^{-12}$ &
	$-1023^{+83\phn}_{-77\phn}$ 
	& $1119 ^{+247\phn}_{-282\phn}$ \\
A611	& $0.565^{+0.050}_{-0.040}$ & $\phn 17.5^{+3.5}_{-3.5}$ &
	$5.91^{+1.06}_{-0.76}$ $\times 10^{-2}$ &
	$2.01^{+0.36}_{-0.26}$ $\times 10^{-12}$ &
	$-\phn 853^{+120}_{-140}$ 
	& $\phn 995 ^{+325\phn}_{-293\phn}$ \\
A697	& $0.540^{+0.045}_{-0.035}$ & $\phn 37.8^{+5.6}_{-4.0}$ &
	$3.24^{+0.22}_{-0.25}$ $\times 10^{-2}$ &
	$1.02^{+0.07}_{-0.08}$ $\times 10^{-12}$ &
	$ -1410^{+160}_{-180}$ 
	& $\phn 998 ^{+298\phn}_{-250\phn}$ \\
A1835	& $0.595^{+0.007}_{-0.005}$ & $\phn 12.2^{+0.6}_{-0.5}$ &
	$1.50^{+0.10}_{-0.07}$ $\times 10^{-0}$ &
	$2.02^{+0.14}_{-0.10}$ $\times 10^{-11}$ &
	$-2502^{+150}_{-175}$ 
	& $1027 ^{+194\phn}_{-198\phn}$ \\
A2261	& $0.516^{+0.014}_{-0.013}$ & $\phn 15.7^{+1.2}_{-1.1}$ &
	$1.37^{+0.08}_{-0.08}$ $\times 10^{-1}$ &
	$4.31^{+0.26}_{-0.26}$ $\times 10^{-12}$ &
	$-1697^{+200}_{-200}$ 
	& $1049 ^{+306\phn}_{-272\phn}$ \\
A773	& $0.597^{+0.064}_{-0.032}$ & $\phn 45.0^{+7.0}_{-5.0}$ &
	$3.05^{+0.24}_{-0.24}$ $\times 10^{-2}$ &
	$8.28^{+0.65}_{-0.65}$ $\times 10^{-13}$ &
	$-1260^{+160}_{-160}$ 
	& $1450 ^{+361\phn}_{-332\phn}$ \\
A2163	& $0.674^{+0.011}_{-0.008}$ & $\phn 87.5^{+2.5}_{-2.0}$ &
	$6.82^{+0.15}_{-0.15}$ $\times 10^{-2}$ &
	$1.36^{+0.03}_{-0.03}$ $\times 10^{-12}$ &
	$-1900^{+140}_{-140}$ 
	& $\phn 828^{+181\phn}_{-205\phn}$ \\
A520	& $0.844^{+0.040}_{-0.040}$ & $123.3^{+8.0}_{-8.0}$ &
	$2.57^{+0.11}_{-0.11}$ $\times 10^{-2}$ &
	$4.08^{+0.18}_{-0.18}$ $\times 10^{-13}$ &
	$-\phn 662^{+95\phn}_{-95\phn}$ 
	& $\phn 723 ^{+270\phn}_{-236\phn}$ \\
A1689	& $0.609^{+0.005}_{-0.005}$ & $\phn 26.6^{+0.7}_{-0.7}$ &
	$4.50^{+0.13}_{-0.11}$ $\times 10^{-1}$ &
	$6.01^{+0.18}_{-0.15}$ $\times 10^{-12}$ &
	$-1729^{+105}_{-120}$ 
	& $\phn 688 ^{+172\phn}_{-163\phn}$ \\
A665	& $0.615^{+0.006}_{-0.006}$ & $\phn 71.7^{+1.5}_{-1.5}$ &
	$4.75^{+0.08}_{-0.08}$ $\times 10^{-2}$ &
	$6.78^{+0.12}_{-0.12}$ $\times 10^{-13}$ &
	$-\phn 728^{+150}_{-150}$ 
	& $\phn 466 ^{+217\phn}_{-179\phn}$ \\
A2218	& $0.692^{+0.008}_{-0.008}$ & $\phn 67.5^{+1.5}_{-1.8}$ &
	$5.14^{+0.12}_{-0.10}$ $\times 10^{-2}$ &
	$7.08^{+0.16}_{-0.14}$ $\times 10^{-13}$ &
	$-\phn 731^{+125}_{-100}$ 
	& $1029 ^{+339\phn}_{-352\phn}$ \\
A1413	& $0.639^{+0.009}_{-0.009}$ & $\phn47.7^{+2.0}_{-2.0}$ &
	$1.52^{+0.07}_{-0.07}$ $\times 10^{-1}$ &
	$2.04^{+0.09}_{-0.09}$ $\times 10^{-12}$ &
	$-\phn 856^{+110}_{-110}$ 
	& $\phn 573 ^{+171\phn}_{-151\phn}$
\enddata
\tablenotetext{a}{Units are cnt s$^{-1}$ arcmin$^{-2}$.}
\tablenotetext{b}{Units are erg s$^{-1}$ cm$^{-2}$ arcmin$^{-2}$.}
%\tablecomment{}

\end{deluxetable}

%%%%%
% Figure of UV radial profiles w/best fit model; this is a comparison
% for with and without the primary beam correction of the model and
% the shifting of the phase center
%%%%%
%%
% Figure 4
%%

%%%%%
% Figure of UV radial profiles w/best fit model;  all point sources
% are subtracted first
%%%%%
%%
% Figure 5
%%

\begin{figure*}[tbph]
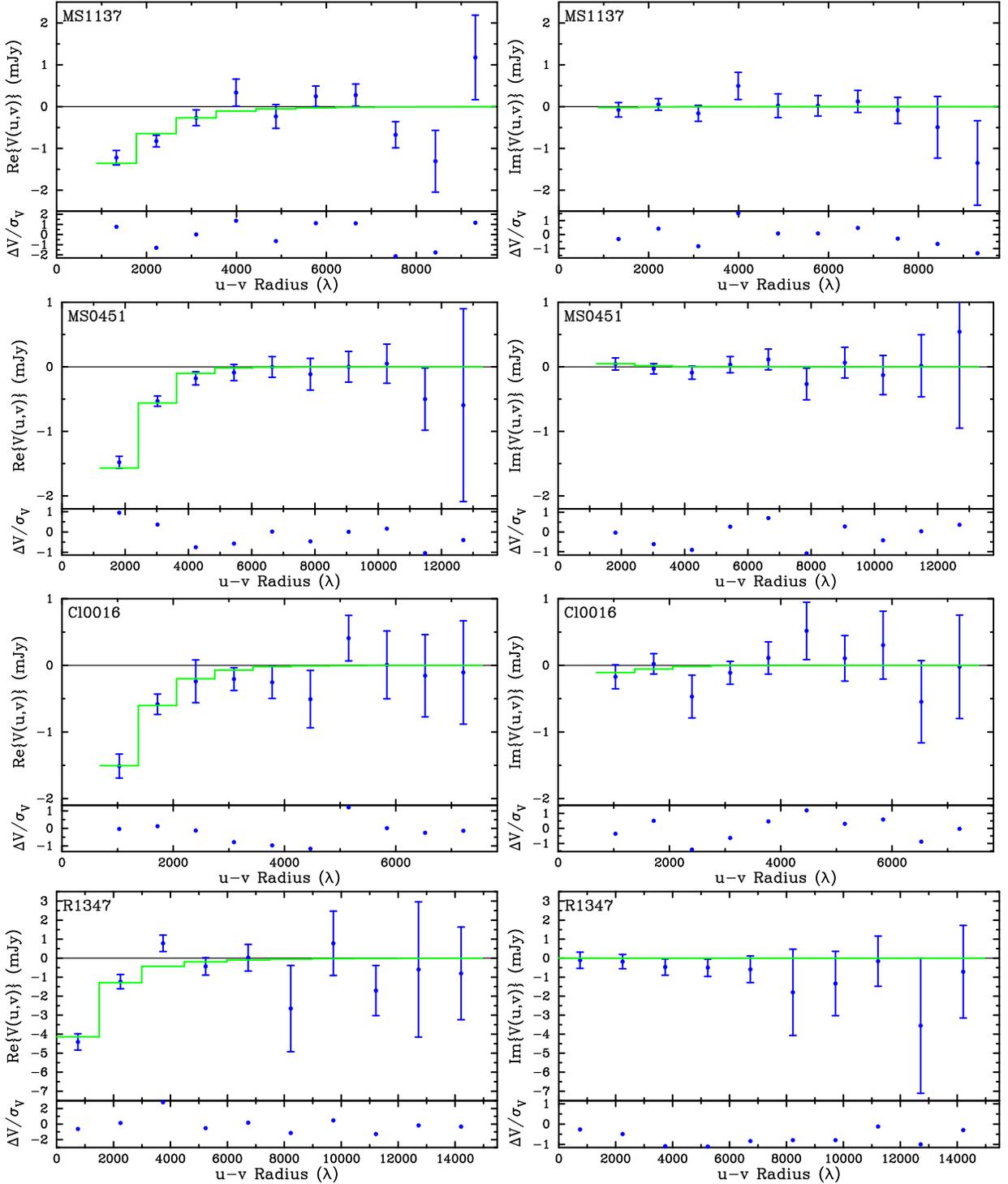

%\epsfxsize = 7.5 in
%%  \epsfxsize = 7.5 in
%%  \epsfysize = 6.0 in
%\centerline{\epsfbox{f?.eps}} 
\centerline{
  \psfig{figure=f5a_color.ps,height=1.9in}
  \psfig{figure=f5b_color.ps,height=1.9in}
}
\centerline{
  \psfig{figure=f5c_color.ps,height=1.9in}
  \psfig{figure=f5d_color.ps,height=1.9in}
}
\centerline{
  \psfig{figure=f5e_color.ps,height=1.9in}
  \psfig{figure=f5f_color.ps,height=1.9in}
}
\centerline{
  \psfig{figure=f5g_color.ps,height=1.9in}
  \psfig{figure=f5h_color.ps,height=1.9in}
}
\caption[SZE \uv\ radial profiles for the real and imaginary parts of
the visibilities]{The real and imaginary components of the complex
visibilities plotted as a function of radius in the \uv-plane.  The
points and error bars are the data and the solid line is the best fit
model.  Residuals are shown in units of the standard deviation.  Any
point sources in the cluster field have been subtracted directly from
the visibilities and the phase center of the map has been shifted to
the center of the cluster before making these radial averages.
Non-zero signal in the imaginary components is due to the asymmetry in
the cluster and the possible asymmetry introduced by the primary beam
correction.  The models provide good fits to the data for all the
clusters in our sample.}
\label{fig:uvradprof}
\end{figure*}

\begin{figure*}[tbph]
\centerline{
  \psfig{figure=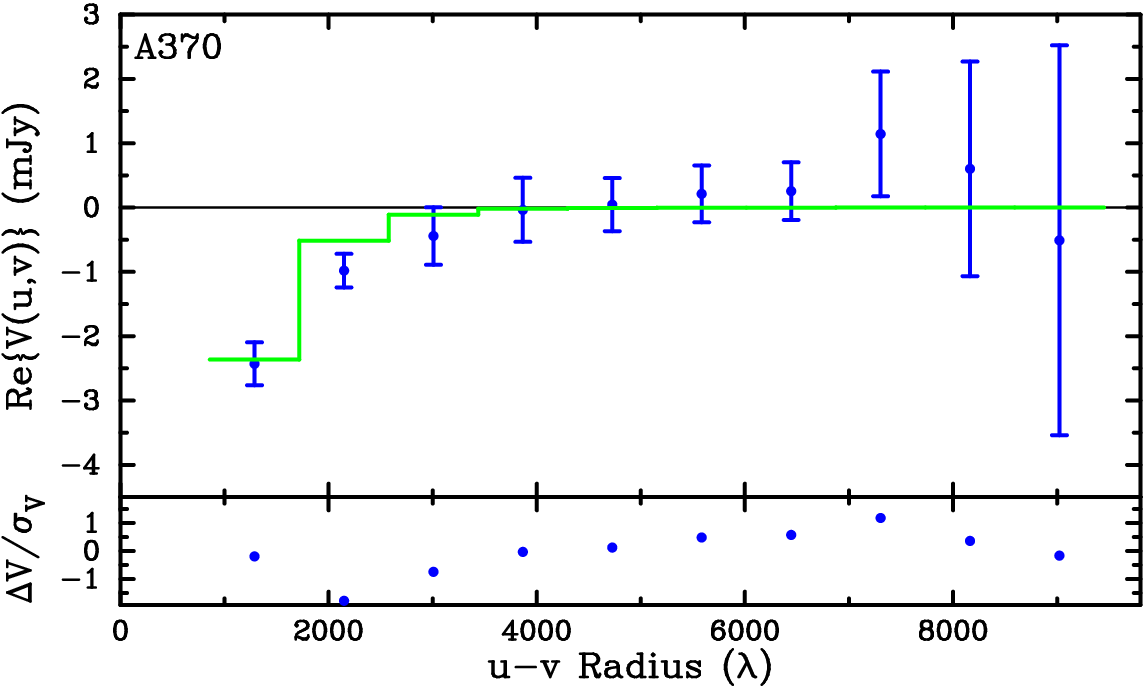,height=1.9in}
  \psfig{figure=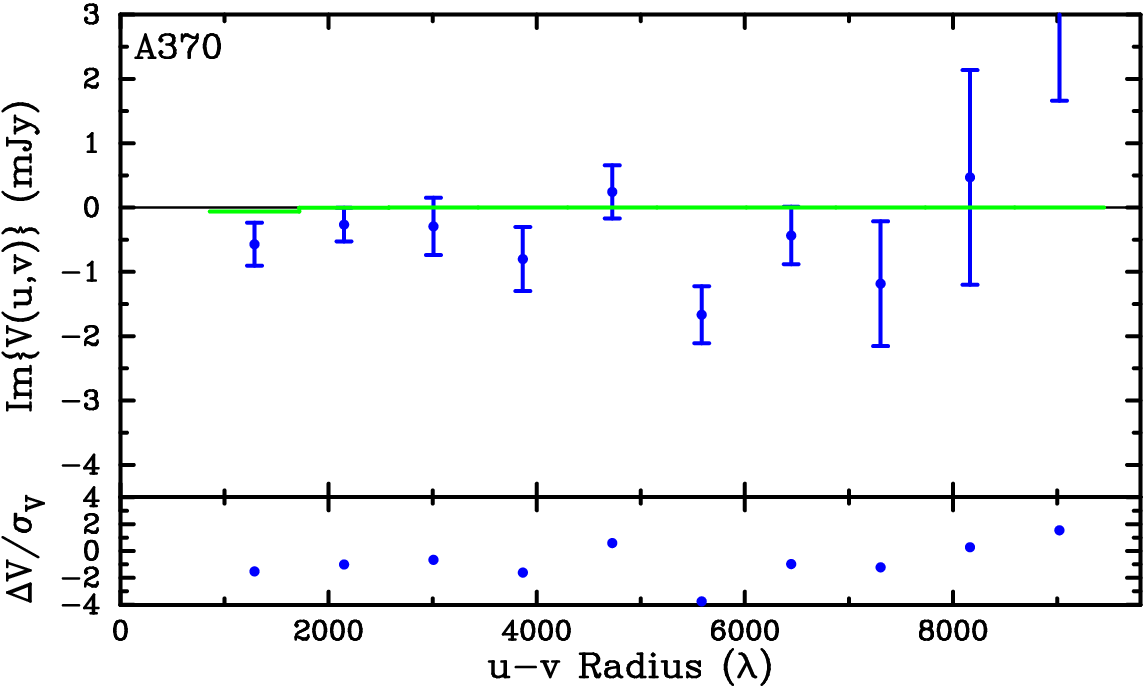,height=1.9in}
}
\centerline{
  \psfig{figure=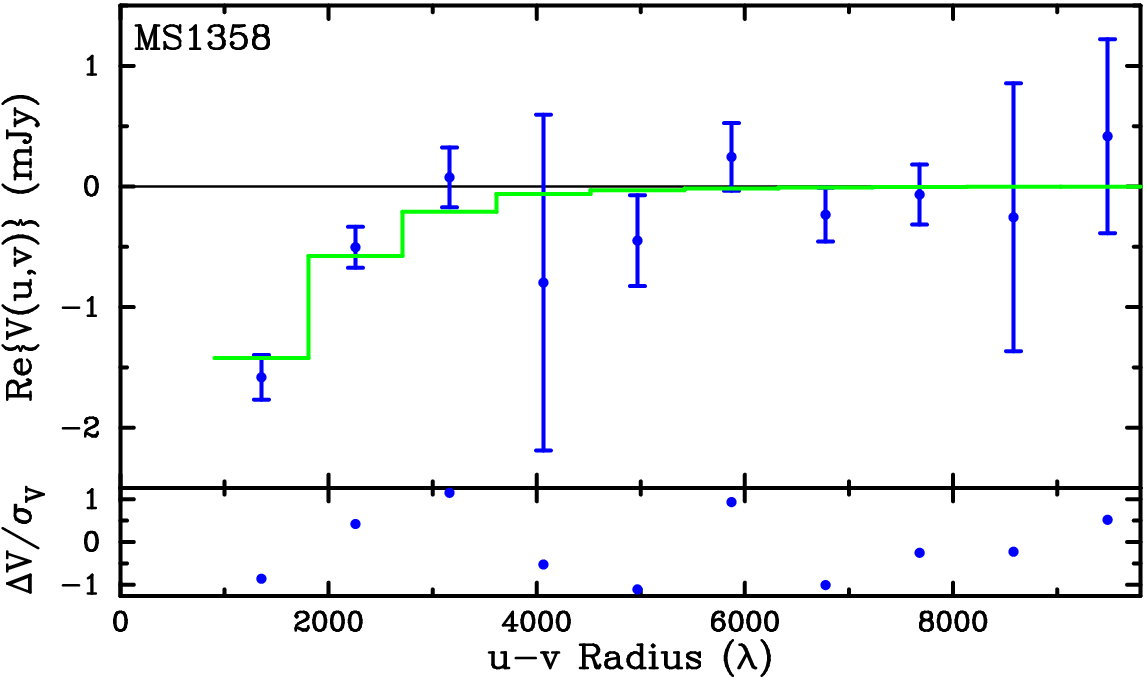,height=1.9in}
  \psfig{figure=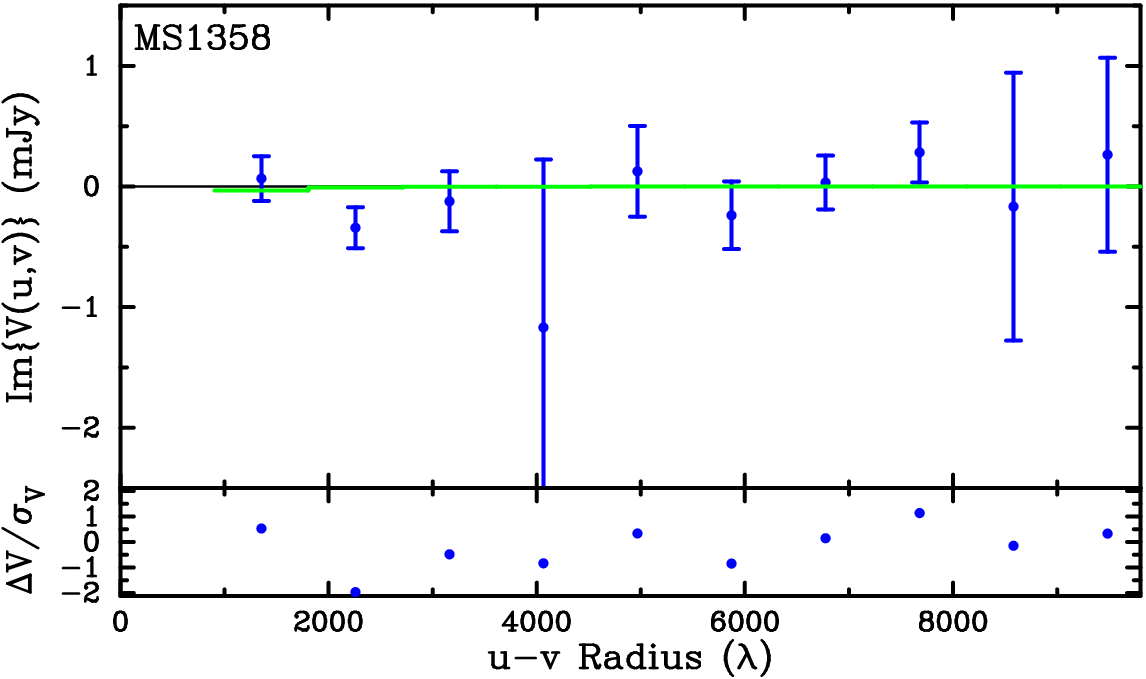,height=1.9in}
}
\centerline{
  \psfig{figure=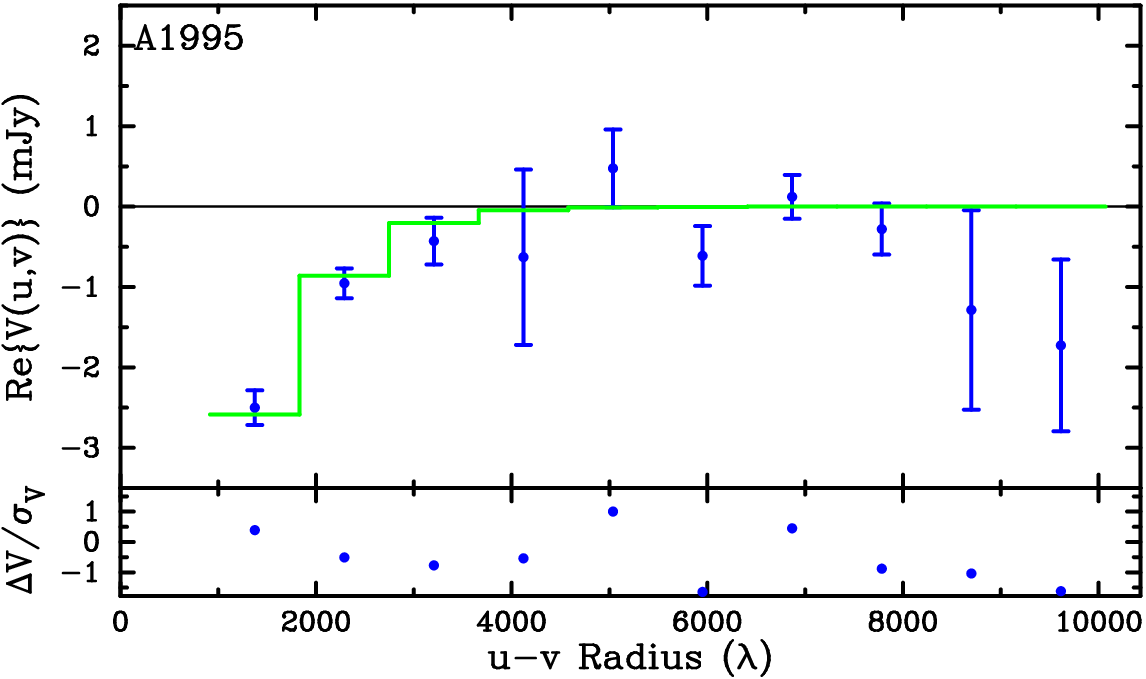,height=1.9in}
  \psfig{figure=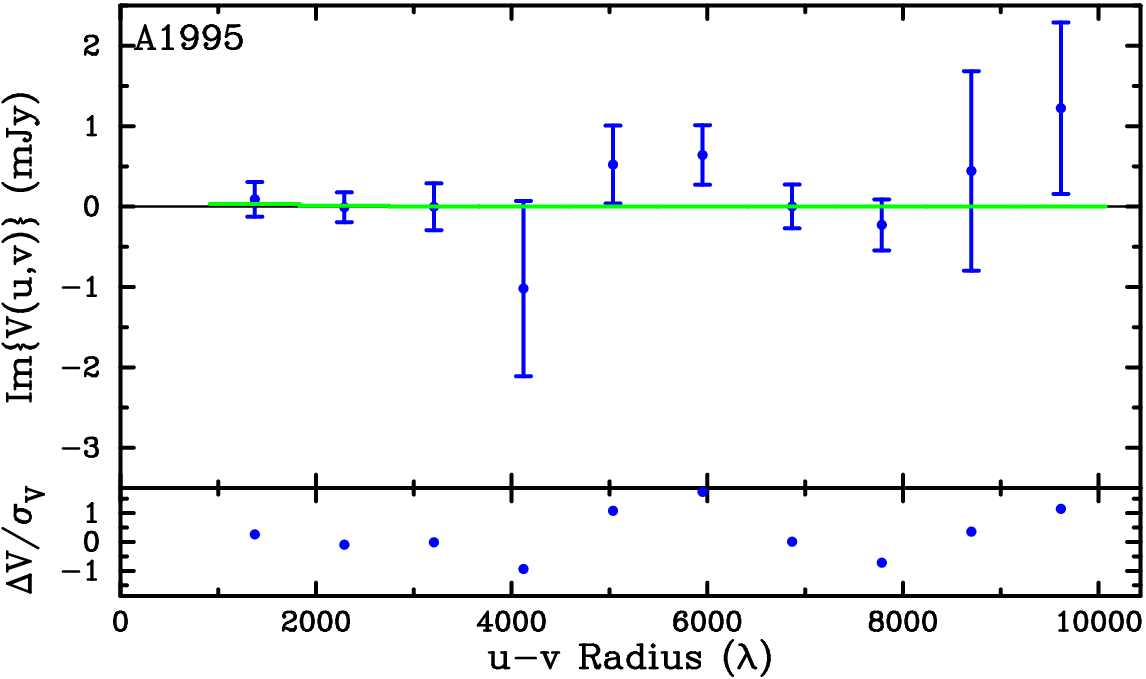,height=1.9in}
}		
\centerline{	
  \psfig{figure=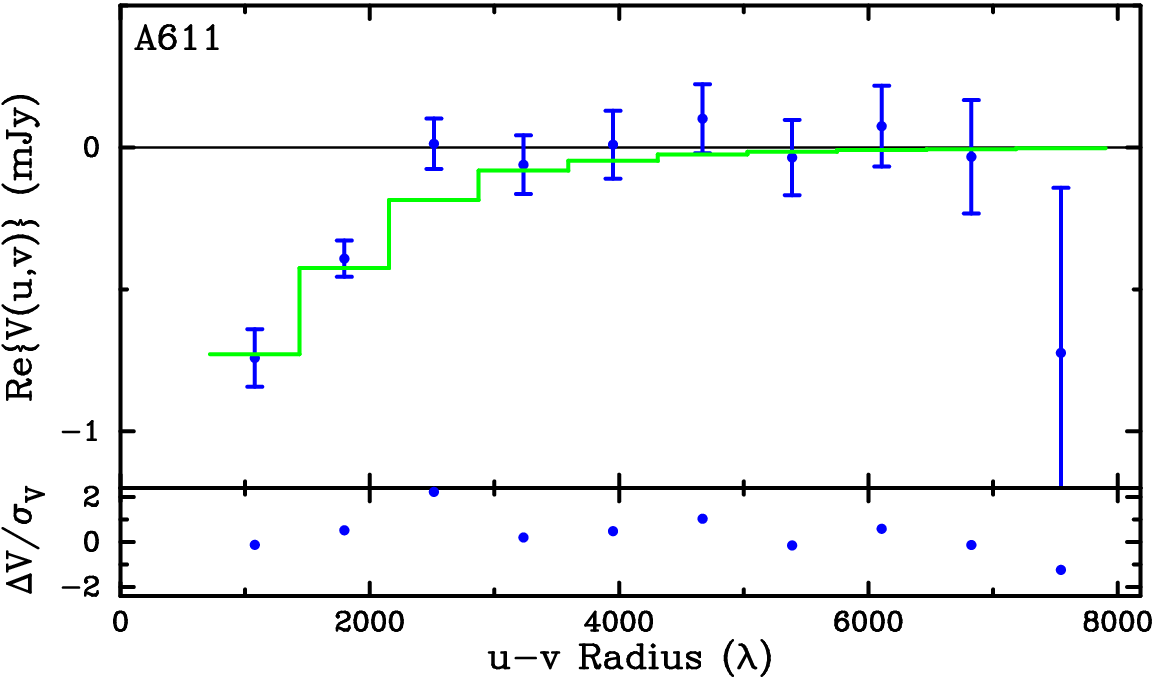,height=1.9in}
  \psfig{figure=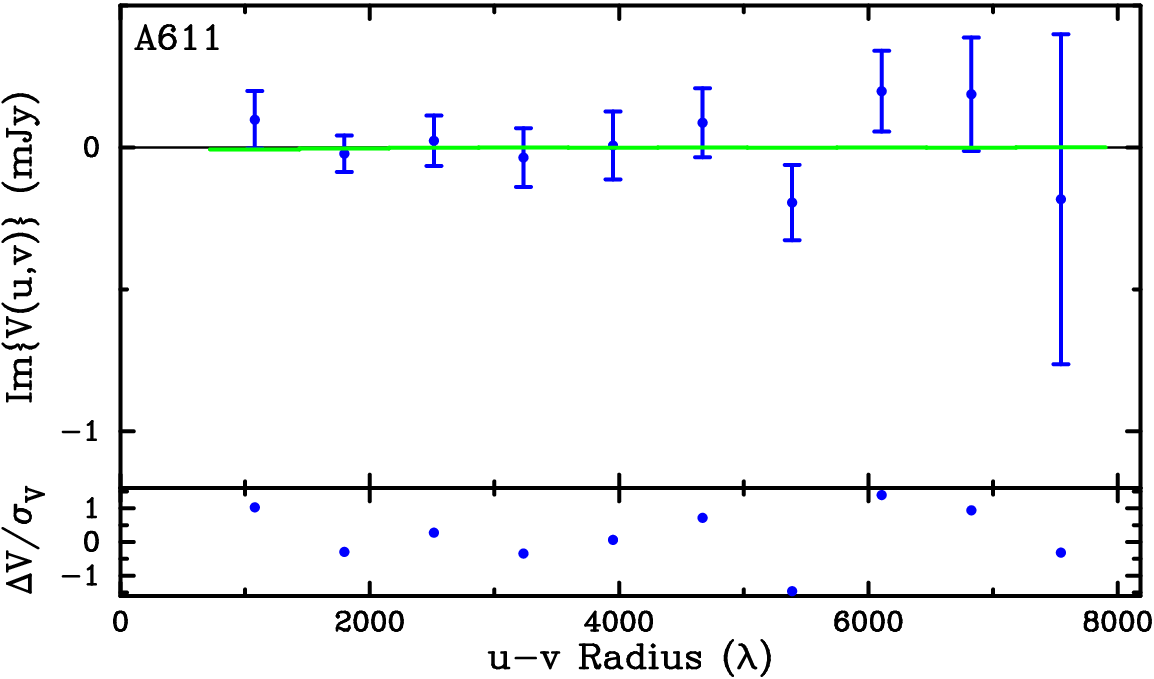,height=1.9in}
}		
\contcaption{Cont.}
\end{figure*}

\begin{figure*}[tbph]
\centerline{	
  \psfig{figure=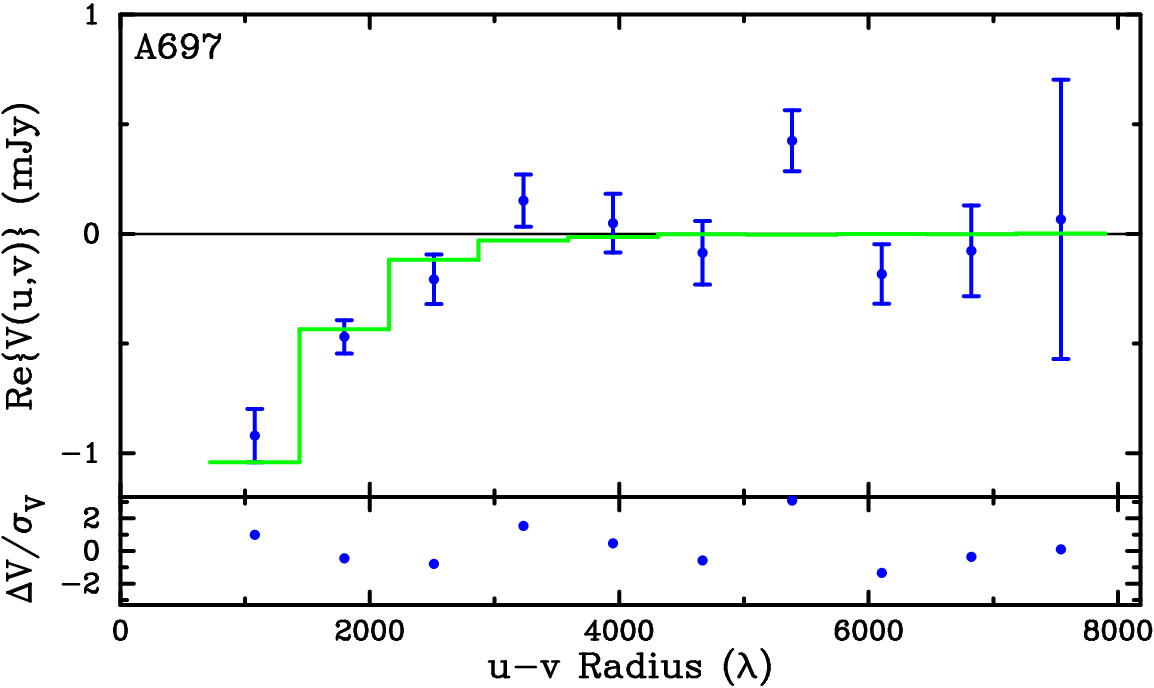,height=1.9in}
  \psfig{figure=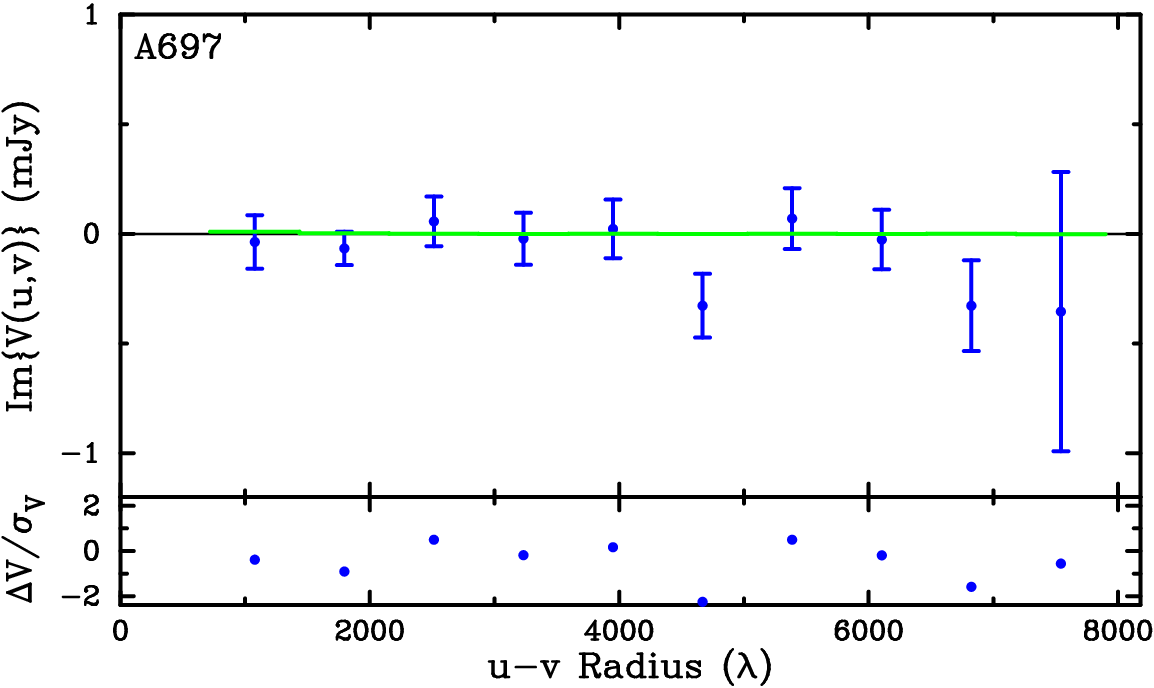,height=1.9in}
}
\centerline{
  \psfig{figure=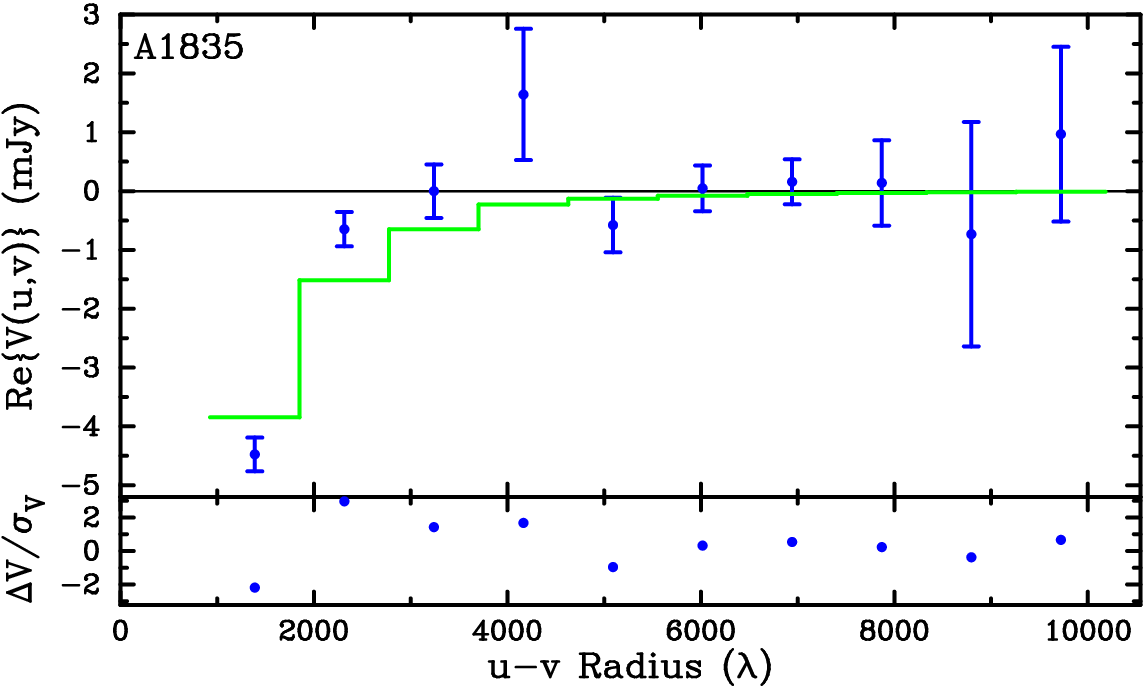,height=1.9in}
  \psfig{figure=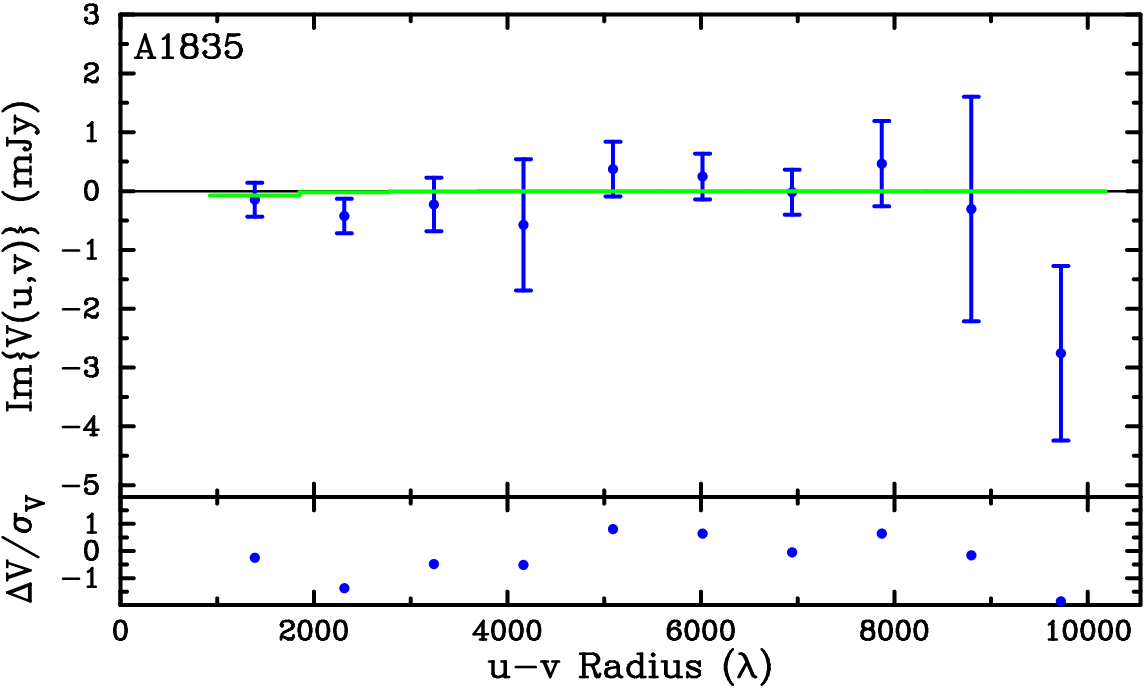,height=1.9in}
}		
\centerline{	
  \psfig{figure=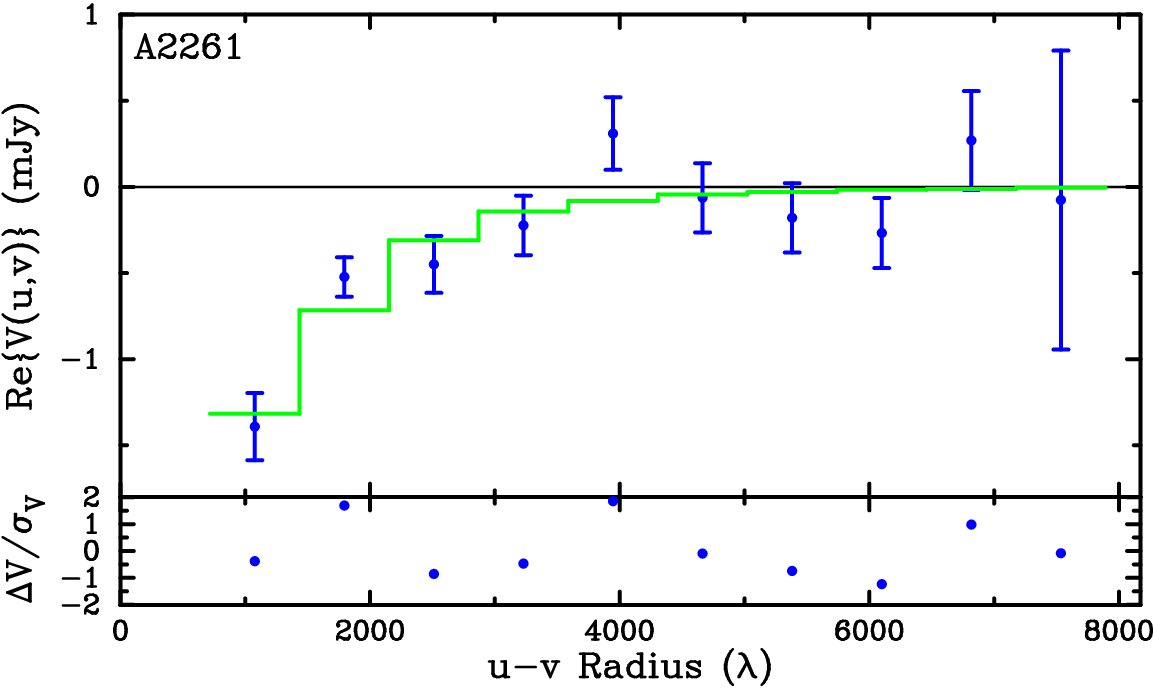,height=1.9in}
  \psfig{figure=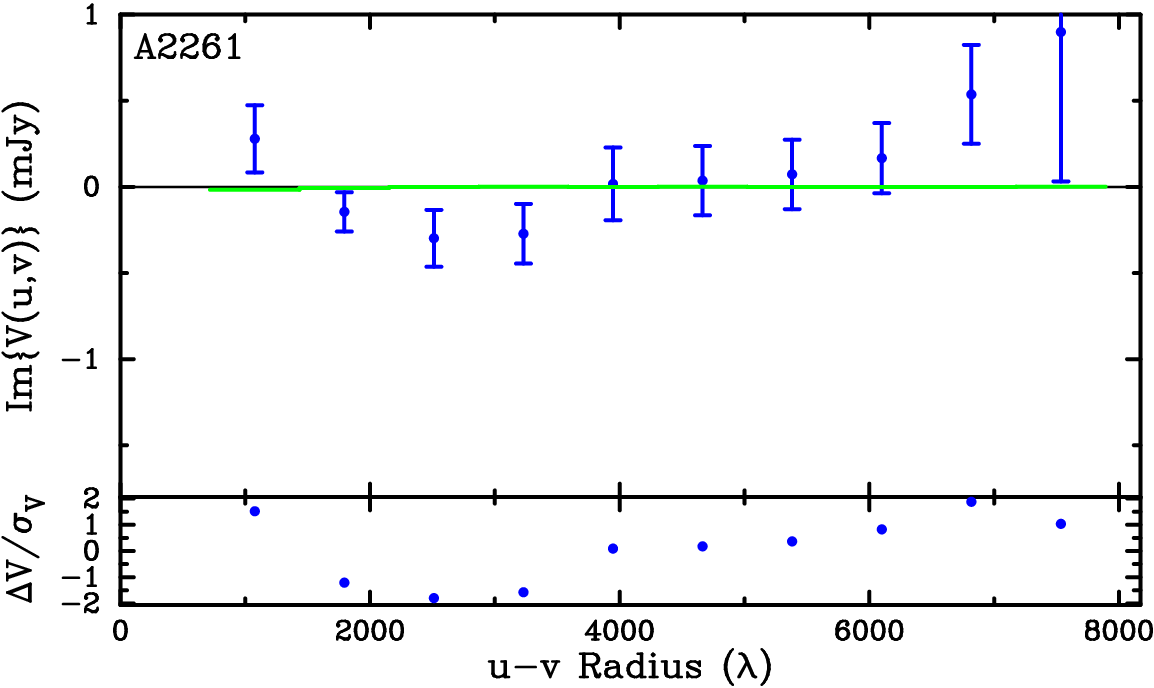,height=1.9in}
}		
\centerline{	
  \psfig{figure=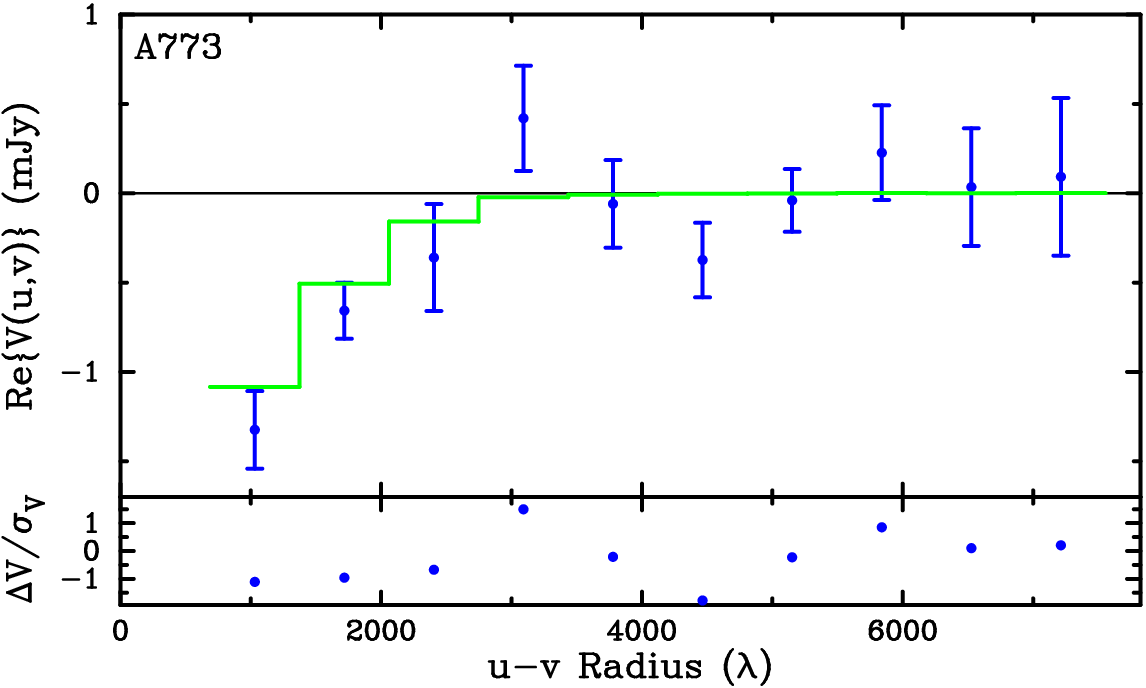,height=1.9in}
  \psfig{figure=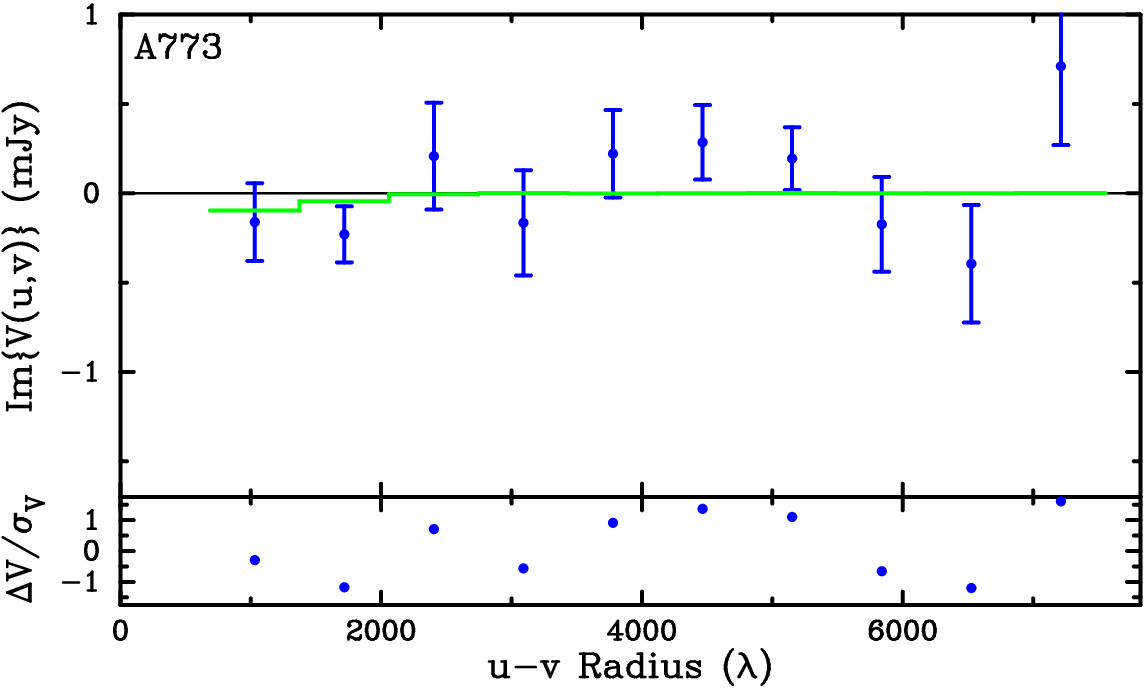,height=1.9in}
}
\contcaption{Cont.}
\end{figure*}

\begin{figure*}[tbph]
\centerline{
  \psfig{figure=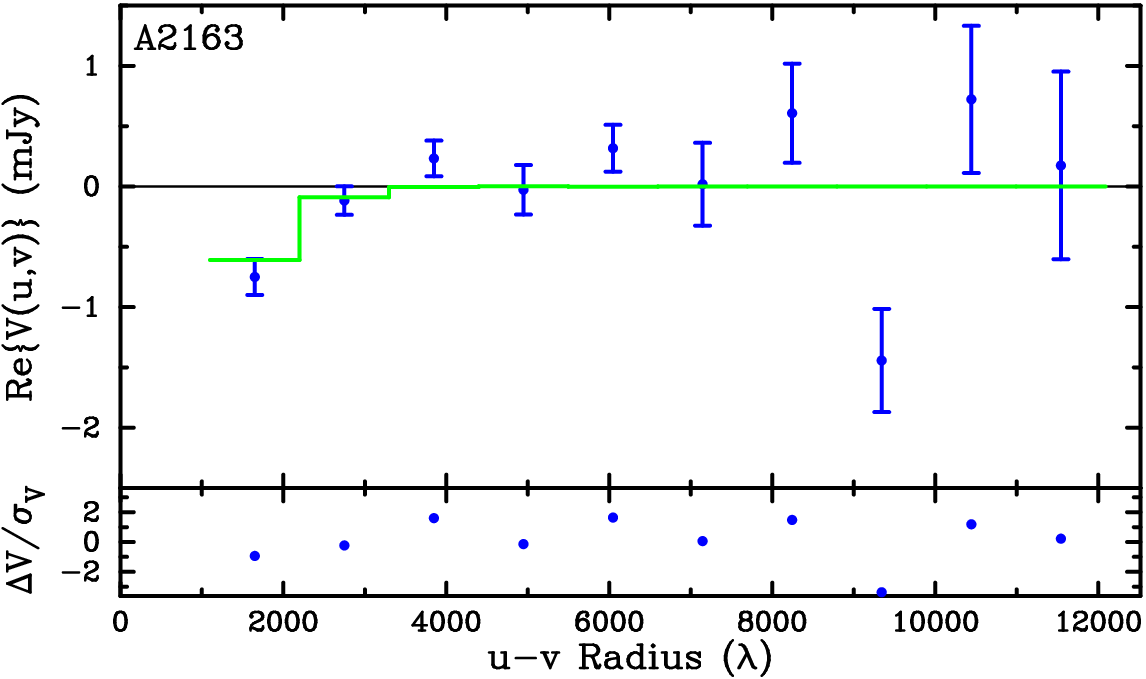,height=1.9in}
  \psfig{figure=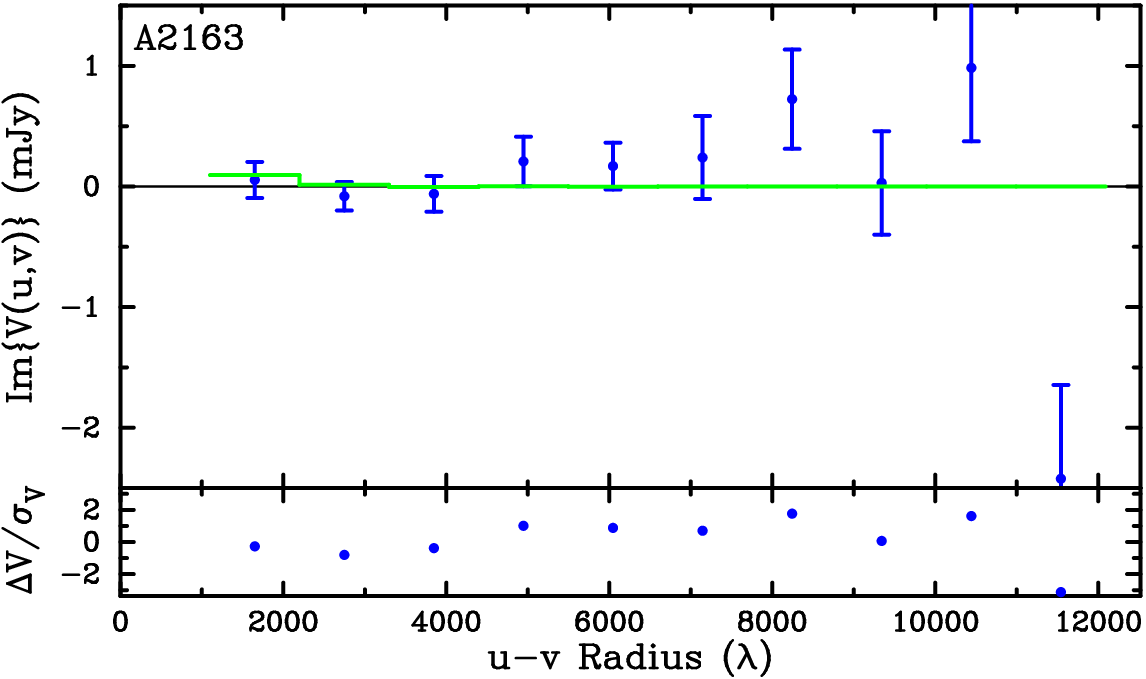,height=1.9in}
}		
\centerline{	
  \psfig{figure=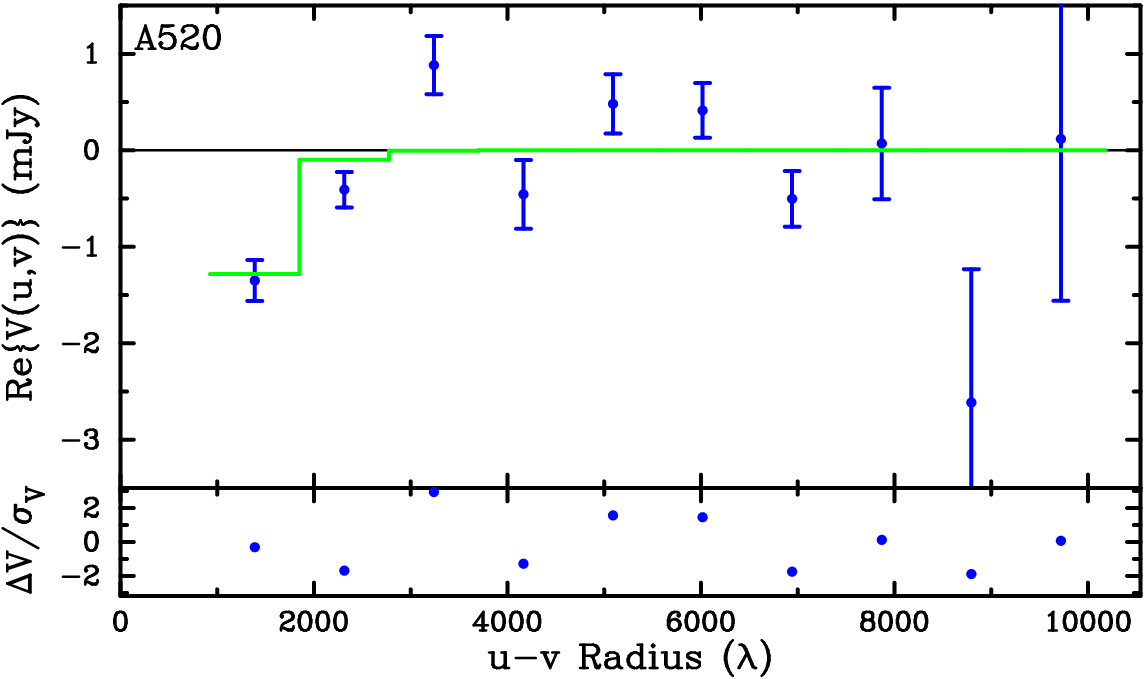,height=1.9in}
  \psfig{figure=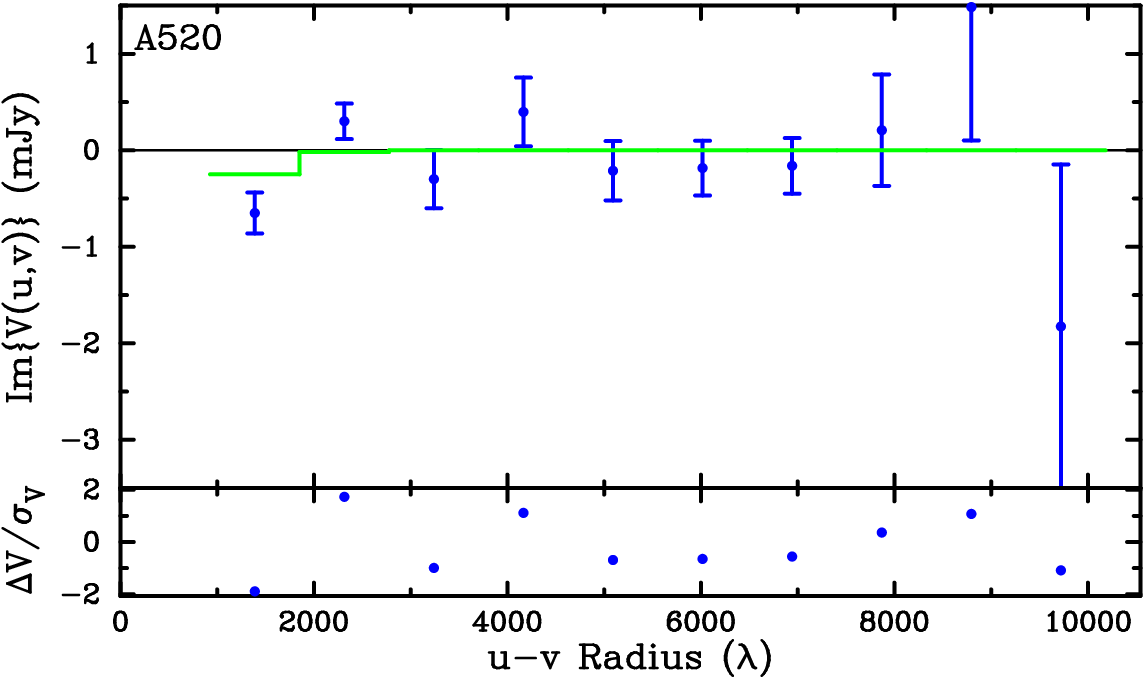,height=1.9in}
}		
\centerline{	
  \psfig{figure=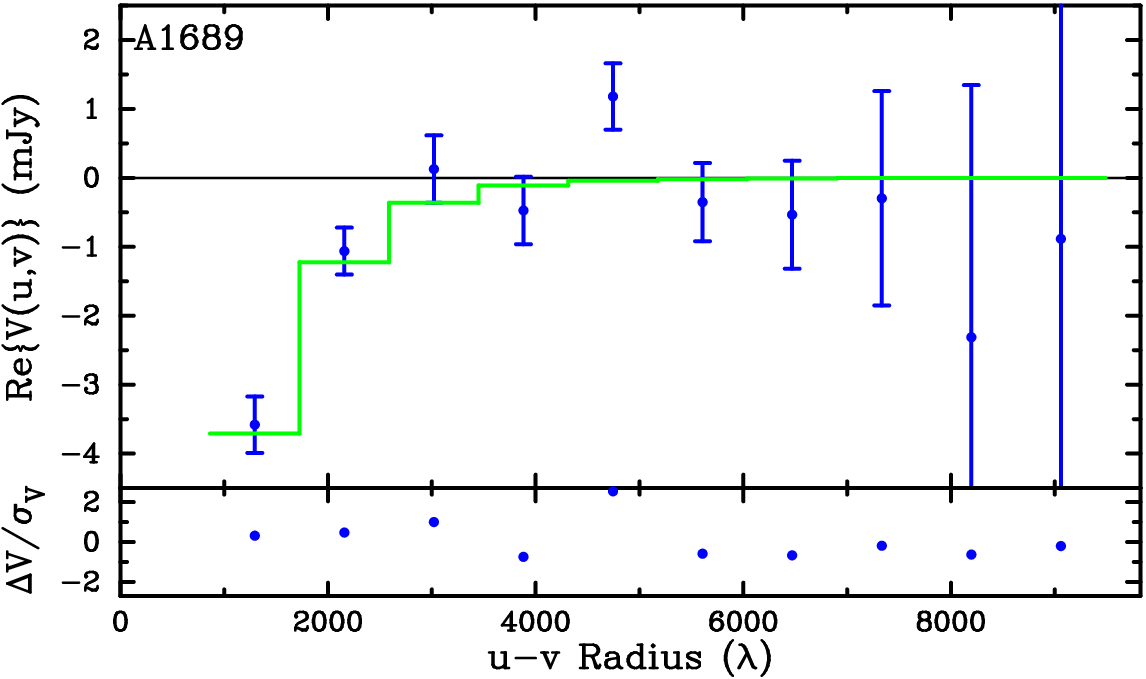,height=1.9in}
  \psfig{figure=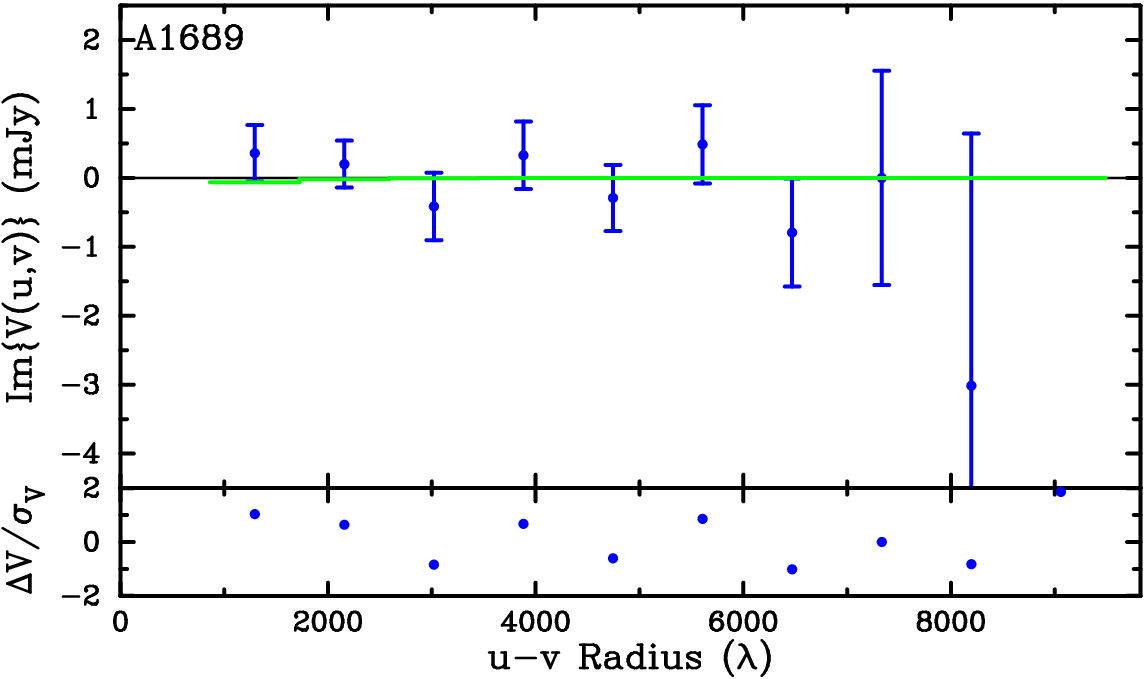,height=1.9in}
}
\centerline{
  \psfig{figure=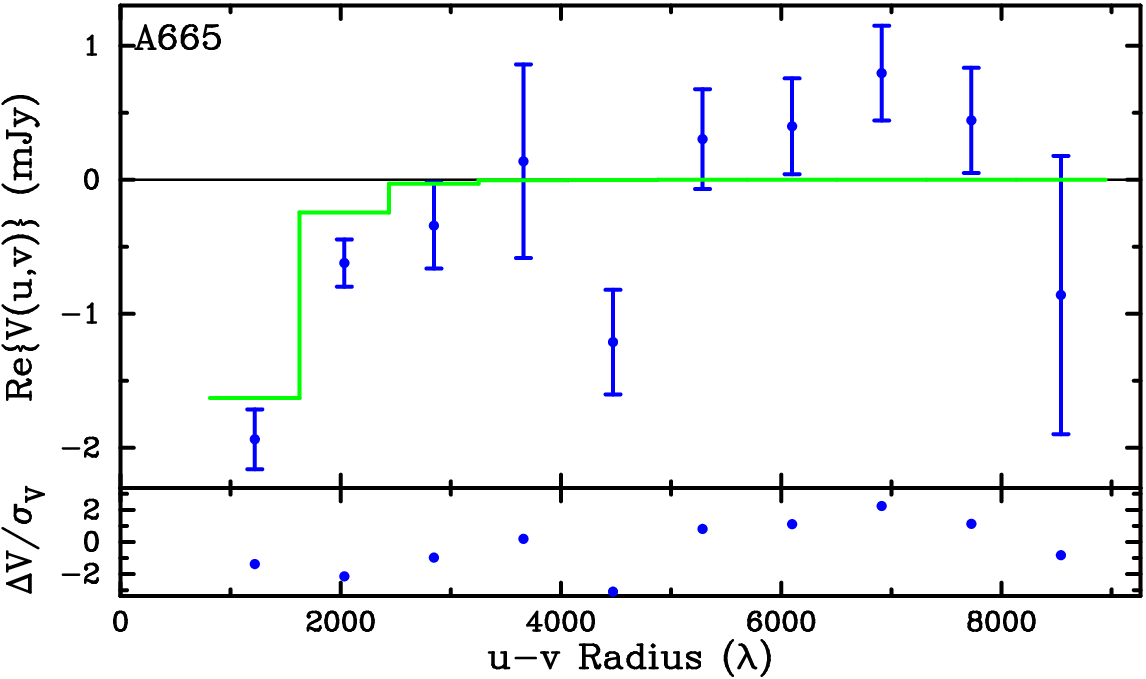,height=1.9in}
  \psfig{figure=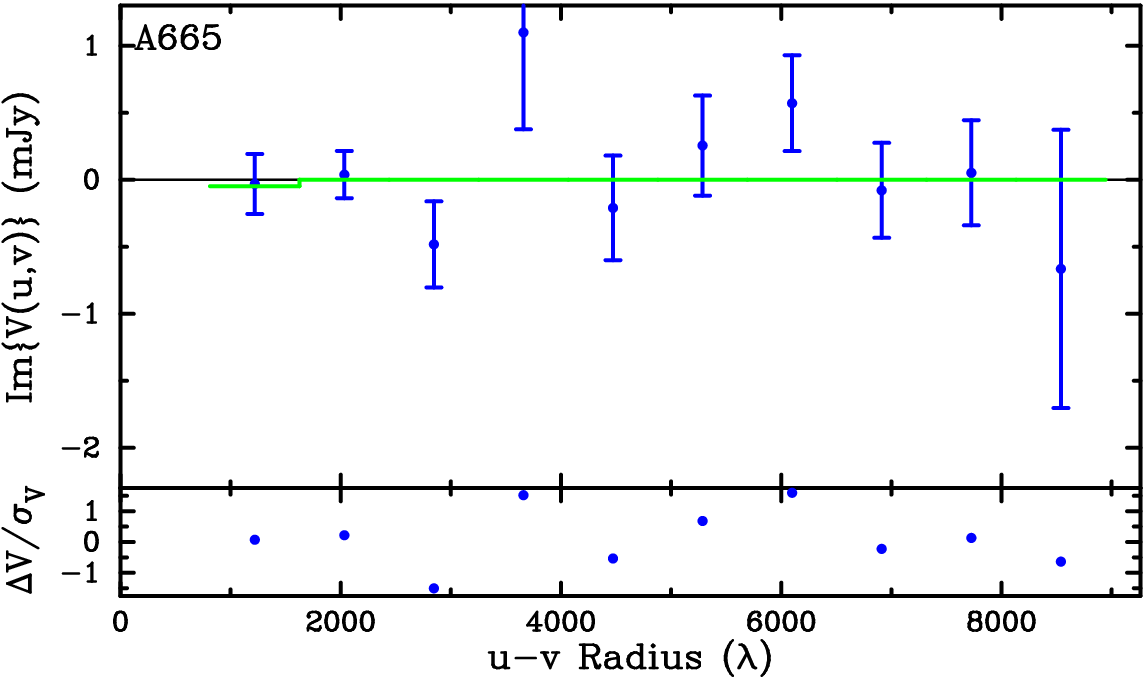,height=1.9in}
}		
\contcaption{Cont.}
\end{figure*}

\begin{figure*}[tbph]
\centerline{	
  \psfig{figure=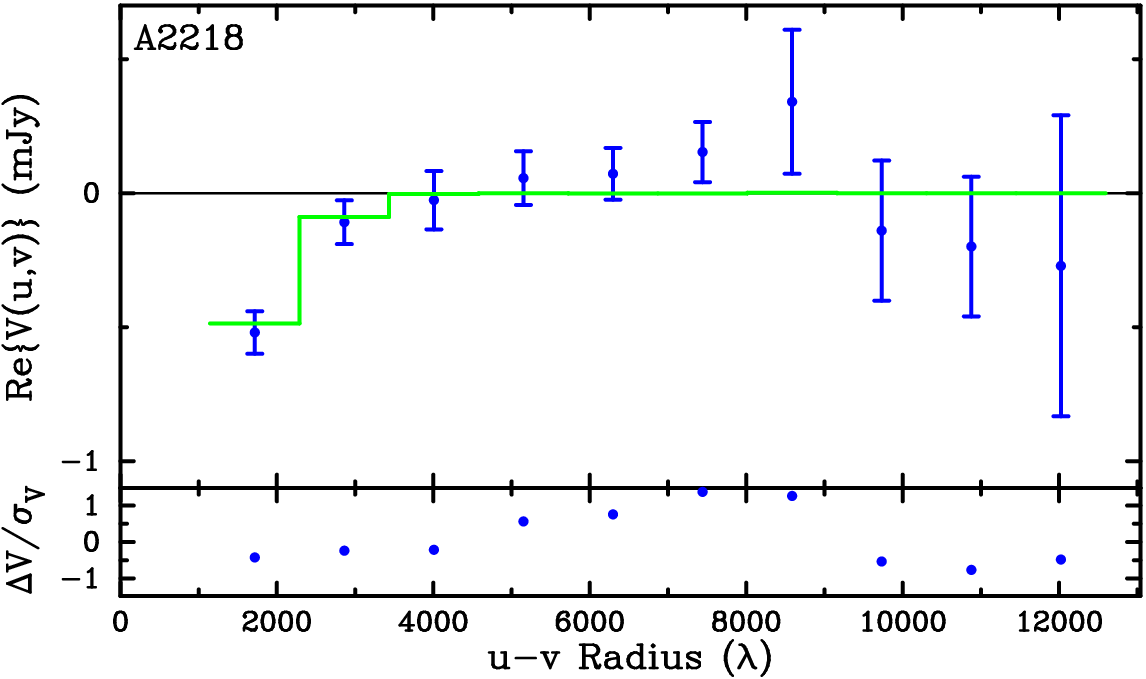,height=1.9in}
  \psfig{figure=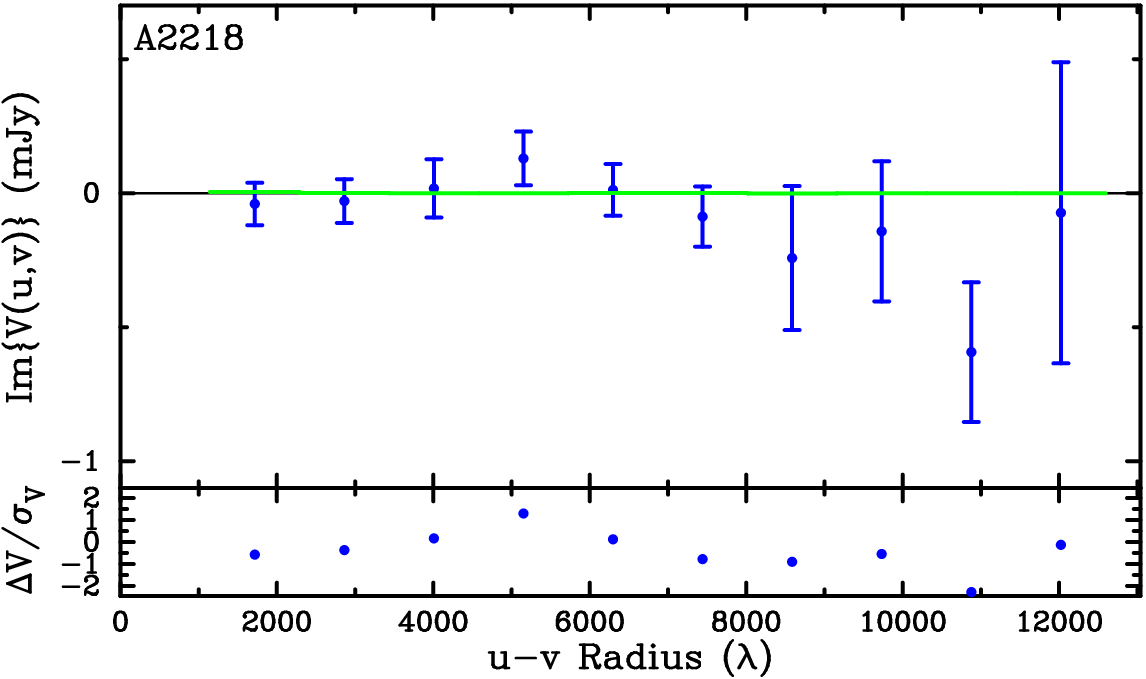,height=1.9in}
}		
\centerline{	
  \psfig{figure=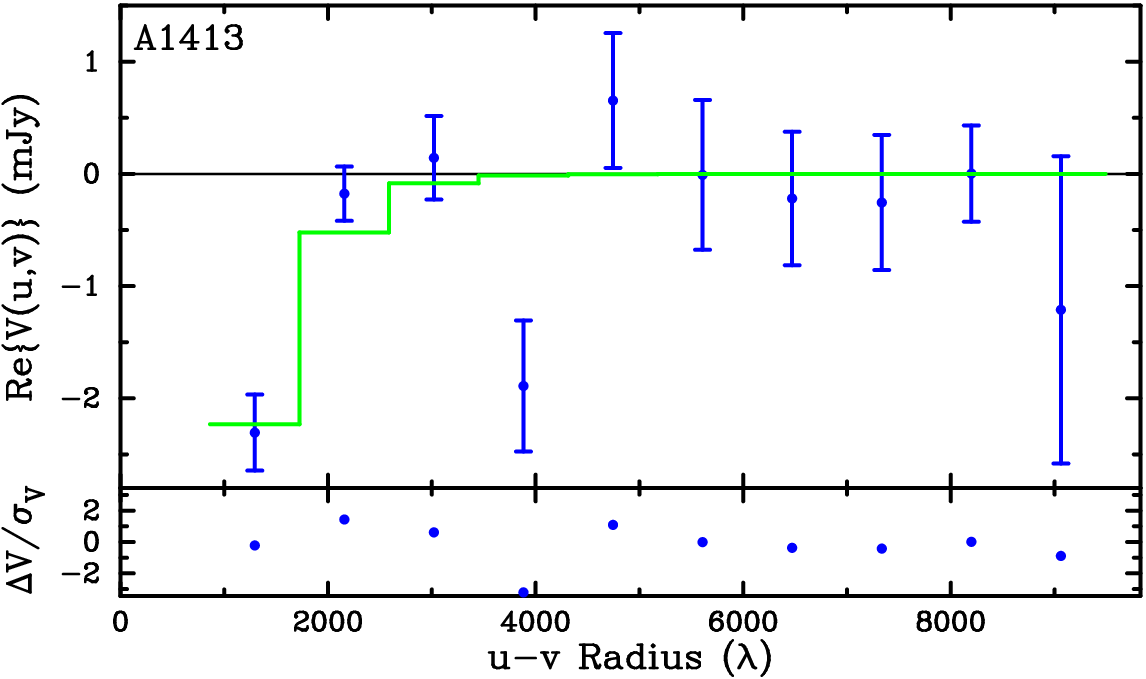,height=1.9in}
  \psfig{figure=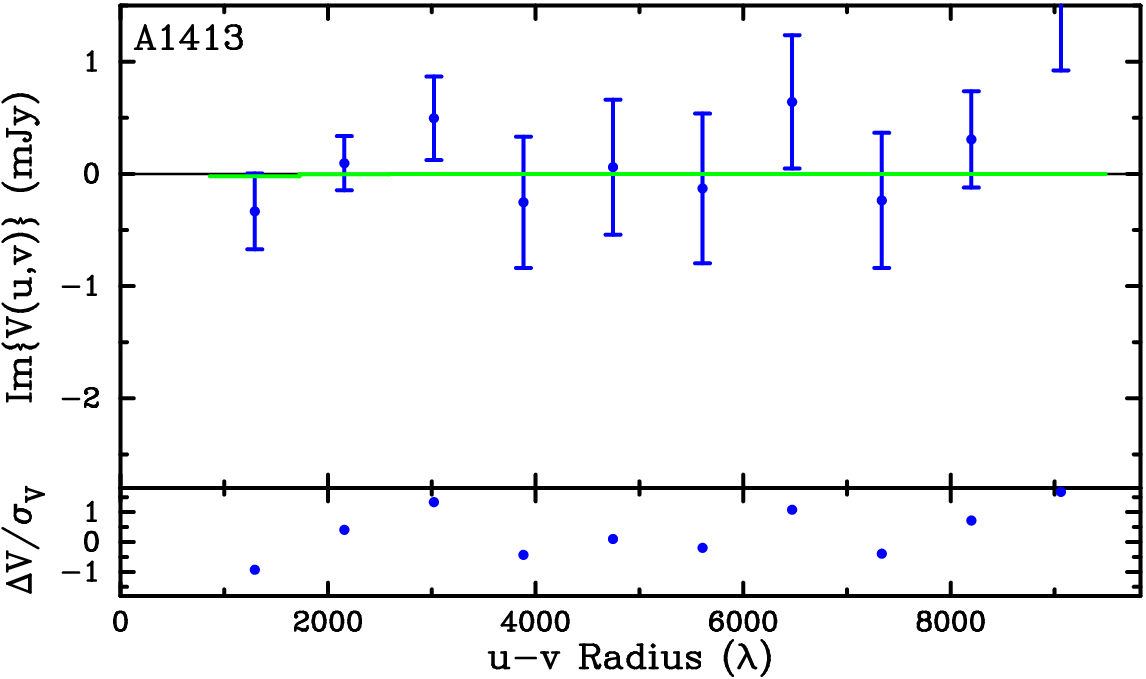,height=1.9in}
}
\contcaption{Cont.}
\end{figure*}

The \uv\ radial profiles for the real and imaginary components of the
complex visibilities for each cluster in our sample are shown in
Figure~\ref{fig:uvradprof}.  Any point sources in the field are
subtracted directly from the visibilities and the phase center is
shifted to the center of the cluster before azimuthally averaging the
real and imaginary components of the complex visibilities.  The points
with error bars are the data and the best fit model from the joint SZE
and X-ray analysis is shown as a solid line, averaged the same way as
the data.  Also shown are the residuals in units of the standard
deviation.  A simple $\chi^2$ analysis of the SZE \uv\ radial profiles
reveals that the models provide a good fit to the data for every
cluster.  The real and imaginary components are shown on the same
scale for each cluster for easy comparison, though the scale on the
residuals may change.  The cluster with the most apparent imaginary
component, A520, is also the cluster with its best fit center the
furthest away from the pointing center, $\sim 65\arcsec$.  The primary
beam attenuation introduces asymmetry and produces a non-zero
imaginary component paralleled in the best fit model for A520.

Figure~\ref{fig:radprof} shows the X-ray radial surface brightness
profiles and the best fit composite models for each cluster in the
sample.  Residuals in units of the standard deviation are also
plotted, $\Delta S_x / \sigma_{S_x}$. A simple $\chi^2$ analysis of
the radial profiles shows that, in general, the models provide a
reasonable fit to the data over a large range of angular radii.  There
are 3 clusters with residuals $\gsim 5\ \sigma$ in a few of the radial
bins; A1835, A665, and A2281.  In A1835 the model systematically
underpredicts the surface brightness for a few intermediate radial
bins.  This cluster contains a strong cooling flow
\citep[e.g.,][]{peres1998}, though shows no sign of a central emission
excess over the best fit model.  The other cooling flow clusters in
our sample (see \S\ref{subsubsec:sys_atmos}) do not exhibit large
residuals in the X-ray model fit.  Both A665
\citep{gomez2000, kalloglyan1990, geller1982} and A2218
\citep{cannon1999, girardi1997, markevitch1997, kneib1995} exhibit
complicated structure, possibly indicating a recent merger.  Though
the worst cases, the best fit models for these 3 clusters still
provide reasonable descriptions of the data.  The only cluster that
shows a marginally significant central X-ray surface brightness excess
is A2218, which has not been identified with a cooling flow.  We also
note that the cooling flow clusters in our sample do not exhibit the
largest residuals.

%%%%%
% X-ray radial profiles
%%%%%
%%
% Figure 6
%%

\begin{figure*}[tbph]
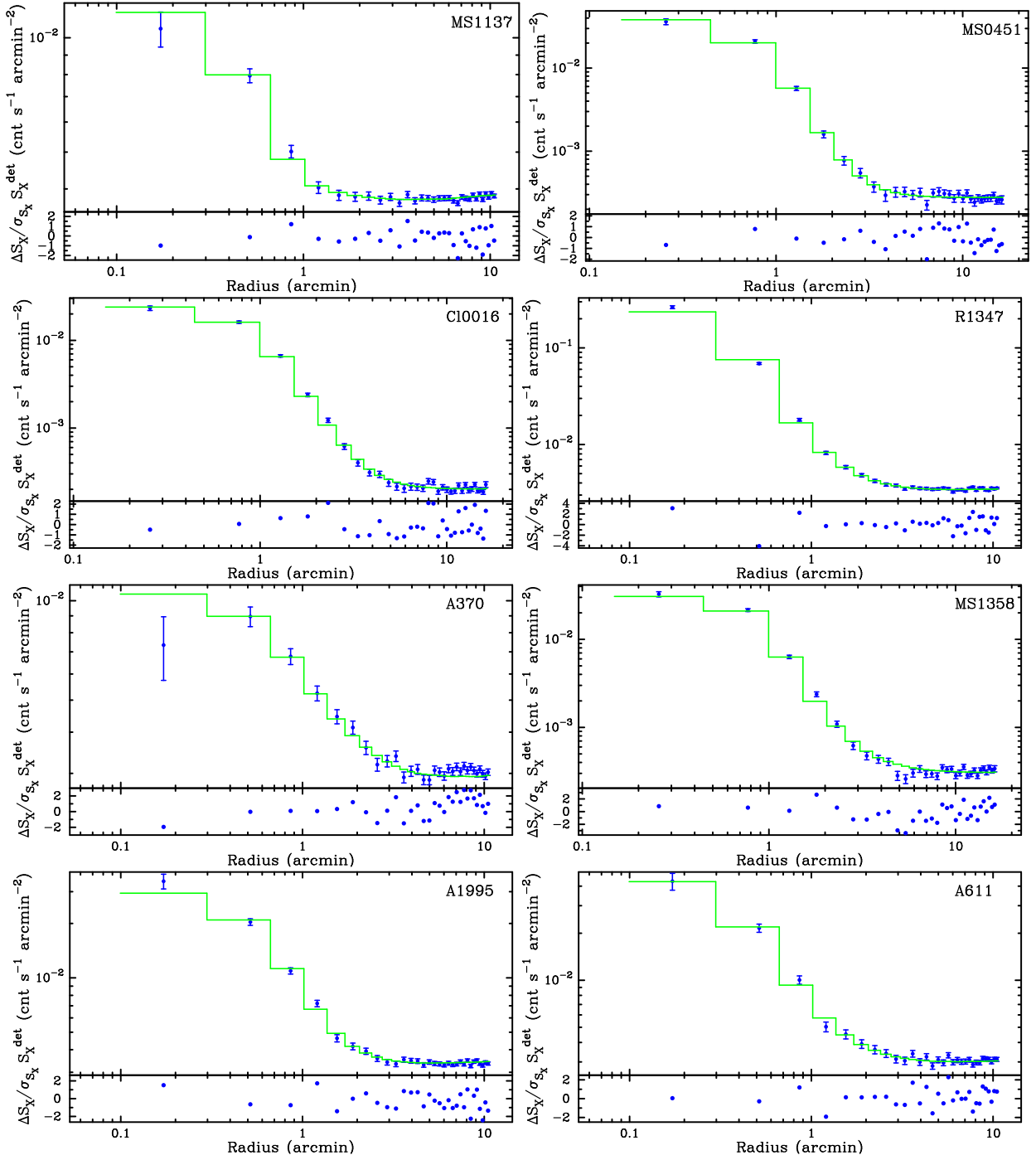

%\epsfxsize = 7.5 in
%%  \epsfxsize = 7.5 in
%%  \epsfysize = 6.0 in
%\centerline{\epsfbox{f?.eps}} 
\centerline{
  \psfig{figure=f6a_color.ps,height=1.85in}
  \psfig{figure=f6b_color.ps,height=1.80in}
}

\centerline{
  \psfig{figure=f6c_color.ps,height=1.80in}
  \psfig{figure=f6d_color.ps,height=1.80in}
}

\centerline{
  \psfig{figure=f6e_color.ps,height=1.80in}
  \psfig{figure=f6f_color.ps,height=1.80in}
}
\centerline{
  \psfig{figure=f6g_color.ps,height=1.80in}
  \psfig{figure=f6h_color.ps,height=1.80in}
}
\caption[X-ray surface brightness radial profiles]{X-ray surface
brightness radial profiles with the best fit model and residuals in
units of the standard deviation.  In general, the models provide good
fits to the data over a large range of radii.}
\label{fig:radprof}
\end{figure*}

\begin{figure*}[tbph]
\centerline{
  \psfig{figure=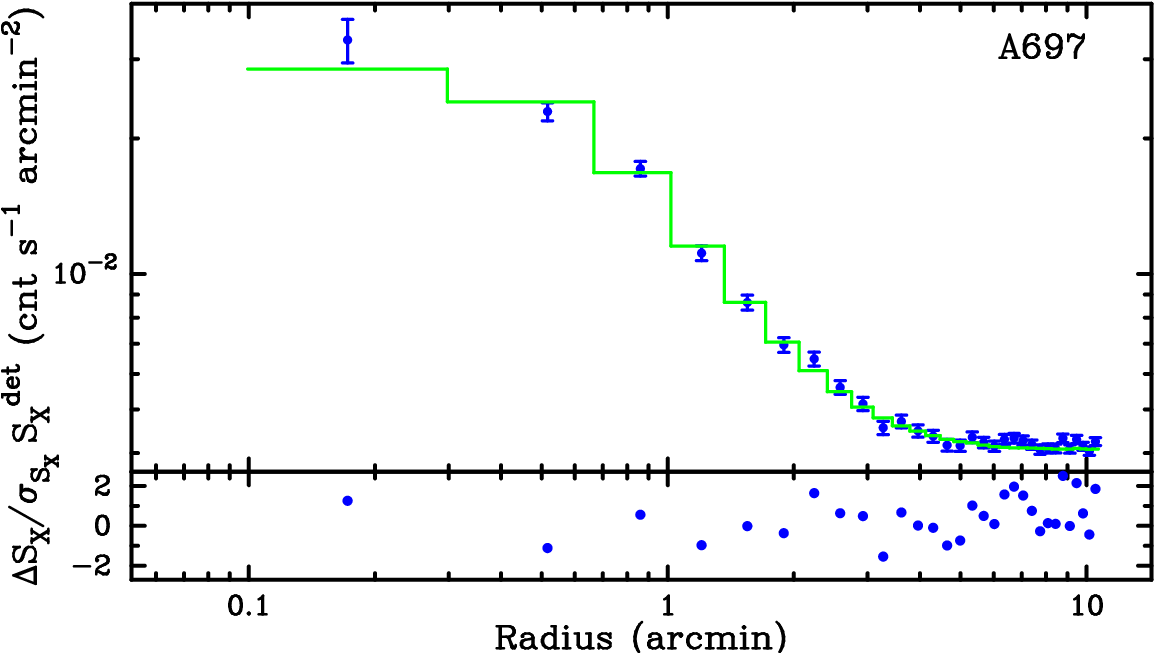,height=1.80in}
  \psfig{figure=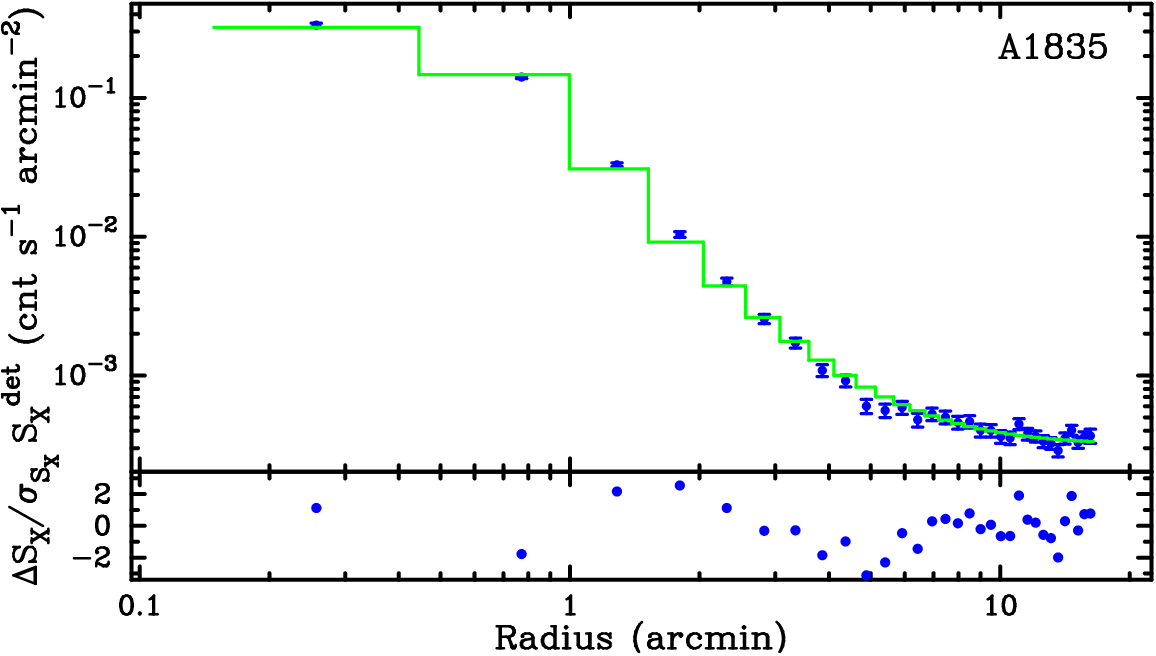,height=1.80in}
}
\centerline{
  \psfig{figure=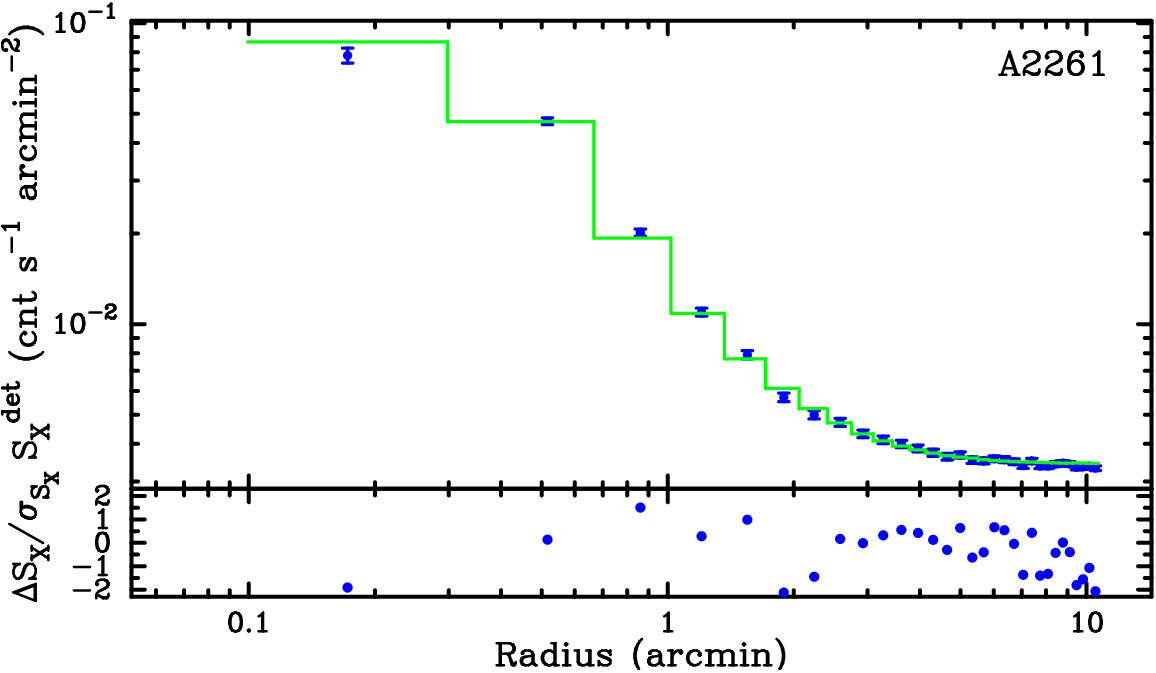,height=1.80in}
  \psfig{figure=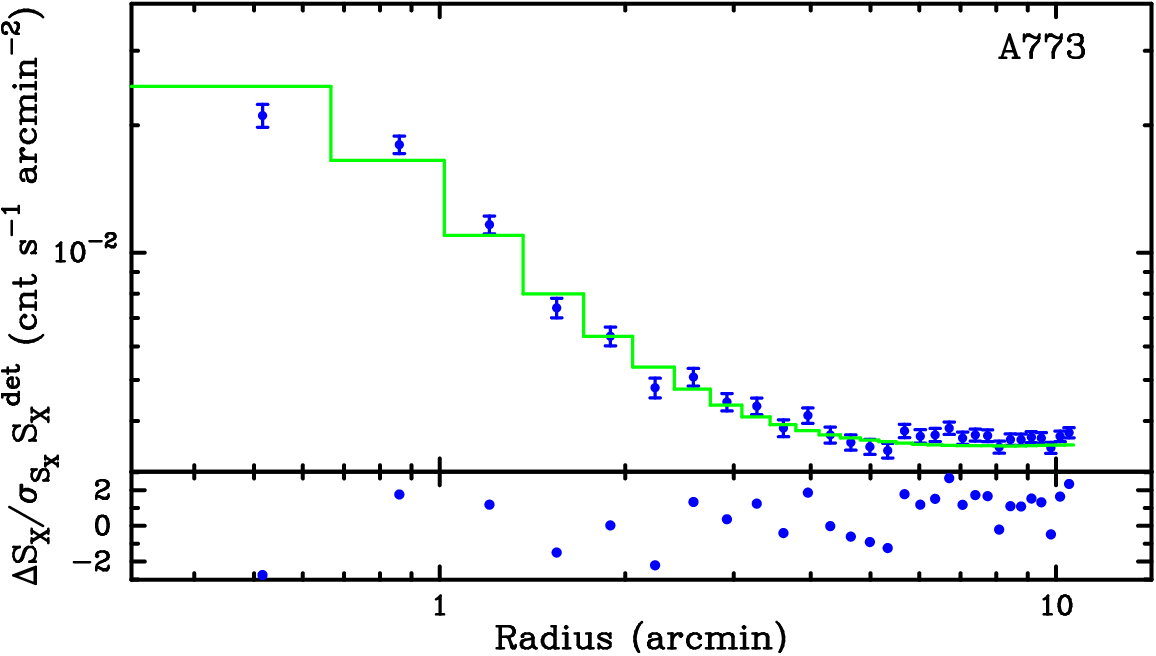,height=1.80in}
}
\centerline{
  \psfig{figure=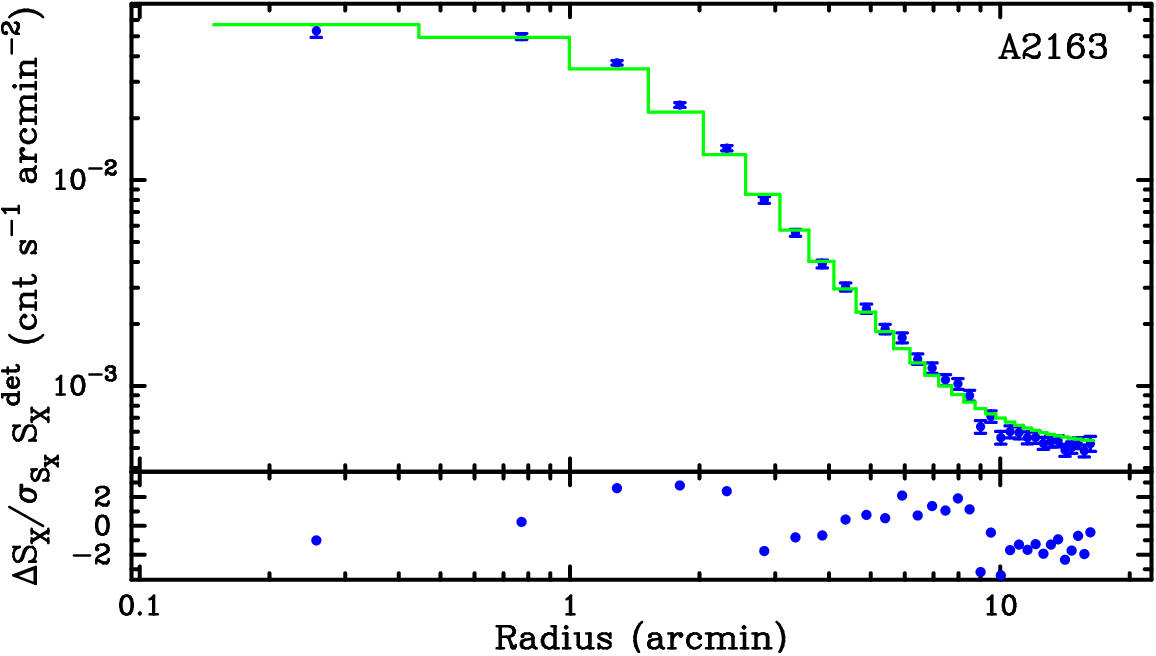,height=1.80in}
  \psfig{figure=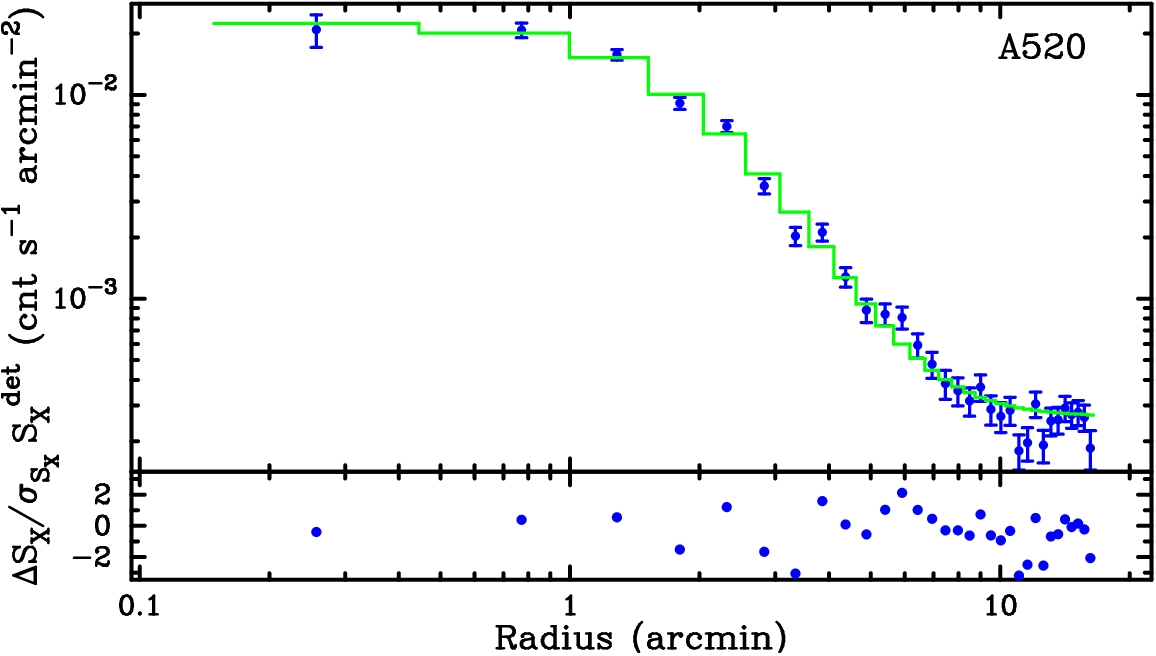,height=1.80in}
}
\centerline{
  \psfig{figure=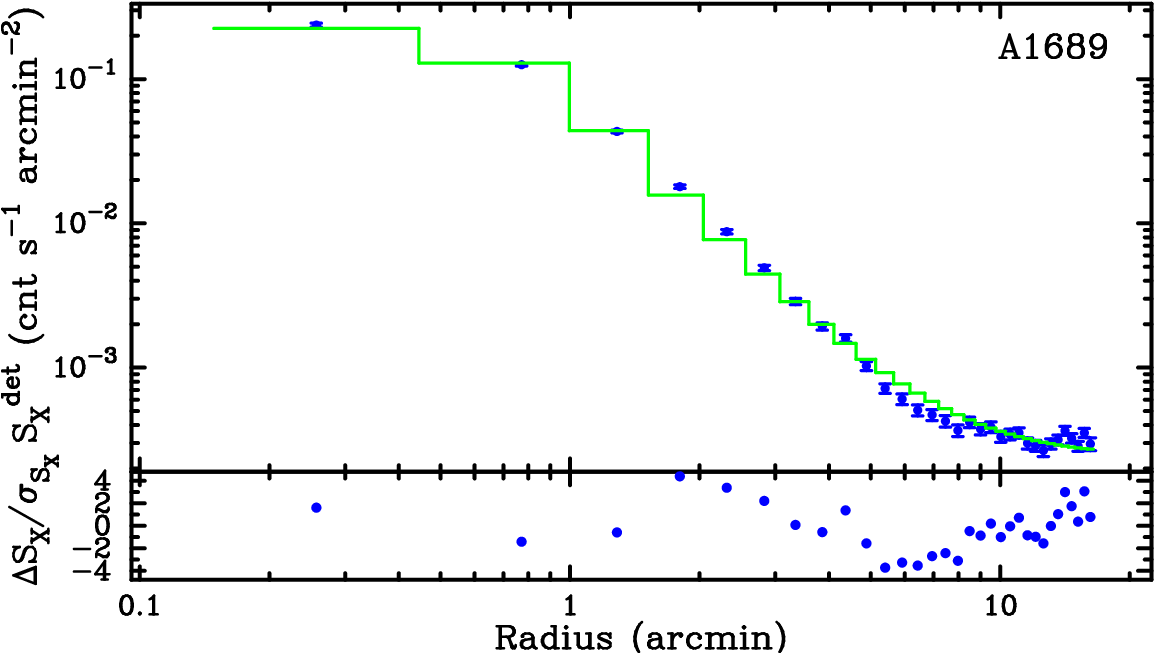,height=1.80in}
  \psfig{figure=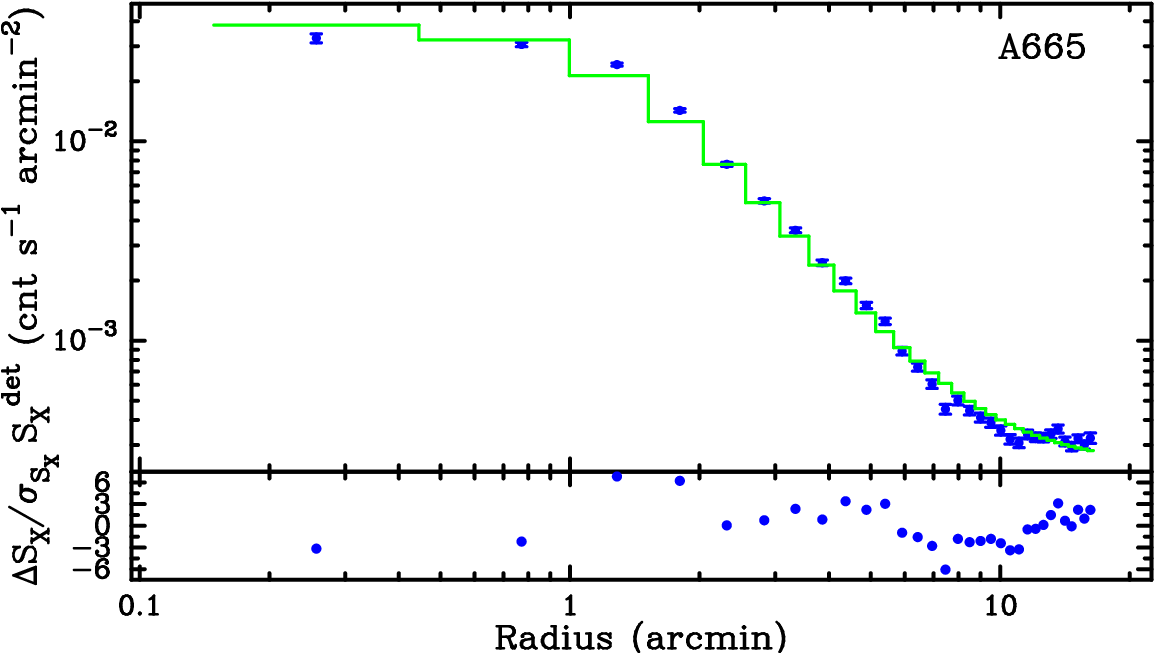,height=1.80in}
}
\contcaption{Cont.}
\end{figure*}

\begin{figure*}[tbph]
\centerline{
  \psfig{figure=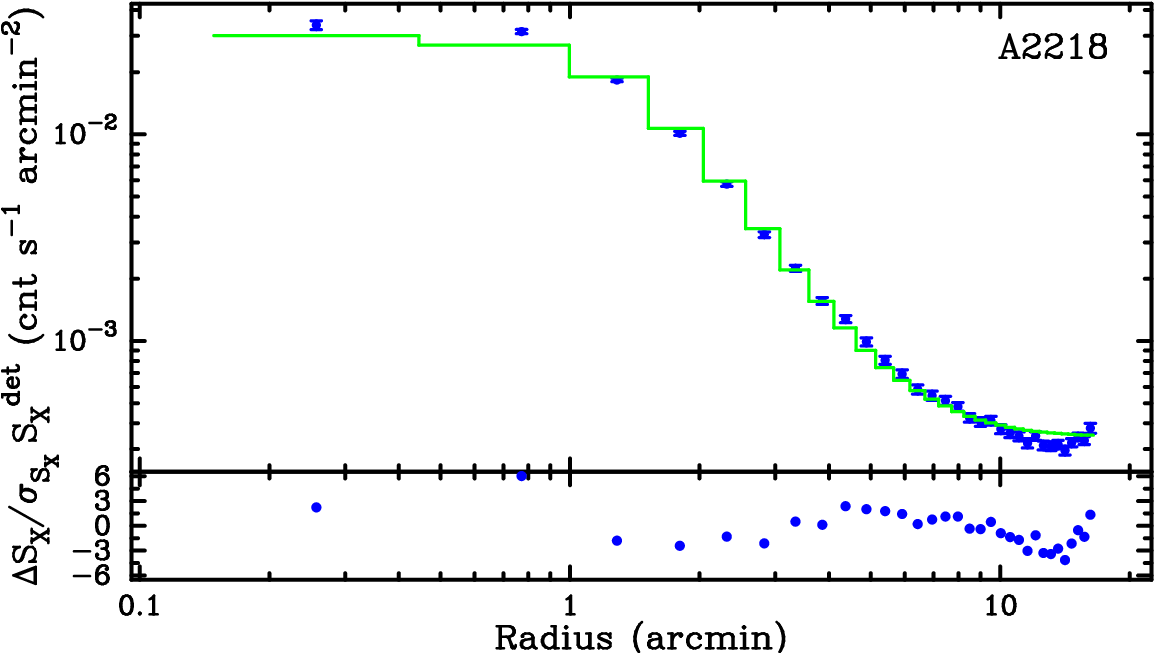,height=1.80in}
  \psfig{figure=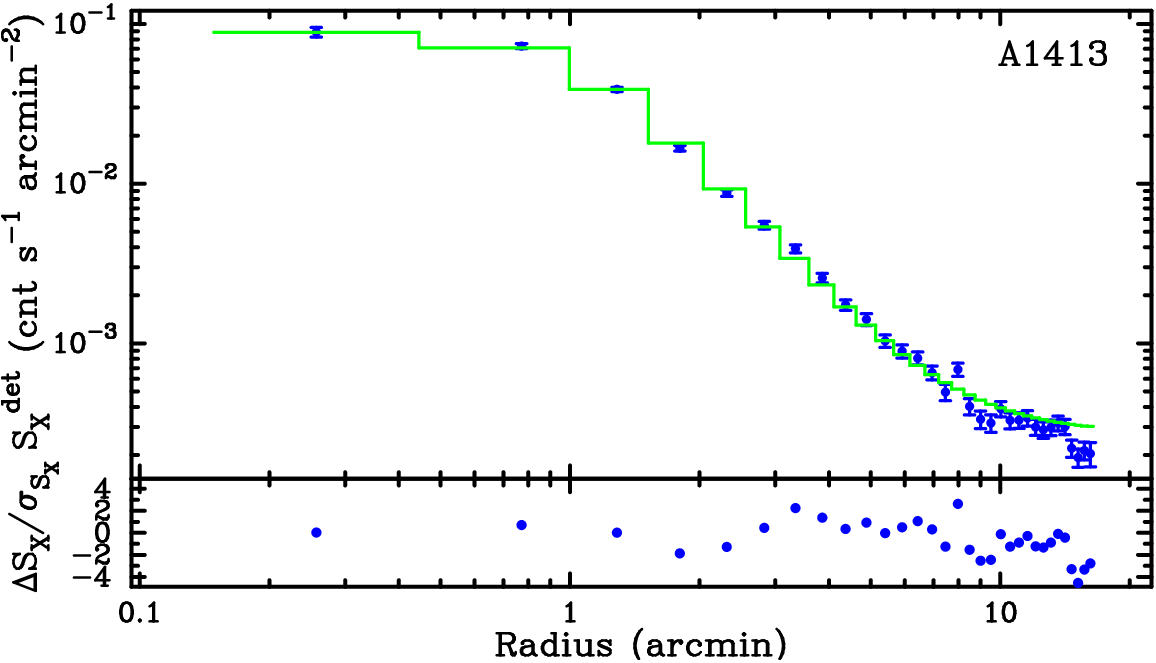,height=1.80in}
}
\contcaption{Cont.}
\end{figure*}

%%%%%%%%%%%%%%%%%%%%%%%%%%%%%%%%%%%%%%%%%%
%% V.   Distances and Hubble Constant
%%%%%%%%%%%%%%%%%%%%%%%%%%%%%%%%%%%%%%%%%%

\section{Distances and the Hubble Constant}
\label{sec:dist_Ho}

We use the results from the maximum likelihood model fitting described
in \S\ref{subsec:model_fit} and \S\ref{subsec:fit_results} to compute
the angular diameter distance to each of our 18 galaxy clusters.
Table~\ref{tab:fitparam} shows the derived angular diameter distances
for each galaxy cluster as well as the best fit ICM shape parameters.
The uncertainties on \Da\ include the entire observational uncertainty
budget, which are shown for each cluster in Table~\ref{tab:Daerr}.
The uncertainties in the fitted parameters come from the procedure
described in \S\ref{subsubsec:uncertain}.

%%%%%
% Table:  \Da Observational Uncertainty Budget
%%%%%
%%Da Observational Uncertainty Budget
%%
%%Cluster Fit N_H Metallicity Te Total
%%
\begin{deluxetable}{lccccccc}
\tablewidth{0pt}
%\tablenum{}
\tablecaption{\Da\ Observational Uncertainty Budget (percent)
	\label{tab:Daerr}}
\tablehead{
\colhead{Cluster} & 
\colhead{Fit\tablenotemark{a}} & 
\colhead{$T_e$\tablenotemark{b}} & 
\colhead{$\LameH (Z)$\tablenotemark{c}} &
%%%%\colhead{$\LameH ([\mbox{Fe}]/[\mbox{H}])$\tablenotemark{c}} &
\colhead{$\LameH (\NH)$\tablenotemark{b}} & 
\colhead{$\LameH (T_e)$} &
\colhead{pt src\tablenotemark{d}} &
\colhead{Total\tablenotemark{e}}
}
\tablecolumns{8}
\startdata
MS1137   & $^{+ 22.3}_{- 22.9}$ & $^{+ 24.6}_{- 45.6}$ &
         $^{+  4.9}_{-  4.7}$ & $^{+\phn  9.0}_{-\phn  6.1}$ &
         $^{+  0.0}_{-  0.8}$ & $^{+\phn  0.0}_{-\phn  0.0}$ &
         $^{+ 34.7}_{- 51.6}$ \\
MS0451   & $^{+ 13.8}_{- 13.1}$ & $^{+ 15.4}_{- 19.2}$ &
         $^{+  1.0}_{-  1.0}$ & $^{+\phn  0.9}_{-\phn  1.2}$ &
         $^{+  0.4}_{-  0.5}$ & $^{+\phn  0.6}_{-\phn  0.6}$ &
         $^{+ 20.7}_{- 23.4}$ \\
CL0016   & $^{+ 17.8}_{- 16.4}$ & $^{+ 15.4}_{- 19.1}$ &
         $^{+  1.9}_{-  1.2}$ & $^{+\phn  1.1}_{-\phn  1.2}$ &
         $^{+  0.1}_{-  0.2}$ & $^{+\phn  1.0}_{-\phn  0.6}$ &
         $^{+ 23.7}_{- 25.2}$ \\
R1347    & $^{+ 19.2}_{- 17.1}$ & $^{+ 12.9}_{- 15.1}$ &
         $^{+  1.0}_{-  1.0}$ & $^{+15.7}_{-11.7}$ &
         $^{+  0.4}_{-  0.5}$ & $^{+11.2}_{-11.4}$ &
         $^{+ 30.1}_{- 28.1}$ \\
A370     & $^{+ 24.4}_{- 16.7}$ & $^{+ 15.2}_{- 21.2}$ &
         $^{+  1.9}_{-  1.9}$ & $^{+ 13.1}_{-\phn  8.2}$ &
         $^{+  0.4}_{-  0.6}$ & $^{+\phn  4.0}_{-\phn  4.0}$ &
         $^{+ 31.9}_{- 28.6}$ \\
MS1358   & $^{+ 24.0}_{- 21.3}$ & $^{+ 11.2}_{- 13.4}$ &
         $^{+  1.5}_{-  1.5}$ & $^{+\phn  3.1}_{-\phn  3.0}$ &
         $^{+  0.3}_{-  0.4}$ & $^{+ 10.0}_{- 25.2}$ &
         $^{+ 28.6}_{- 35.8}$ \\
A1995    & $^{+ 10.3}_{- 11.5}$ & $^{+ 15.6}_{- 20.0}$ &
         $^{+  1.1}_{-  1.1}$ & $^{+\phn  8.5}_{-\phn  6.5}$ &
         $^{+  0.5}_{-  0.7}$ & $^{+\phn  8.2}_{-\phn  7.8}$ &
         $^{+ 22.1}_{- 25.2}$ \\
A611     & $^{+ 22.2}_{- 20.5}$ & $^{+ 18.2}_{- 18.2}$ &
         $^{+  3.7}_{-  3.7}$ & $^{+ 15.3}_{-\phn  9.8}$ &
         $^{+  0.4}_{-  0.5}$ & $^{+\phn  0.0}_{-\phn  0.0}$ &
         $^{+ 32.7}_{- 29.4}$ \\
A697     & $^{+ 22.7}_{- 18.7}$ & $^{+ 14.3}_{- 14.3}$ &
         $^{+  3.0}_{-  3.0}$ & $^{+ 12.9}_{-\phn  8.2}$ &
         $^{+  0.7}_{-  0.7}$ & $^{+\phn  0.0}_{-\phn  0.0}$ &
         $^{+ 29.9}_{- 25.1}$ \\
A1835    & $^{+ 13.2}_{- 12.6}$ & $^{+\phn  4.1}_{-\phn  4.6}$ &
         $^{+  0.7}_{-  0.5}$ & $^{+\phn  3.7}_{-\phn  3.5}$ &
         $^{+  0.2}_{-  0.2}$ & $^{+ 12.4}_{- 13.4}$ &
         $^{+ 18.9}_{- 19.3}$ \\
A2261    & $^{+ 25.0}_{- 22.9}$ & $^{+\phn  7.3}_{-\phn  8.4}$ &
         $^{+  0.9}_{-  0.8}$ & $^{+ 12.7}_{-\phn  8.1}$ &
         $^{+  0.4}_{-  0.5}$ & $^{+\phn  3.4}_{-\phn  3.4}$ &
         $^{+ 29.2}_{- 25.9}$ \\
A773     & $^{+ 21.8}_{- 20.2}$ & $^{+\phn  7.8}_{-\phn  8.8}$ &
         $^{+  0.7}_{-  0.7}$ & $^{+\phn  9.3}_{-\phn  6.2}$ &
         $^{+  0.5}_{-  0.5}$ & $^{+\phn  0.0}_{-\phn  0.0}$ &
         $^{+ 24.9}_{- 22.9}$ \\
A2163    & $^{+ 16.3}_{- 14.4}$ & $^{+ 11.5}_{- 18.0}$ &
         $^{+  1.2}_{-  1.1}$ & $^{+\phn  2.9}_{-\phn  2.2}$ &
         $^{+  0.7}_{-  1.0}$ & $^{+\phn  8.4}_{-\phn  8.6}$ &
         $^{+ 21.8}_{- 24.8}$ \\
A520     & $^{+ 32.4}_{- 27.2}$ & $^{+\phn  9.6}_{- 11.0}$ &
         $^{+  1.0}_{-  1.0}$ & $^{+ 12.1}_{- 10.2}$ &
         $^{+  0.3}_{-  0.4}$ & $^{+ 10.2}_{- 10.2}$ &
         $^{+ 37.3}_{- 32.7}$ \\
A1689    & $^{+ 14.4}_{- 12.2}$ & $^{+\phn  4.1}_{-\phn  4.6}$ &
         $^{+  0.5}_{-  0.5}$ & $^{+\phn  2.9}_{-\phn  2.8}$ &
         $^{+  0.2}_{-  0.2}$ & $^{+ 19.8}_{- 19.6}$ &
         $^{+ 25.0}_{- 23.7}$ \\
A665     & $^{+ 45.5}_{- 37.3}$ & $^{+\phn  6.9}_{-\phn  7.8}$ &
         $^{+  0.6}_{-  0.8}$ & $^{+\phn  6.7}_{-\phn  6.1}$ &
         $^{+  0.3}_{-  0.4}$ & $^{+\phn  0.4}_{-\phn  0.6}$ &
         $^{+ 46.5}_{- 38.5}$ \\
A2218    & $^{+ 30.3}_{- 31.7}$ & $^{+\phn  6.0}_{-\phn  6.2}$ &
         $^{+  0.7}_{-  0.7}$ & $^{+\phn  5.2}_{-\phn  4.8}$ &
         $^{+  0.2}_{-  0.2}$ & $^{+\phn 10.0}_{-\phn 10.0}$ &
         $^{+ 32.9}_{- 34.2}$ \\
A1413    & $^{+ 28.5}_{- 24.8}$ & $^{+\phn  4.2}_{-\phn  4.5}$ &
         $^{+  0.5}_{-  0.5}$ & $^{+\phn  3.5}_{-\phn  3.3}$ &
         $^{+  0.2}_{-  0.2}$ & $^{+\phn  7.2}_{-\phn  7.2}$ &
         $^{+ 29.9}_{- 26.4}$ \\
\enddata
\tablenotetext{a}{The 68.3\% uncertainties over the
	four-dimensional error surface for $\beta$, $\theta_c$, \Xo,
	and \dTo.}
\tablenotetext{b}{\Da\ decreases as parameter increases.}
\tablenotetext{c}{Metallicity relative to solar.}
\tablenotetext{d}{Maximum effect from detected point sources.}
\tablenotetext{e}{Combined in quadrature.}
%\tablecomment{}
\end{deluxetable}

The only other parameter that enters directly into the \Da\
calculation is \Teo.  Since $\Da \propto \Teo^{\!\!\!\!\! -2}$, the
uncertainty in \Da\ due to \Teo\ is listed as twice the fractional
uncertainty on \Teo.  The other parameters, column density and
metallicity, as well as \Teo, affect the X-ray cooling function.  We
estimate the uncertainties in \Da\ due to these parameters by taking
their 68.3\% ranges and seeing how much they affect the cooling
function.  The uncertainty on \Da\ due to observations is dominated by
the uncertainty in the electron temperature and the SZE central
decrement.  Note that changes of factors of two in metallicity result
in a $\sim 1$\% effect on \Da.  

%%%%%
% Da versus z for our H0 sample
%%%%%
%%
% Figure 7
%%

\begin{figure*}[!tbph]
%\epsfxsize = 7.5 in
%%  \epsfxsize = 7.5 in
%%  \epsfysize = 6.0 in
%\centerline{\epsfbox{f?.eps}} 
\centerline{
\psfig{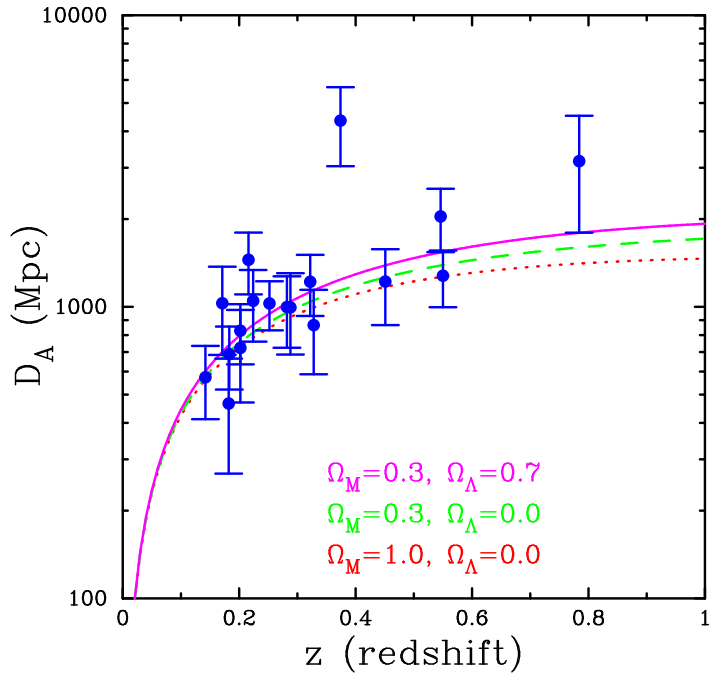}
%?%\psfig{figure=/home/piglet/reese/latex/thesis/apj_version/submit/da_z_thesis_bw.ps,height=3.5in}
%?%\psfig{figure=/home/piglet/reese/latex/thesis/apj_version/submit/da_z_thesis_log_bw.ps,height=3.5in}
%\psfig{figure=/home/piglet/reese/latex/thesis/da_z_thesis.ps,height=5.0in}
%\psfig{figure=/home/piglet/reese/latex/thesis/apj_version/submit/da_z_thesis_log.ps,height=5.0in}
%%\psfig{figure=/home/piglet/reese/latex/sz/hubble/sample/da_z_Hosample.ps,height=3.5in}
}
\caption[\Da\ versus $z$ for our 18 cluster sample]{SZE determined
distances as a function of redshift.  The error bars are 68.3\%
statistical uncertainties only.  Also plotted are the theoretical
angular diameter distance relations assuming $\Ho = 60$ \ksM\ for three
different cosmological models; the currently favored $\Lambda$ cosmology
$\Om=0.3$, $\Ol=0.7$ %(magenta)
(solid) 
cosmology;
an open $\Om=0.3$ %(green) 
(dashed) 
universe; and a flat $\Om=1$ %(red)
(dotted)
cosmology.}
\label{fig:da_z}
\end{figure*}

The column densities measured from the X-ray spectra are different
from those from \ion{H}{1} surveys \citep{dickey1990}.  We use the
column densities from X-ray spectral fits when possible since that
includes contributions from non-neutral hydrogen and other elements
which absorb X-rays.  For MS0451 and Cl0016, using the survey derived
column densities instead of the fitted values changes the angular
diameter distance by $\sim \pm 5$\% \citep{reese2000}, which we
include as a systematic uncertainty (see \S~\ref{sec:disc_concl}).

Figure~\ref{fig:da_z} shows the SZE determined distances for each
cluster as a function of redshift.  Also plotted are the theoretical
angular diameter distance relations assuming $\Ho = 60$ \ksM\ for
three different cosmologies: the currently favored $\Lambda$ cosmology
$\Om=0.3$, $\Ol=0.7$ %(magenta)
(solid) 
cosmology;
an open $\Om=0.3$ %(green) 
(dashed) 
universe; and a flat $\Om=1$ %(red)
(dotted) 
cosmology.  The SZE distances are beginning to probe the angular
diameter distance relation.  The uncertainties on \Da\ in
Figure~\ref{fig:da_z} are the 68.3\% statistical uncertainties only,
including all of the statistical uncertainties in the calculation
outlined above.  We refer the reader to \citet{carroll1992},
\citet{kolb1990}, and \citet{peacock1999} for derivations of the
theoretical angular diameter distance relation.

\begin{figure*}[tbp]
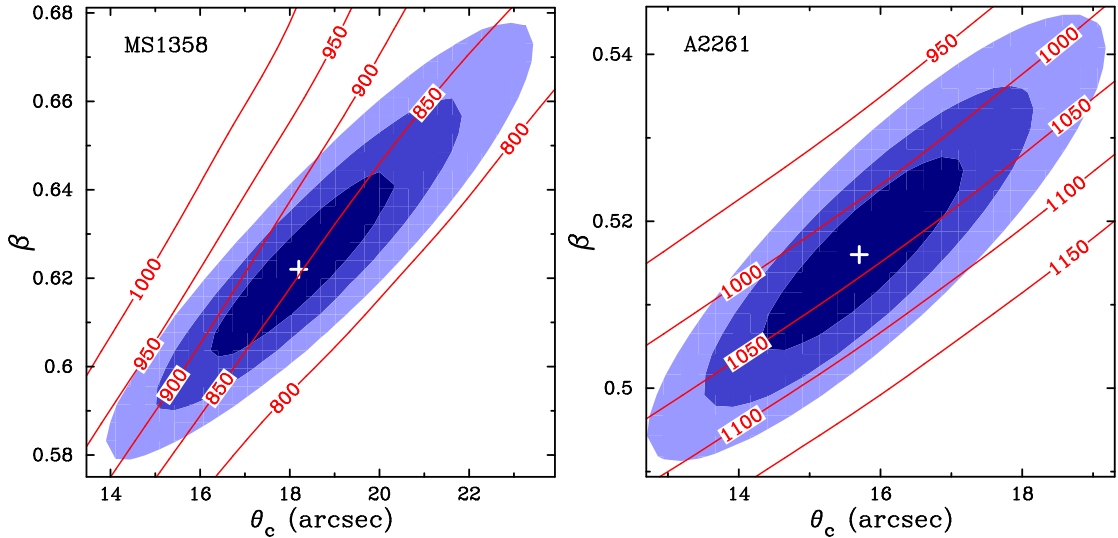

%\epsfxsize = 7.5 in
%%  \epsfxsize = 7.5 in
%%  \epsfysize = 6.0 in
%\centerline{\epsfbox{}} 
\centerline{
\psfig{figure=f8a_color.ps,height=2.8in}
\psfig{figure=f8b_color.ps,height=2.8in}
}
\caption[Confidence regions in $\theta_c$-$\beta$ plane with \Da\
contours]{Confidence regions from the joint SZE and X-ray fit for
MS1358 and A2261.  The filled regions are 1, 2, and 3 $\sigma$
confidence regions for $\beta$ and $\theta_c$ jointly ($\Delta S =
2.3$, 6.2, 11.8), and the cross marks the best-fit $\beta$ and
$\theta_c$.  Solid lines are contours of angular diameter distance in
megaparsecs.  The \Da\ contours lie roughly parallel to the
$\beta$-$\theta_c$ correlation, minimizing the effect of this
correlation on the uncertainties of \Da.}
\label{fig:conf_region}
\end{figure*}

There is a known correlation between the $\beta$ and $\theta_c$
parameters of the $\beta$ model.  Figure~\ref{fig:conf_region}
illustrates this correlation and its effect on \Da\ for MS1358, and
A2261.  The filled contours are the 1, 2, and 3 $\sigma$ $\Delta S$
confidence regions for $\beta$ and $\theta_c$ jointly with the plus
marking the best fit for each cluster.  The lines are contours of
constant \Da\ in megaparsecs.  With our interferometric SZE data, the
contours of constant \Da\ lie roughly parallel to the
$\beta$-$\theta_c$ correlation, minimizing the effect of this
correlation on the uncertainties of \Da.  Similar figures for MS0451
and Cl0016 appear in \citet{reese2000}, which show similar behavior.
The alignment of the \Da\ contours with the $\beta$-$\theta_c$
correlation is a general feature of our observing strategy.  Different
observing techniques will result in different behavior.  Contours of
constant \Da\ have been found to be roughly orthogonal to this
$\beta$-$\theta_c$ correlation for some single dish SZE observations
\citep{birkinshaw1994, birkinshaw1991}.

To determine the Hubble Constant, we perform a $\chi^2$ fit to our
calculated \Da's versus $z$ for three different cosmologies.  To
estimate statistical uncertainties, we combine the uncertainties on
\Da\ listed in Table~\ref{tab:Daerr} in quadrature, which is only
strictly valid for Gaussian distributions.  This combined statistical
uncertainty is symmetrized (averaged) and used in the fit.  We find
\begin{equation}
\Ho = \cases{
	60 ^{+4}_{-4} \ \ksM; &\Om=0.3, \Ol=0.7, \cr
	56 ^{+4}_{-4} \ \ksM; &\Om=0.3, \Ol=0.0, \cr
	54 ^{+4}_{-3} \ \ksM; &\Om=1.0, \Ol=0.0, \cr
}
\label{eq:Horesult}
\end{equation}
where the uncertainties are statistical only at 68.3\% confidence.
The statistical error comes from the $\chi^2$ analysis and includes
uncertainties from $T_e$, the parameter fitting, metallicity, \NH, and
detected radio point sources (see Table~\ref{tab:Daerr}).  We have
chosen three cosmologies encompassing the currently favored models.
With this sample of clusters, there is a $\sim 10$\% range in our
inferred \Ho\ due to the geometry of the universe.  For the $\Lambda$
cosmology, $\chi^2 = 16.5$ with a corresponding reduced chi-squared of
$\chi^2_{red} = 0.97$.  The difference in $\chi^2$ between the
$\Om=0.3, \Ol=0.7$ and the flat $\Om=1$ universes is roughly $\Delta
\chi^2 \sim 0.3$, with the $\Lambda$ cosmology having the lowest
$\chi^2$.  Clearly a larger sample of high redshift ($z\sim 1$)
clusters is required for a determination of the geometry of the
universe from SZE and X-ray determined direct distances to galaxy
clusters (see \S\ref{sec:disc_concl}).

%%%%%
% Confidence regions plots with Da contours for MS1358 & A2261
%%%%%
%%
% Figure 8
%%

%%%%%%%%%%%%%%%%%%%%%%%%%%%%%%%%%%%%%%%%%%
%% VI.   Sources of Possible Systematic Uncertainty
%%%%%%%%%%%%%%%%%%%%%%%%%%%%%%%%%%%%%%%%%%

\section{Sources of Possible Systematic Uncertainty}
\label{subsec:sys_sources}

%%%%%
% Table: \Ho Systematic Uncertainty Budget
%%%%%

%%H0 Systematic Uncertainty Budget
%%
%%Systematic %Effect
%%
\begin{deluxetable}{lr}
\tablewidth{0pt}
%\tablenum{}
\tablecaption{\Ho\ Systematic Uncertainty Budget (\%)
	\label{tab:Hosyserr}}
\tablehead{
\colhead{Systematic} 	& \colhead{Effect}
}
\tablecolumns{2}
\startdata
SZE calibration			&$\pm \phn 8$\\   % conservative
X-ray calibration		&$\pm 10$\\	  % who knows
\NH				&$\pm \phn 5$\\
Asphericity\tablenotemark{a}	&$\pm \phn 5$\\
Isothermality 			&$\pm 10$\\	  % conservative (?)
Clumping			&$-   20$\\
Undetected radio sources\tablenotemark{b}
				&$\pm 12$\\
Kinetic SZE\tablenotemark{a}	&$\pm \phn 2$\\
Primary CMB\tablenotemark{a}	&$< \pm 1$\\   % 68%
Radio Halos			&$- \phn 3$\\   % worst case
Primary Beam			&$\pm \phn 3$\\   % max effect (avg within 0.2\arcmin)
%\hline
\ \ Total\tablenotemark{c}		&$^{+22}_{-30}$
\enddata
\tablenotetext{a}{Includes a $1/\sqrt{18}$ factor for our 18 cluster sample.}
\tablenotetext{b}{Average of effect from the 18 cluster fields.}
\tablenotetext{c}{Combined in quadrature.}
%\tablecomment{}
\end{deluxetable}

The absolute calibration of both the SZE observations and the PSPC and
HRI directly affects the distance determinations.  The absolute
calibration of the interferometric observations is conservatively
known to about 4\% at 68.3\% confidence, corresponding to a 8\%
uncertainty in \Ho\ ($\propto \Delta T_0^{-2}$).  The effective
areas of the PSPC and HRI are thought to be known to about 10\%,
introducing a 10\% uncertainty into the \Ho\ determination through
the calculation of $\Sigma$.  In addition to the absolute calibration
uncertainty from the observations, there are possible sources of
systematic uncertainty that depend on the physical state of the ICM
and other sources that can contaminate the cluster SZE emission.
Table~\ref{tab:Hosyserr} summarizes the systematic uncertainties in
the Hubble constant determined from our 18 cluster sample.

%%%%%
%%  VI.1 Cluster Atmospheres and Morphology
%%%%%

\subsection{Cluster Atmospheres and Morphology}
\label{subsubsec:sys_atmos}

\subsubsection{Asphericity}
\label{subsubsec:asphericity}

Most clusters do not appear circular in radio, X-rays, or optical.
Fitting a projected elliptical isothermal $\beta$ model gives typical
axial ratios that are close to the local average of $0.80$
\citep{mohr1995}.  Under the assumption of axisymmetric clusters, the
combined effect of cluster asphericity and its orientation on the sky
conspires to introduce a $\sim\pm 20$\% random uncertainty in \Ho\
determined from one galaxy cluster \citep{hughes1998}.  When one
considers a large, unbiased sample of clusters, with random
orientations, the errors due to imposing a spherical model are
expected to cancel, resulting in a precise determination of \Ho.
Recently, Sulkanen (1999)\nocite{sulkanen1999} studied projection
effects using triaxial $\beta$ models.  Fitting these with spherical
models he found that the Hubble constant estimated from the sample
was within 5\% of the input value.  We are in the process of using
N-body and smoothed particle hydrodynamics (SPH) simulations of 48
clusters to quantify the effects of complex cluster structure on our
results.

A 20\% effect from one cluster implies a 5\% ($=20/\sqrt{18}$) effect
for a sample of 18 clusters.  Therefore, we include at 5\% effect from
asphericity for our cluster sample.

\subsubsection{Temperature Gradients}
\label{subsubsec:Tgrad}

Departures from isothermality in the cluster atmosphere may result in
a large error in the distance determination from an isothermal
analysis; moreover, an isothermal analysis of a large cluster sample
could lead to systematic errors in the derived Hubble parameter if
most clusters have similar departures from isothermality
\citep{birkinshaw1994, inagaki1995, holzapfel1997}.  The \rosat\ band
is fairly insensitive to temperature variations, showing a $\sim 10$\%
change in the PSPC count rate for a factor of 2 change in temperature
for $T_e > 1.5$ keV gas \citep{mohr1999a}.  In theory, cluster
temperature profiles may significantly affect the distance
determinations through the SZE since $\Delta T \propto \int n_e T_e
d\ell$.  The spatial filtering of the interferometer makes our SZE
observations insensitive to angular scales larger than a few
arcminutes.  Therefore we are relatively insensitive to large scale
temperature gradients.  However, we are sensitive to temperature
gradients on smaller scales, for example at the center of cooling flow
clusters.

A mixture of simulations and studies of nearby clusters suggests a
10\% effect on the Hubble parameter \citep{inagaki1995, roettiger1997}
due to departures from isothermality.  The spatial filtering of the
interferometer is not accounted for in these studies and thus provides
a conservative estimate.  We include a conservative $\pm 10$\% effect
on the inferred Hubble parameter due to departures of isothermality,
consistent with both cooling flow (see \S\ref{subsubsec:cool_flow})
and non-cooling flow departures.

\subsubsection{Cooling Flows}
\label{subsubsec:cool_flow}

Cooling flows affect the emission weighted mean temperature and
enhance the X-ray central surface brightness \citep[see,
e.g.,][]{fabian1994, nagai2000}.  When the cooling time at the center
of the cluster is less than the age of the cluster, then the central
gas has time to cool.  This is known as a cooling flow
\citep[e.g.,][]{fabian1994}.  The cluster temperature is expected to
decrease towards the center of the cluster, which has recently been
seen with both Chandra \citep[e.g.,][]{markevitch2000,
nevalainen2000b} and XMM-Newton \citep[e.g.,][]{tamura2001}.

%%%%%
% Table: Ratio of t_cool/t_H(z)
%%%%%

%%t_cool/t_H ratio's
%%
%%ratios for 3 cosmologies
%%
\begin{deluxetable}{lccc}
\tablewidth{0pt}
%\tablenum{}
\tablecaption{Ratio of $t_{cool}/t_H(z)$ \label{tab:tcool}}
\tablehead{
\colhead{} &
\multicolumn{3}{c}{Cosmology (\Om, \Ol)}
\\
\cline{2-4}
\colhead{Cluster} &
\colhead{(0.3, 0.7)} &
\colhead{(0.3, 0.0)} &
\colhead{(1.0, 0.0)}
}
\tablecolumns{4}
\startdata
MS1137 	     & 1.1 & 1.9 & 1.4 \\
MS0451 	     & 1.0 & 1.6 & 1.2 \\
Cl0016 	     & 1.5 & 2.5 & 1.9 \\
R1347 	     & 0.1 & 0.2 & 0.1 \\
A370         & 3.6 & 5.8 & 4.4 \\
MS1358       & 0.4 & 0.7 & 0.5 \\
A1995  	     & 0.9 & 1.5 & 1.2 \\
A611         & 0.4 & 0.7 & 0.5 \\
A697         & 1.1 & 1.7 & 1.3 \\
A1835  	     & 0.1 & 0.2 & 0.2 \\
A2261  	     & 0.3 & 0.5 & 0.4 \\
A773         & 1.5 & 2.4 & 1.8 \\
A2163  	     & 1.3 & 2.0 & 1.6 \\
A520         & 2.1 & 3.3 & 2.6 \\
A1689  	     & 0.3 & 0.5 & 0.4 \\
A665         & 1.2 & 1.8 & 1.4 \\
A2218  	     & 1.4 & 2.2 & 1.7 \\
A1413  	     & 0.6 & 0.9 & 0.7 \\
\enddata
%\tablenotetext{a}{}
%\tablenotetext{b}{} 
%\tablecomment{}
\end{deluxetable}

A characteristic cooling time for the ICM is the available radiative
energy divided by its luminosity given by
\begin{equation}
t_{cool} \sim \frac{3 k T_e n_{tot}}{2 \Lambda n_e \nH} 
	=  \frac{3 k T_e}{2 \Lambda n_e} \frac{\muH}{\mu_{tot}},
\label{eq:tcool1}
\end{equation}
where $\Lambda$ is the bolometric cooling function of the cluster and
all quantities are evaluated at the center of the cluster.  Cooling
flows may occur if the cooling time is less than the age of the
cluster, which we conservatively estimate to be the age of the
universe at the redshift of observation, $t_{cool} < t_{H}(z)$.

As a check, we calculate $t_{cool}/t_H$ ratios for each cluster
analyzed by Mohr \etal\ (1999)\nocite{mohr1999a}.  We check our
cooling flow and non-cooling flow determinations versus those of Peres
\etal\ (1998)\nocite{peres1998} and Fabian (1994)\nocite{fabian1994}.
Of the 45 clusters in the Mohr sample, 41 have published mass
deposition rates.  We assume the cluster does not contain a cooling
flow if its mass deposition rate is consistent with zero, otherwise it
is designated as a cooling flow cluster.  We are able to predict
whether a cluster has a cooling flow or not with a 90\% success rate,
suggesting that the ratio $t_{cool}/t_H$ presented in
equation~(\ref{eq:tcool1}) is a good predictor for the presence of a
cooling flow.

The ratio $t_{cool}/t_H(z)$ for each cluster is summarized in
Table~\ref{tab:tcool} for the same three cosmologies used to determine
the Hubble constant.  The central densities, \no, used in this
calculation are determined by eliminating \Da\ in
equations~(\ref{eq:szsignal}) and (\ref{eq:xsignal}) in favor of \no.
From this analysis, the seven clusters R1347, MS1358, A611, A1835,
A2261, A1689, and A1413 are cooling flow clusters.  Both A1995 and
A1413 are borderline cases.  Such clusters are expected to have
falling temperatures towards the center of the cluster.

In principle, the multiphase medium expected in cooling flow clusters
could introduce large biases in isothermal beta model SZE and X-ray
distances \citep{nagai2002}.  However, recent observations with
Chandra and XMM-Newton suggest that cooling flows are not as strong as
previously expected \citep[e.g.,][]{fabian2001,peterson2001}.  To
estimate the possible effects of cooling flow-like temperature
profiles, we adopt the deprojected temperature profile of A1835 from
an analysis of XMM-Newton observations and determine the change in the
angular diameter distance introduced by this profile.  The inclusion
of the A1835 temperature profile reduces the angular diameter distance
by $\sim 10$\%, causing a $\sim 10$\% underestimate in the Hubble
constant from cooling flow clusters when an isothermal analysis is
performed (A.~D.\ Miller 2001, private communication).  In addition, a
theoretical examination of the effects of cooling flows on SZE and
X-ray determined distances suggests a $\sim 10$\% underestimate of
\Ho\ from an isothermal analysis \citep{majumdar2000}.  Assuming all
seven cooling flow clusters in our sample produce a similar 10\% bias
in \Ho, the average underestimate in \Ho\ for our 18 cluster sample is
$\sim 4$\%.

We combine the uncertainty from cooling flow and non-cooling flow (see
\S\ref{subsubsec:Tgrad}) departures from isothermality into a
conservative $\pm 10$\% effect on the inferred Hubble parameter.

\subsubsection{Clumping; Small Scale Structure}
\label{subsubsec:clumping}

Clumping of the intracluster gas is a potentially serious source of
systematic error in the determination of the Hubble constant.
Unresolved clumps in an isothermal intracluster plasma will enhance
the X-ray emission by the factor
\begin{equation}
C \equiv \frac{\left <n_e^2 \right >}{\left <n_e \right >^2}.
\label{eq:clump_iso}
\end{equation}
The cluster generates more X-ray emission than expected from a uniform
ICM leading to an underestimate of the angular diameter distance ($\Da
\propto S_x^{-1}$) and therefore an overestimate of the Hubble
parameter for $C > 1$.  Unlike the orientation bias which averages
down for a large sample of clusters, clumping must be measured in each
cluster or estimated for an average cluster.  Theoretical estimates of
$C$ are difficult because they must account for the complicated
processes that both generate and damp density enhancements, such as
preheating and gas-dynamical processes.

There is currently no observational evidence of significant clumping
in galaxy clusters.  If clumping were significant and had large
variations from cluster to cluster, we might expect larger scatter
than is seen in the Hubble diagrams from SZE and X-ray distances
(Figure~\ref{fig:da_z}; see also \citealt{birkinshaw1999}).
Gas-dynamical cluster simulations provide an opportunity to test the
effects of observing strategy and cluster structure on our distance
determinations.  These simulated clusters exhibit X-ray merger
signatures consistent with those observed in real clusters and,
presumably, they exhibit the appropriate complexities in their
temperature structure as well.  Preliminary work indicates that
temperature profiles do not introduce a large error on our distances
but that clumping in the ICM may bias distances low by up to $\sim
20$\%.  As mentioned above, there is no observational evidence of
clumping within the ICM.  However, merger signatures are common
\citep{mohr1995}, and the mergers are the driving mechanism behind
these fluctuations in the simulated clusters \citep{mathiesen1999}.

Clumping causes an overestimate of \Ho, so we include a conservative
one-sided $-20$\% possible systematic due to clumping.

\vspace{1in}
%%%%%
%%  VI.2 Possible SZE Contaminants
%%%%%
\subsection{Possible SZE Contaminants}
\label{subsubsec:sys_sze}

\subsubsection{Possible Undetected Point Sources in the Field}
\label{subsubsec:sys_ptsrc}

Undetected point sources near the cluster center mask the central
decrement, causing an underestimate in the magnitude of the decrement
and therefore an underestimate of the angular diameter distance.  The
synthesized beam shapes, which include negative sidelobes, allow both
underestimates and overestimates in the magnitude of the decrement.
As a conservative estimate of our detection threshold, we use 3 times
the rms of the high reslution map, applying a $\geq 2000\ \lambda$ cut
on the baselines for each cluster data set.  Placing a point source
with flux equal to our detection limit near the cluster center and
re-analyzing to find the change in the central decrement provides an
estimate of the upper bound of the effects of undetected radio point
sources.

We have additional information on the distribution of point sources in
all our cluster fields from observations at lower frequencies.
Sources with flux densities greater than 2 mJy at 1.4 GHz appear in
the NVSS catalog \citep{condon1998}.  We use the NVSS catalog to find
point sources within 400\arcsec\ of each cluster center.  We
extrapolate the NVSS sources in our fields to 28.5 GHz using the
average spectral index of radio sources in galaxy clusters $\alpha =
0.77$ \citep{cooray1998a}, where $S_\nu \propto \nu^{-\alpha}$.
Extrapolated NVSS sources with fluxes greater than our 3 $\sigma$
threshold are ruled out by the 30 GHz data and their fluxes are fixed
at the maximal 30 GHz 3 $\sigma$ value.  The extrapolated NVSS sources
are added to to the 30 GHz visibilities data, which is re-analyzed,
not accounting for the additional point sources.  The uncertainty on
the angular diameter distance from undetected point sources is
summarized in Table~\ref{tab:nodetpt} for each cluster field.  The
average over the 18 cluster fields yields a $\sim 12$\% uncertainty on
the Hubble parameter.

We know that clusters have central dominant (cD) galaxies, which are
often radio bright.  Therefore it is likely that there is a radio
point source near the center of each cluster.  To estimate the effects
of cD galaxies on the central decrement we pick three clusters for
which we do not detect a central radio point source, A697, A2261, and
A1413.  We add a point source fixed at the optical position of the cD
\citep{crawford1999} and vary both the flux of the cD galaxy and the
central decrement, keeping the ICM shape parameters fixed at their
best-fit values.  The cD fluxes are all consistent with zero and the
corresponding changes in the central decrement are $\lsim 2$\%.  This
suggests that undetected cD galaxies do not contribute significantly
to the uncertainty on the Hubble constant, $\lsim 4$\%, within our
uncertainty budget for possible undetected point sources.

%%%%%
% Table: Effects of Undetected Point Sources on \Da
%%%%%

%% Effects of undetected point sources
%%
%%	        BIMA          OVRO
%%          -----------    ----------  --------
%% cluster, 3sig, nvss,    3sig, nvss  Best(min)
%%
%%

\begin{deluxetable}{lc}
%\singlespace
%\footnotesize
%\rotate
\tablewidth{0pt}
%\tablenum{}
\tablecolumns{7}
%\tableheadfrac{}
\tablecaption{Effects of Undetected Point Sources on \Da\
              (\%)\label{tab:nodetpt}} 
\tablehead{
\colhead{Cluster} &
\colhead{Effect (\%)}
}
\startdata
MS1137			& \phn 0\\
MS0451			& \phn 0\\
CL0016			& \phn 2\\
R1347\tablenotemark{a}  &     32\\
A370			& \phn 6\\
MS1358\tablenotemark{b}	& \phn 0\\
A1995			& \phn 6\\
A611\tablenotemark{b}	& \phn 0\\
A697			& \phn 6\\
A1835			& \phn 0\\
A2261\tablenotemark{a}  &     46\\
A773			& \phn 6\\
A2163\tablenotemark{a}  &     22\\
A520\tablenotemark{a}	&     16\\
A1689\tablenotemark{a}	&     18\\
A665\tablenotemark{a}   &     12\\
A2218			& \phn 0\\
A1413			&     42\\
%%\cline{1-8}		 		      
\ \ \ Average & \phn 12\\
\enddata
\tablenotetext{a}{Required 30 GHz 3 $\sigma$ truncation.}
\tablenotetext{b}{No NVSS sources in the cluster field.}
%\tablecomment{}
\end{deluxetable}

\vspace{1in}
\subsubsection{Kinetic SZE}
\label{subsubsec:kinetic_sze}

Cluster peculiar velocities with respect to the CMB introduce an
additional CMB spectral distortion known as the kinetic SZE.  The
kinetic SZE is proportional to the thermal effect but has a different
spectral signature so it can be distinguished from the thermal SZE
with spectral SZE observations.  For a 10 keV cluster with a
line-of-sight peculiar velocity of 1000 \kms, the kinetic SZE is $\sim
11$\% of the thermal SZE at 30 GHz.  Watkins
(1997)\nocite{watkins1997} presented observational evidence suggesting
a one-dimensional rms peculiar velocity of $\sim 300$ \kms\ for
clusters, and recent simulations found similar results
\citep{colberg2000}.  With a line-of-sight peculiar velocity of 300
\kms\ and a more typical 8 keV cluster, the kinetic SZE is $\sim 4$\%
of the thermal effect, introducing up to a $\sim \pm 8$\% correction
to the angular diameter distance computed from one cluster.  When
averaged over an ensemble of clusters, the effect from peculiar
velocities should cancel, manifesting itself as an additional
statistical uncertainty similar to the effects of asphericity.
Therefore, we include a 2\% ($=8/\sqrt{18}$) effect from the kinetic
SZE for our 18 cluster sample.

\subsubsection{CMB Primary Anisotropies}
\label{subsubsec:cmb}

CMB primary anisotropies have the same spectral signature as the
kinetic SZE.  Recent BIMA observations provide limits on primary
anisotropies on the scales of the observations presented here
\citep{dawson2001, holzapfel2000}.  They place a 95\% confidence upper
limit to the primary CMB anisotropies of $\Delta T < 19$ $\mu$K at
$\ell \sim 5500$ ($\sim 2\arcmin$ scales).  Thus primary CMB
anisotropies are an unimportant ($\lsim 2$\%) source of uncertainty
for our observations.  At 68.3\% confidence, primary CMB anisotropies
contribute a $\lsim 1$\% uncertainty in the measured Hubble parameter.
In addition, CMB primary anisotropy effects on the inferred \Ho\
should average out over the sample; with an 18 cluster sample CMB
primary anisotropy contributes $< 1$\% uncertainty to \Ho.

% Note:
%      \Delta T = sqrt(12/5) Q_flat
%

\subsubsection{Radio Halos}
\label{subsubsec:radio_halos}

The SZE decrement may be masked by large scale diffuse non-thermal
radio emission in clusters of galaxies, known as radio halos.  If
present, radio halos are located at the cluster centers, have sizes
typical of galaxy clusters, and a steep radio spectrum $\alpha \sim
1-3$ \citep{kempner2001, giovannini1999, moffet1989, hanisch1982}.
Similar structures at the cluster periphery, usually with an irregular
shape, are called relics.  In general, radio halos and relics are rare
phenomena that are present in rich, massive clusters, characterized by
high X-ray luminosity and high temperature \citep{giovannini2000,
giovannini1999}.  Cooling flow clusters rarely contain radio halos.
Because halos and relics are rare, little is known about their nature
and origin but they are thought to be produced by synchrotron emission
from an accelerated or reaccelerated population of relativistic
electrons \citep[e.g.,][]{jaffe1977, dennison1980, roland1981,
schlickeiser1987}.  Shocks from cluster mergers may be the
acceleration mechanism though there are numerous theories
\citep[e.g.,][]{jaffe1977, jaffe1980, dennison1980, roland1981,
schlickeiser1987, ensslin1998, blasi1999, dolag2000, liang2000}.

According to the literature \citep{kempner2001, giovannini2000,
giovannini1999}, the following clusters in our sample exhibit radio
halos: Cl0016, A773, A2163, A520, A665, A2218.  In
Figure~\ref{fig:nvss_halo}, we show NVSS 1.4 GHz image contours
\citep{condon1998} overlaid on color scale images of the SZE cluster
emission.  Contours are multiples of twice the rms in the NVSS maps
(rms $\sim 0.45$ \mJybm).  It is apparent that many of these halos are
at the 2 $\sigma \approx 0.9$ \mJybm\ level.  The brightest known halo
is seen in A2163 with a peak brightness of $\sim 5.4$ \mJybm.  

%%%%%
% NVSS overlay on high resolution 30 GHz images
%%%%%
%%
% Figure 9
%%

\begin{figure*}[!ptbh]
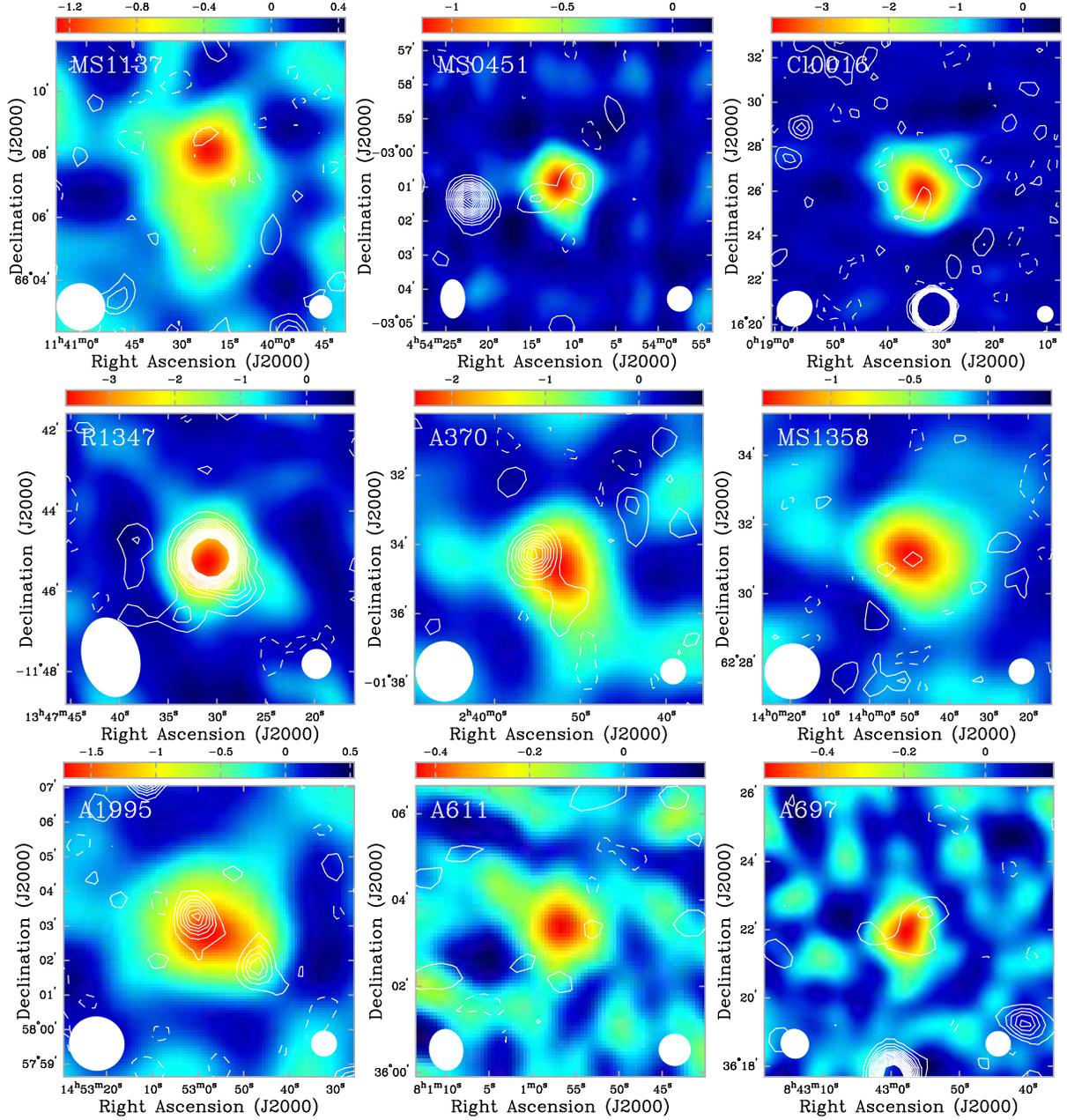

%\epsfxsize = 7.5 in
%%  \epsfxsize = 7.5 in
%%  \epsfysize = 6.0 in
%\centerline{\epsfbox{f?.eps}} 
\centerline{
  \psfig{figure=f9a_color.ps,height=2.2in}
  \psfig{figure=f9b_color.ps,height=2.2in}
  \psfig{figure=f9c_color.ps,height=2.2in}
}
\centerline{
  \psfig{figure=f9d_color.ps,height=2.2in}
  \psfig{figure=f9e_color.ps,height=2.2in}
  \psfig{figure=f9f_color.ps,height=2.2in}
}
\centerline{
  \psfig{figure=f9g_color.ps,height=2.2in}
  \psfig{figure=f9h_color.ps,height=2.2in}
  \psfig{figure=f9i_color.ps,height=2.2in}
}
\caption[30 GHz SZE and 1.4 GHz overlays]{SZE from our own 30 GHz
observations (color scale) with NVSS 1.4 GHz contours.  The 30 GHz
images are the color scale version of the SZE shown in
Figure~\ref{fig:image}.  The color scale wedge above each image shows
the range in the flux density of the 30 GHz map in units of mJy
beam$^{-1}$.  Contours are multiples of twice the rms and the NVSS rms
is $\sim 0.45$ \mJybm.  The FWHM of the 30 GHz synthesized beam is
shown in the lower left hand corner of each panel and the 45\arcsec\
FWHM beam of the NVSS survey is shown in the lower right hand corner
of each panel.  A2163 exhibits the brightest 1.4 GHz radio halo in our
18 cluster sample.}
\label{fig:nvss_halo}
\end{figure*}

\begin{figure*}[!ptbh]
\centerline{
  \psfig{figure=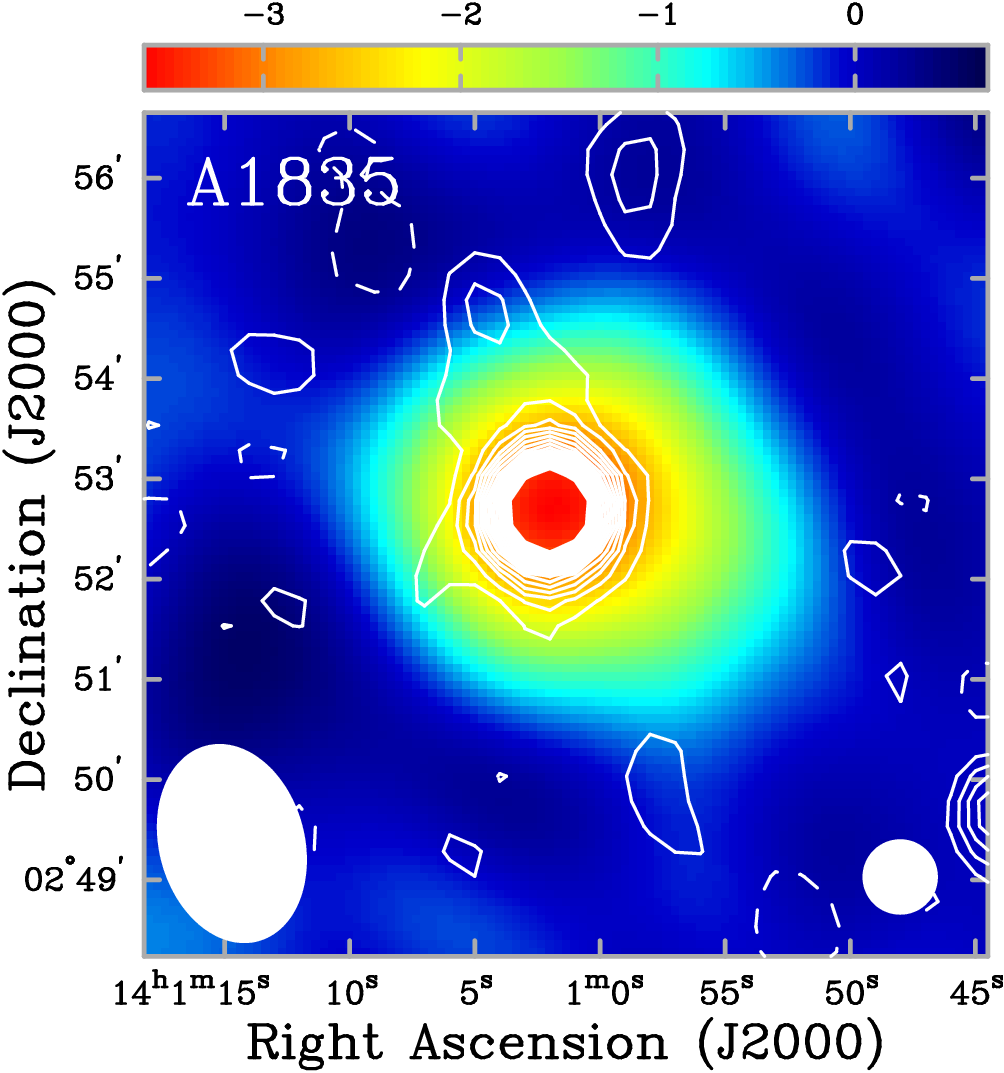,height=2.2in}
  \psfig{figure=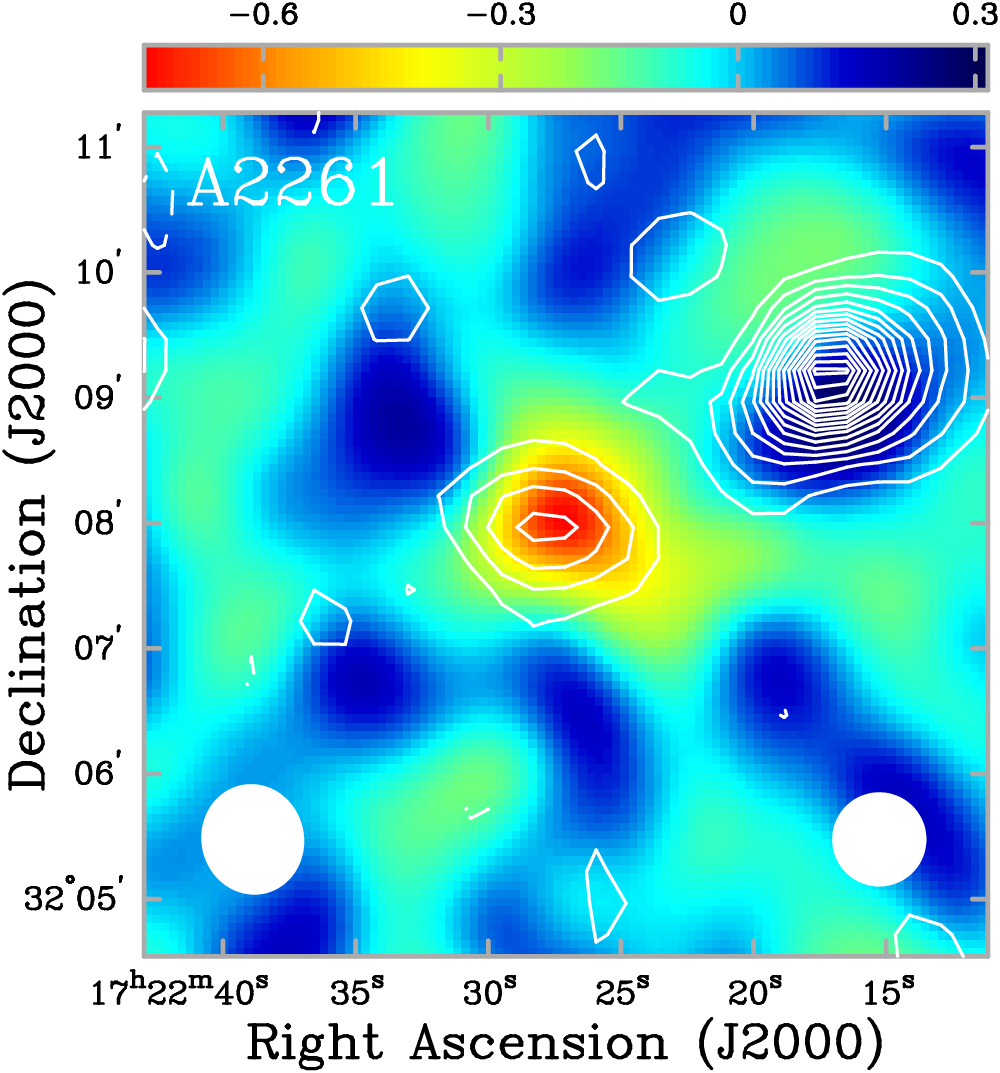,height=2.2in}
  \psfig{figure=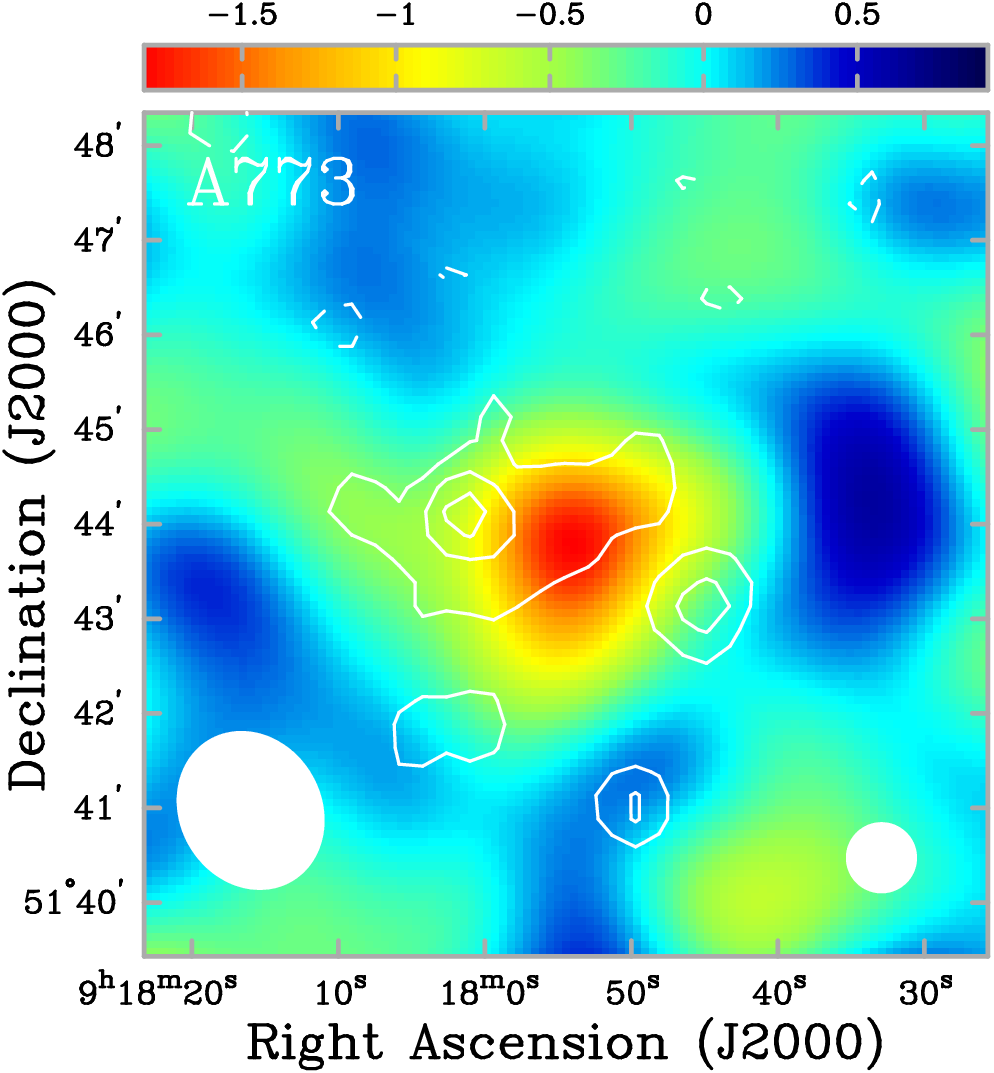,height=2.2in}
}
\centerline{
  \psfig{figure=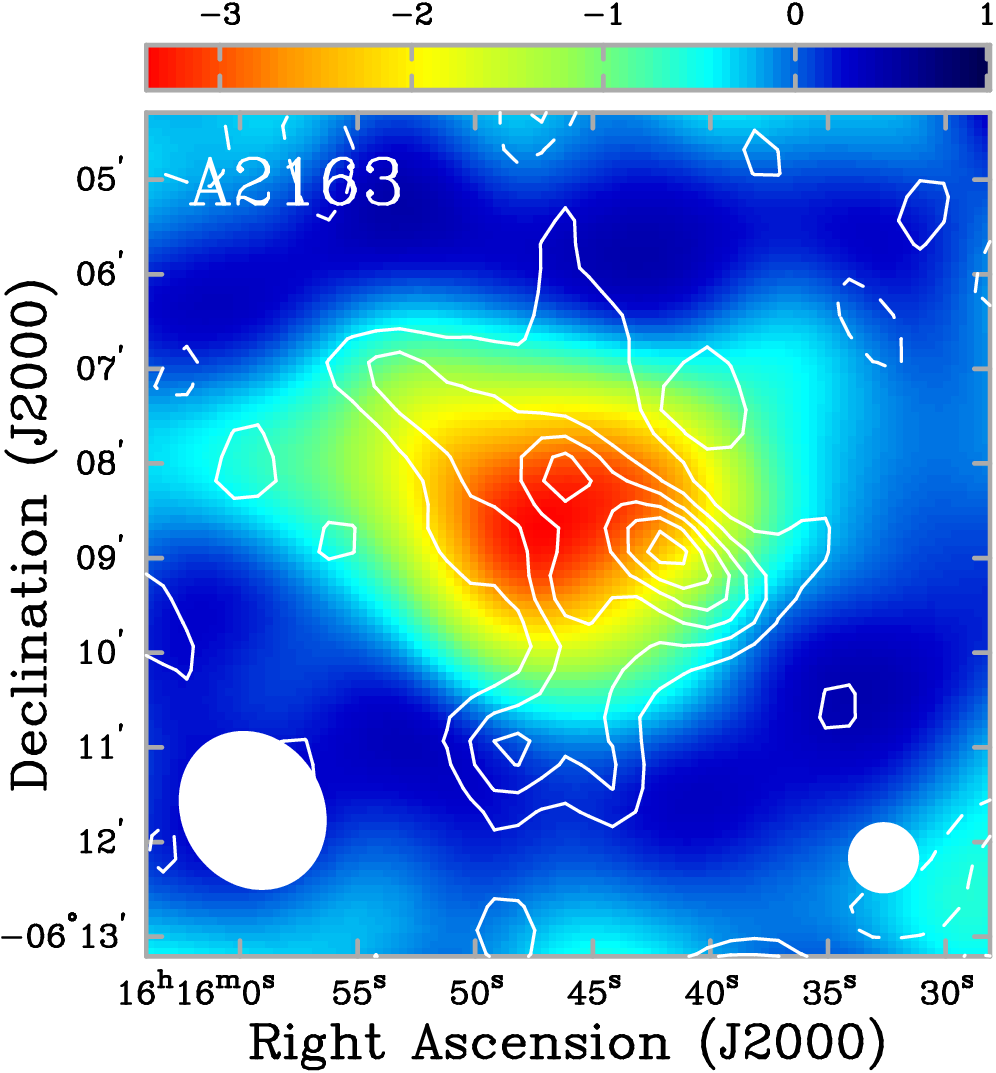,height=2.2in}
  \psfig{figure=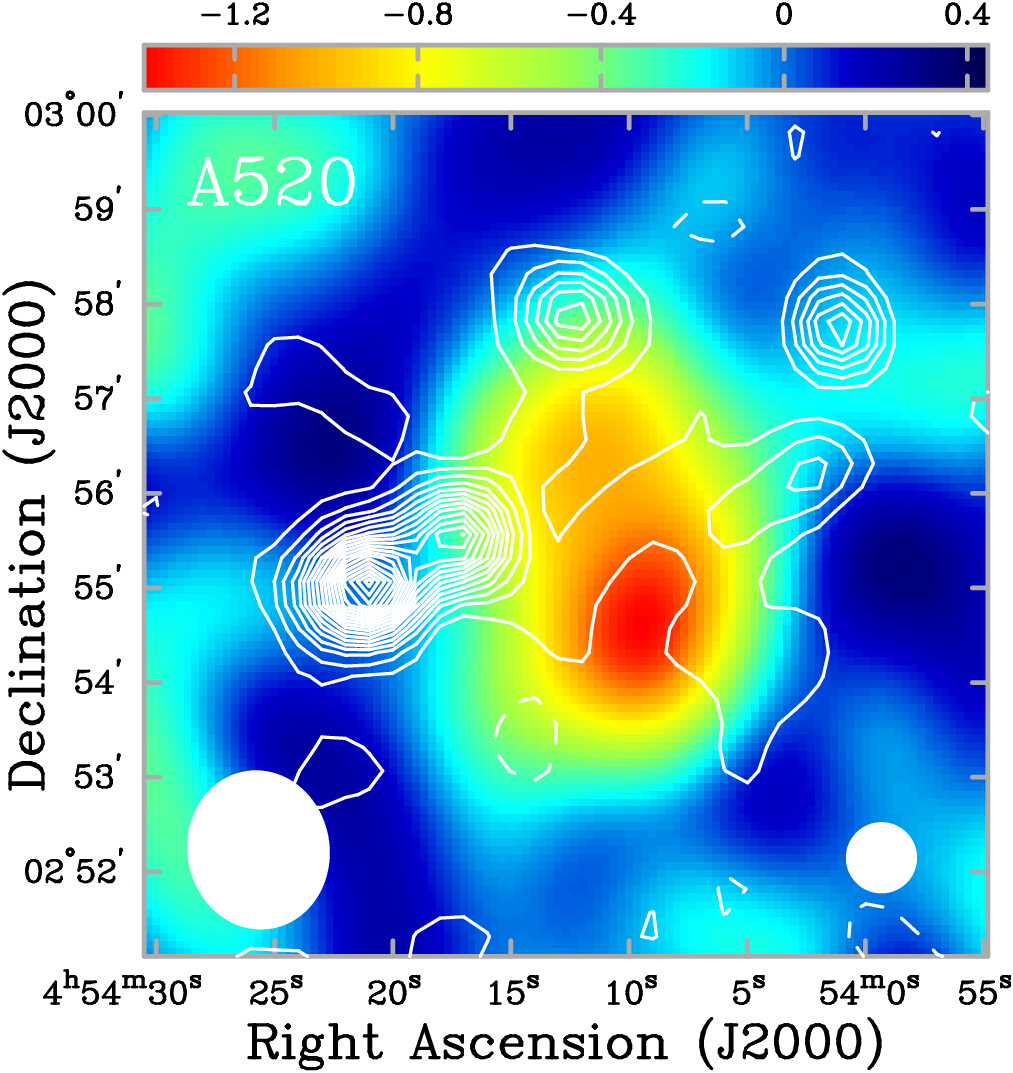,height=2.2in}
  \psfig{figure=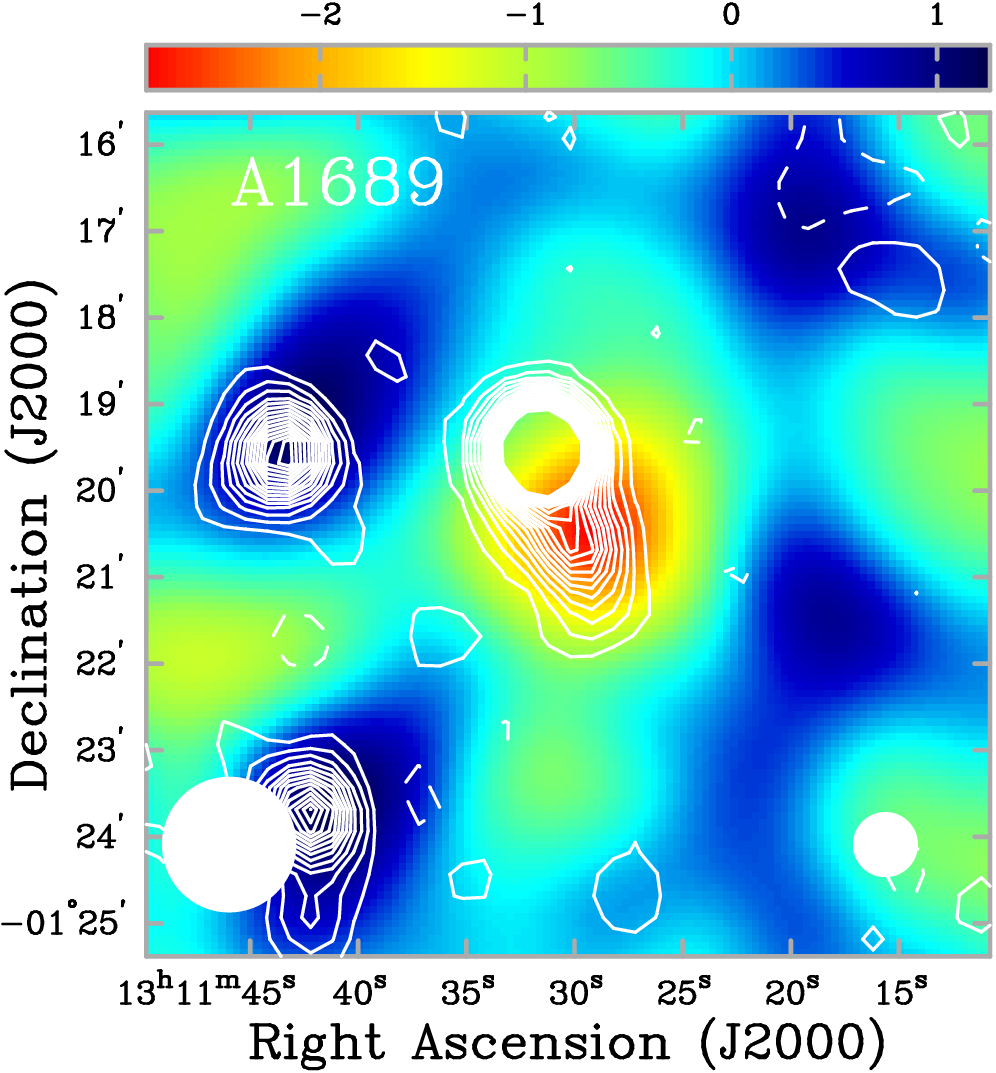,height=2.2in}
}
\centerline{
  \psfig{figure=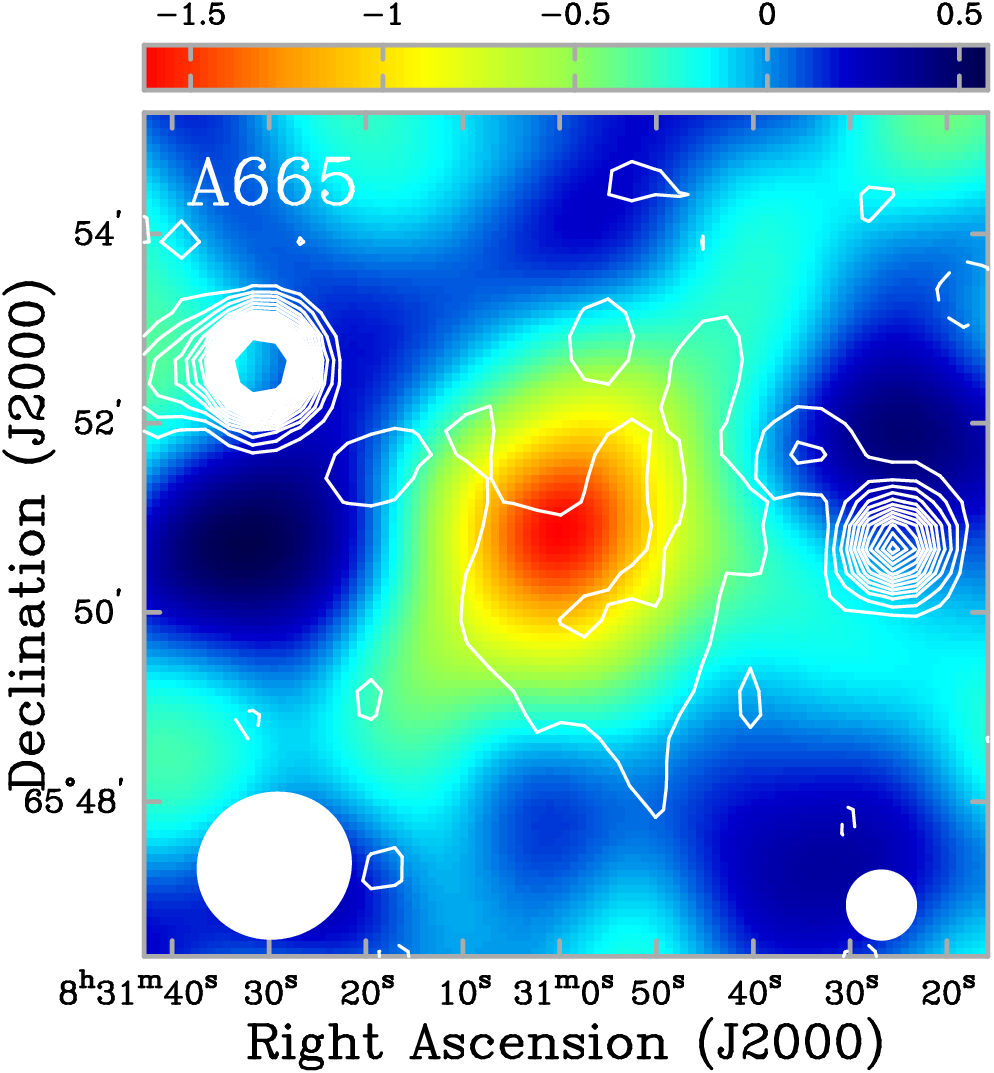,height=2.2in}
  \psfig{figure=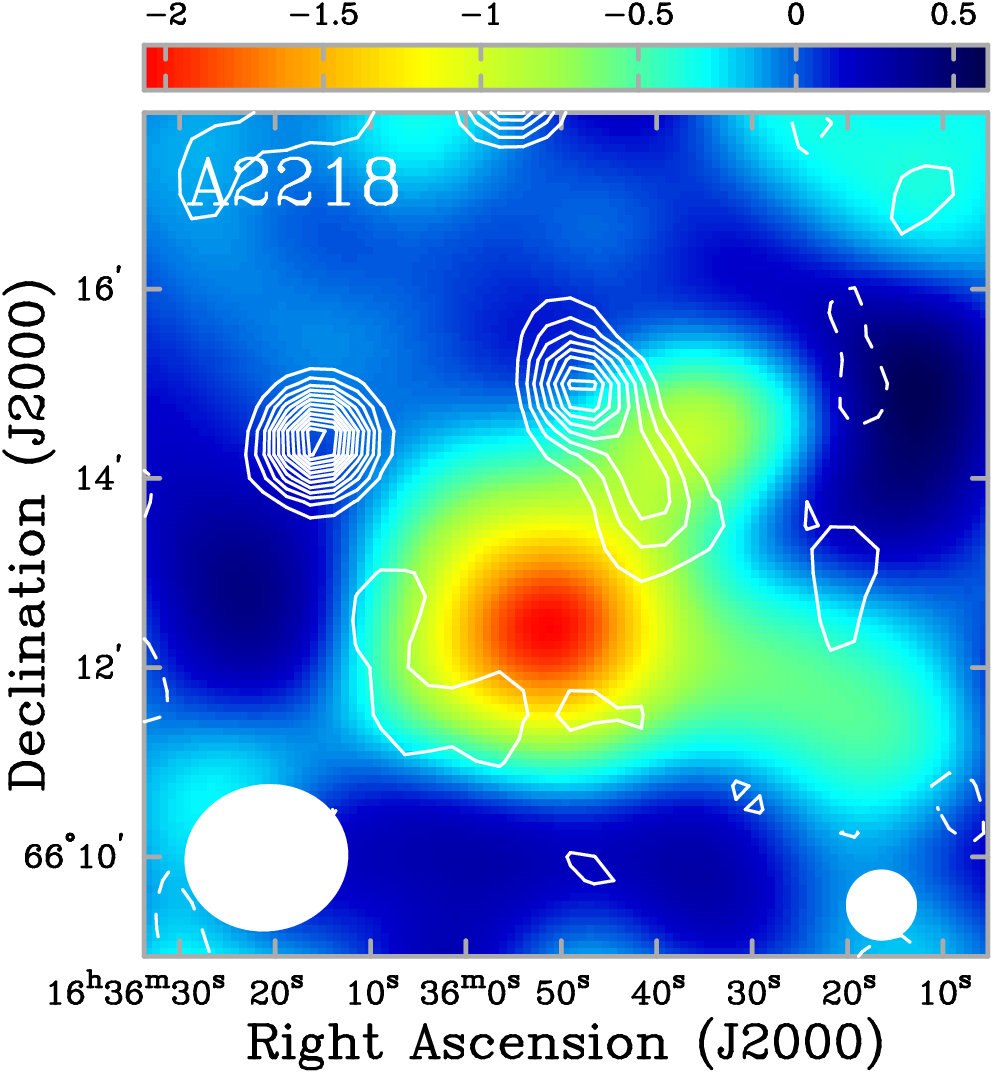,height=2.2in}
  \psfig{figure=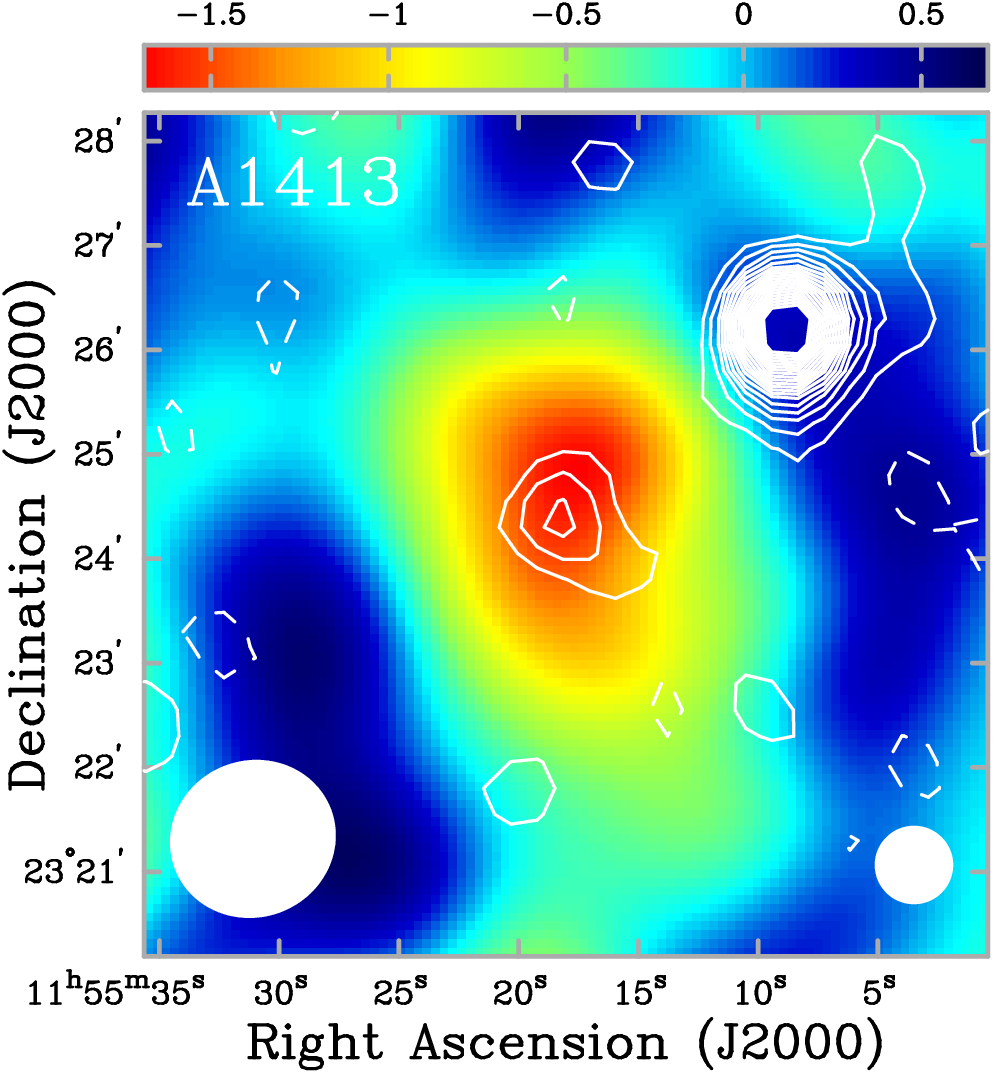,height=2.2in}
}
\contcaption{Cont.}
\end{figure*}

For each of the halo clusters, we conservatively model the halo as a
point source at the cluster center with flux from an $\alpha = 1$
extrapolation of the peak NVSS halo flux and re-analyze to determine
the effect on the central decrement.  The average effect on the
central decrement from the clusters with radio halos in our sample is
4\%, excluding A2163 which shows a $\sim 10$\% effect on the central
decrement.  Radio halos typically have spectral indices $\alpha \geq
1.5$, making the $\alpha = 1$ extrapolation a conservative upper bound
for the effects of radio halos.  Averaged over the entire 18 cluster
sample, the $\alpha = 1$ extrapolation results imply a $\sim 3$\%
overestimate ($-3$\% effect) on our inferred Hubble parameter ($\Ho
\propto \Delta T^{-2}$) from radio halos.

\subsubsection{Imprecisely Measured Primary Beam}
\label{subsubsec:primary_beam}

The primary beam is determined from holography measurements at both
OVRO and BIMA.  The effect of the primary beam on the
interferometric observations is a convolution in the Fourier plane
(see \S\ref{subsubsec:cmsystem}) or equivalently, an attenuation
across the field of view.  Therefore, differences in primary beam
shape simply alter the smoothing kernel in the \uv\ plane only
slightly, having a small affect on the derived distances.

To assess the effects of the primary beam quantitatively, we fit to an
OVRO and BIMA data set using a cluster model attenuated by a Gaussian
beam with different assumed FWHMs.  There is no significant change in
the central decrements when using a Gaussian approximation for the
primary beam instead of the measured beam.  Even with an
unrealistically large $\pm 0.2\arcmin$ uncertainty in the primary beam
FWHM, the uncertainty introduced in the Hubble constant is $\lsim 3$\%
($\lsim 2$\% in \dTo).  Artificially broadening the wings of the real
beam has a negligible effect on the derived central decrements.  We
adopt a 3\% uncertainty in \Ho\ as a conservative estimate of the
maximum effects from an imprecisely measured primary beam.

%%%%%%%%%%%%%%%%%%%%%%%%%%%%%%%%%%%%%%%%%%
%% VI.  Discussion and Conclusion
%%%%%%%%%%%%%%%%%%%%%%%%%%%%%%%%%%%%%%%%%%

\vspace{1in}
\section{Discussion and Conclusion}
\label{sec:disc_concl}

We perform a maximum-likelihood joint fit to centimeter-wave
interferometric SZE and \rosat\ X-ray (PSPC and HRI) data to constrain
the ICM parameters for a sample of high redshift clusters of galaxies.
We model the ICM as a spherical, isothermal $\beta$ model.  From this
analysis we determine the distances to 18 galaxy clusters.  Together,
these distances imply a Hubble constant of
\begin{equation}
\Ho = \cases{
	60 ^{+4}_{-4} \, ^{+13}_{-18} \ \ksM; &\Om=0.3, \Ol=0.7, \cr
	56 ^{+4}_{-4} \, ^{+12}_{-17} \ \ksM; &\Om=0.3, \Ol=0.0, \cr
	54 ^{+4}_{-3} \, ^{+12}_{-16} \ \ksM; &\Om=1.0, \Ol=0.0, \cr
}
\label{eq:Horesult_sys}
\end{equation}
where the uncertainties are statistical followed by systematic at
68.3\% confidence.  The systematic uncertainties have been added in
quadrature and include an 8\% (4\% in \dTo) uncertainty from the
absolute calibration of the SZE data, a 10\% effective area
uncertainty for the PSPC and HRI, a 5\% uncertainty from the column
density, a 5\% ($\simeq 20/\sqrt{18}$) uncertainty due to asphericity,
a 10\% effect for our assumptions of isothermality, a one-sided
$-20$\% effect from possible small-scale clumping in the ICM, a 12\%
uncertainty from undetected radio sources, a 2\% ($\simeq
8/\sqrt{18}$) uncertainty from the kinetic SZE, a 1\% uncertainty from
primary CMB anisotropies, a $-3$\% effect from radio halos, and a 3\%
effect from an imprecisely measured primary beam.  These systematic
uncertainties are summarized in Table~\ref{tab:Hosyserr}.  We adopt
conservative assumptions when gauging the effects of possible
systematics.  The contributions from asphericity, kinetic SZE, and
primary CMB are expected to average out for a large sample.

The measured distances plotted in Figure~\ref{fig:da_z} with the
theoretical relation show that A370 is the largest outlier from the
theoretical angular diameter distance relation and MS1137 has the
largest distance uncertainty.  A370 exhibits an almost 2-to-1 axial
ratio in the knotty N-S elongation of its X-ray image (see
Fig~\ref{fig:image}).  An optical study of this cluster and its member
galaxies shows that the cluster is dominated by two giant elliptical
galaxies with a projected separation of about 40\arcsec, roughly in
the north-south direction \citep{mellier1988}.  In addition,
gravitational lens models suggest that A370 has a bimodal mass
distribution with the two components separated in a roughly
north-south direction \citep{kneib1993, soucail1988, smail1996}.  The
spherical model used is clearly insufficient for the complex structure
of this cluster.  The uncertainties on the distance to MS1137 are
particularly large since MS1137 resides in the distant universe ($z =
0.78$), making it difficult to collect large numbers of X-ray photons.
Therefore, the uncertainties on the X-ray driven quantities are large;
in particular, the uncertainty on the measured X-ray temperature is
the main contributor to the large uncertainty on the angular diameter
distance to MS1137.  The large uncertainties on the distances to both
A370 and MS1137 mean that those clusters contribute little weight in
the determination of the Hubble constant.

As discussed in \S\ref{sec:sample}, target clusters were originally
chosen from a limited sample of known X-ray clusters.  We construct
subsamples of our cluster sample to explore the robustness of our
result and to look for possible biases in our \Ho\ determination.  A
description of the subsamples, the number of clusters in each
subsample $N$, the Hubble constant from each subsample, and the
$\chi^2$ and reduced $\chi^2$ for the Hubble parameter are summarized
in Table~\ref{tab:subsamples} for each subsample considered.  Only the
$\Om = 0.3$, $\Ol = 0.7$ cosmology is considered in this study.  Other
cosmologies will have similar changes to the best fit Hubble
parameter.  Table~\ref{tab:subsamples} shows that excluding the
largest outlier, A370, has a negligible effect on the determined Hubble
parameter.  We also split the sample up based on having a cooling
flow, by redshift, presence of a radio halo, based on point sources,
based on right ascension and declination, based on X-ray luminosity,
and based on membership in the EMSS survey or being an Abell cluster.

%%%%%
% Table: \Ho from different Subsamples for the \Labmda Cosmology
%%%%%

%%H0 from Different Subsamples
%%
%%Subsample  H0
%%
\begin{deluxetable}{lcccc}

\tablewidth{0pt}
%\tablenum{}
\tablecaption{\Ho\ from Different Subsamples for the $\Lambda$ Cosmology
	\label{tab:subsamples}}
\tablehead{
\colhead{} &
\colhead{} &
\colhead{\Ho} &     
\colhead{} &
\colhead{}
\\
\colhead{Subsample} & 
\colhead{$N$} & 
\colhead{(\ksM)}  &
\colhead{$\chi^2$} &
\colhead{$\chi^2_{red}$\tablenotemark{a}}
}
\tablecolumns{5}
\startdata
All			&  18	&$60^{+4\phn}_{-4\phn}$	& 16.5	   & 0.97\\
\cline{1-5}
No A370			&  17	&$61^{+4\phn}_{-4\phn}$	& 10.8	   & 0.68\\
%%No MS1137		&  17	&$61^{+4\phn}_{-4\phn}$	& 15.4	   & 0.96\\
No MS1137 or A370	&  16	&$61^{+5\phn}_{-4\phn}$	&\phn 9.8  & 0.65\\
\cline{1-5}
No cooling flow		&  11	&$59^{+6\phn}_{-5\phn}$	& 14.5	   & 1.45\\
Only cooling flow	&\phn 7	&$63^{+7\phn}_{-6\phn}$	&\phn 1.7  & 0.28\\
\cline{1-5}
%%$z<0.2$			&\phn 4	&$67^{+12}_{-9\phn}$	&\phn 2.4  & 0.80\\
%%$0.2<z<0.3$		&\phn 7	&$56^{+6\phn}_{-5\phn}$	&\phn 3.3  & 0.55\\
%%$z>0.3$			&\phn 7	&$62^{+8\phn}_{-6\phn}$	&\phn 9.7  & 1.62\\
$z>0.27$		&\phn 9	&$62^{+7\phn}_{-6\phn}$	&\phn 9.7  & 1.21\\
$z<0.27$		&\phn 9	&$59^{+6\phn}_{-5\phn}$	&\phn 6.6  & 0.83\\
\cline{1-5}
No point sources	&\phn 4	&$53^{+9\phn}_{-7\phn}$	&\phn 3.4  & 1.13\\
Only point sources	&  14	&$62^{+5\phn}_{-4\phn}$	& 12.2	   & 0.94\\
\cline{1-5}
Only radio halo		&\phn 6 &$56^{+8\phn}_{-6\phn}$ &\phn 6.7  & 1.34\\
No halo			&  12   &$62^{+5\phn}_{-5\phn}$ &\phn 9.3  & 0.85\\
\cline{1-5}
RA$<11.5^{\mbox{h}}$	&\phn 9	&$61^{+7\phn}_{-6\phn}$	& 13.6     & 1.70\\
RA$>11.5^{\mbox{h}}$	&\phn 9	&$60^{+6\phn}_{-5\phn}$	&\phn 2.9  & 0.36\\
Dec$>30^\circ$		&\phn 9	&$60^{+7\phn}_{-6\phn}$	&\phn 8.3  & 1.04\\
Dec$<30^\circ$		&\phn 9	&$61^{+6\phn}_{-5\phn}$	&\phn 8.1  & 1.01\\
\cline{1-5}
$\Lx>14.5 \times 10 ^{44}$ erg s$^{-1}$
			&\phn 9	&$63^{+6\phn}_{-5\phn}$	&\phn 4.4  & 0.55\\
$\Lx\leq 14.5 \times 10 ^{44}$ erg s$^{-1}$
			&\phn 9	&$57^{+7\phn}_{-5\phn}$	& 11.5     & 1.44\\
%%$\Lx \geq 20 \times 10 ^{44}$ erg s$^{-1}$
%%			&\phn 6	&$62^{+7\phn}_{-5\phn}$	&\phn 1.7  & 0.34\\
%%$(10<\Lx< 20) \times 10 ^{44}$ erg s$^{-1}$
%%			&\phn 9	&$60^{+7\phn}_{-6\phn}$	& 12.8     & 1.60\\
%%$\Lx< 10 \times 10 ^{44}$ erg s$^{-1}$
%%			&\phn 3	&$52^{+14}_{-9\phn}$	&\phn 1.5  & 0.75\\
\cline{1-5}
EMSS clusters		&\phn 5	&$66^{+10}_{-8\phn}$	&\phn 3.5  & 0.88\\
\cline{1-5}
Abell clusters		&  13	&$58^{+5\phn}_{-4\phn}$	& 12.2     & 1.02\\
\cline{1-5}
Orientation unbiased	&\phn 6	&$58^{+8\phn}_{-6\phn}$	&  6.2     & 1.24\\
\enddata
\tablenotetext{a}{$\chi^2_{red}=\chi^2 / (N-1)$, where $N-1$ is the 
degrees of freedom.}
%\tablecomment{}

\end{deluxetable}

\citet{jones2001} constructed an orientation unbiased sample of galaxy
clusters for a SZE and X-ray determination of the Hubble parameter.
Most importantly, they drew clusters present in both the BCS
\citep{ebeling2000a, crawford1999, ebeling1998, ebeling1997} and NORAS
\citep{bohringer2000} surveys with $\Lx > 8 \times 10^{44}$ erg
s$^{-1}$ and well above the survey flux limits ($> 5 \times 10^{-12}$
erg s$^{-1}$ cm$^{-2}$).  Eleven clusters satisfy these criterion as
well as their redshift range choice ($0.14 \leq z \leq 0.30$) and
declination constraint ($\geq 2^\circ$).  They find five of the eleven
to be sufficiently free from point sources at 15 GHz for SZE
measurements.  Six of these eleven are part of our sample, which we
call our orientation unbiased subsample in Table~\ref{tab:subsamples}.

All of these subsamples yield a Hubble constant with 1
$\sigma$ statistical uncertainties consistent with the \Ho\ from the
entire 18 cluster sample.  This argues in favor of a robust \Ho\
determination.

We compare our results with other SZE determined distances to clusters
in our sample in Table~\ref{tab:da_compare}.  Only statistical
uncertainties are included.  There are nine clusters in our sample
that also have previously determined SZE distances.  All of the 1
$\sigma$ confidence regions agree with our own, with the exception of
Cl0016, R1347, A773, and A665.  The systematic uncertainties on the
angular diameter distance are $\gsim 30$\% for one galaxy cluster.
Therefore, all of the distances are in reasonably good agreement, even
after accounting for shared systematics (namely most use ROSAT X-ray
data and ASCA X-ray temperatures).

%%%%%
% Table: Comparison of SZE Determined Distances
%%%%%

%%SZE Determined Distance Comparison
%%
%%
%%
\begin{deluxetable}{lcl}
\tablewidth{0pt}
%\tablenum{}
\tablecaption{Comparison of SZE Determined Distances
\label{tab:da_compare}}
\tablehead{
\colhead{} &
\colhead{\Da\tablenotemark{a}} &
\colhead{}\\
\colhead{Cluster} &
\colhead{(Mpc)} &
\colhead{reference}
}
\tablecolumns{3}
\startdata
Cl0016    & $2041 \pm 499$	& This work\\ %^{+484}_{-514}
	  & $1788 \pm 664$      & H98\\
	  & $1100 \pm 295$      & G02\\  % need to update uncertainty
R1347     & $1221 \pm 356$	& This work\\ %^{+368}_{-343}
	  & $1890 \pm 644$      & P01\\
	  & $1897 \pm 401$      & K99\\
A697	  & $\phn 998 \pm 274$	& This work\\ %^{+298}_{-250}
	  & $1044 \pm 239$      & J02\\  % need to update uncertainty
A1835     & $1027 \pm 196$	& This work\\ %^{+194}_{-198}
	  & $\phn 867 \pm 411$  & M00\\
A773 	  & $1450 \pm 347$	& This work\\ %^{+361}_{-332}
	  & $1002 \pm 257$      & S99\\
A2163     & $\phn 828 \pm 193$	& This work\\ %^{+181}_{-205}
	  & $\phn 728 \pm 387$  & H97\\
	  & $\phn 615 \pm 327$  & L98\\
A665 	  & $\phn 466 \pm 198$	& This work\\ %^{+217}_{-179}
	  & $1017 \pm 229$      & B91\\
A2218     & $1029 \pm 346$	& This work\\ %^{+339}_{-352}
	  & $\phn 616 \pm 118$  & B94\\  % need to update uncertainty
	  & $\phn 720 \pm 422$  & T98\\
	  & $1201 \pm 343$      & J02\\  % need to update uncertainty
A1413     & $\phn 573 \pm 161$  & This work\\
	  & $\phn 565 \pm 164$  & G02b
\enddata 
\tablenotetext{a}{Includes approximate 68.3\% confidence statistical
uncertainties only.} 
\tablenotetext{\phn}{REF: 	
			B91-\citealt{birkinshaw1991};
			B94-\citealt{birkinshaw1994};
			G02-\citealt{grainge2002};
			G02b-\citealt{grainge1999}			
			H97-\citealt{holzapfel1997};
			H98-\citealt{hughes1998};
			J02-\citealt{jones2001};
			K99-\citealt{komatsu1999};
			L98-\citealt{lamarre1998};
			M00-\citealt{mauskopf2000}; 
			P01-\citealt{pointecouteau2001};
			S99-\citealt{saunders1999};
			T98-\citealt{tsuboi1998};
}
\end{deluxetable}

Many of the systematics can be approached and reduced through improved
observations.  For example, Chandra and XMM-Newton are now producing
temperature profiles of galaxy clusters
\citep[e.g.,][]{markevitch2000, nevalainen2000b, tamura2001}.  The
unprecedented angular resolution of Chandra will provide insight into
possible clumping in clusters.  The effects of undetected point
sources are being addressed with multi-wavelength (5 and 8 GHz) VLA
observations of many of our cluster fields.  In addition, there is a
project to produce a $\sim 1$\% calibration of the modified OVRO and
BIMA SZE systems and the current generation of X-ray satellites will
reduce the X-ray absolute calibration uncertainty to the few percent
level.

The 18 cluster distances presented here are beginning to probe the
shape of the angular diameter distance relation.  Moreover,
constructing subsamples from our 18 cluster sample based on such
considerations as cooling flows, redshift, and X-ray luminosity, does
not significantly affect the best fit \Ho, suggesting a robust
determination of the Hubble parameter.  Systematics currently dominate
the uncertainty in our determination of the Hubble parameter.  These
systematics can and will be addressed with current radio observatories
(OVRO, BIMA, and VLA) and X-ray satellites (Chandra and XMM-Newton).
With a sample of high redshift galaxy clusters, this method can be
used to constrain the geometry of the universe, providing a valuable
independent check of the recent supernovae type Ia \citep{schmidt1998,
riess1998, perlmutter1999, goobar2000} and primary CMB power spectrum
results \citep{pryke2002, debernardis2002, stompor2001}.  We emphasize
that SZE and X-ray determined distances are independent of the
extragalactic distance latter and do not rely on clusters being
standard candles or rulers.

A complete review of other distance determination methods is beyond
the scope of this paper.  We will just touch on a few methods that
complement SZE determined distances.  The SZE derived distances are
direct, making them an interesting check of the cosmological distance
ladder.  Our measurement of \Ho\ in the distant universe, agrees
within the uncertainties with the Hubble Space Telescope (HST) \Ho\
Key Project results, which probes the nearby universe.  The HST \Ho\
Key Project finds $\Ho = 72 \pm 3 \pm 7$ \ksM\ \citep{freedman2001},
where the uncertainties are statistical followed by systematic at
68.3\% confidence.  Though few in number, there are other methods that
yield distances that are independent of the extragalactic distance
ladder.  Recent observations of masers orbiting the nucleus of the
nearby galaxy NGC 4258 \citep{herrnstein1999} illustrate a method of
determining direct distances in the nearby universe.  Time delays
produced by lensing of QSO's by galaxies are another direct distance
indicator that can probe the high-redshift universe \citep[for recent
examples, see][]{fassnacht1999, biggs1999, lovell1998, barkana1997,
schechter1997}.

SZE surveys provide a promising method of detecting high redshift
galaxy clusters \citep[e.g.,][]{holder2000, barbosa1996}.  These
surveys will provide large catalogs of high redshift galaxy clusters
required to determine the geometry of the universe from SZE and X-ray
determined direct distances.

\acknowledgments

This work is the thesis work of E.~D.~R and benefited from useful
discussions from very many people.  In particular, we thank Gilbert
Holder, Jack Hughes, Carlo Graziani, Sandeep Patel, and Thomas
Crawford for useful and stimulating discussions.  This project would
not have been possible without all of the help and support of both the
OVRO and BIMA staff over many years.  In particular, E.~D.~R would
like to thank D.\ Woody, S.\ Scott, R.\ Lawrence, J.\ R.\ Forster
(Rick), R. Plambeck (Dick), J.\ Wirth (Red), M.\ Warnock, and M.\
Masters for both helping to keep our observations going and for
teaching E.~D.~R. all about the details of each observatory.  Dave,
Rick, and Dick have been especially patient teachers.  We would also
like to thank the staff of the High Energy Astrophysics Science
Archive Research Center (HEASARC).  They always responded promptly,
politely, and informatively to our many inquiries over the past few
years.  In particular, Michael Corcoran and Michael Arida have
received the brunt of our questions and have been most helpful.

This work is supported by NASA LTSA grant NAG5-7986. E.~D.~R
acknowledges support from NASA GSRP Fellowship NGT5-50173 and the NASA
Chandra Postdoctoral Fellowship PF1-20020.  This research has made use
of data obtained through the High Energy Astrophysics Science Archive
Research Center Online Service, provided by the NASA/Goddard Space
Flight Center.  This work has also made use of the online NVSS and
FIRST catalogs operated by the NRAO as well as the NASA/IPAC
Extragalactic Database (NED), which is operated by the Jet Propulsion
Laboratory, California Institute of Technology, under contract with
the National Aeronautics and Space Administration.

%BiBTeX stuff

\bibliographystyle{/home/piglet/reese/latex/astronat/apj}
\bibliography{/home/piglet/reese/latex/astronat/apj-jour,/home/piglet/reese/latex/clusters}

\end{document}